\documentclass[11pt, a4paper]{article}

\usepackage{amssymb, amsmath, amsfonts, eurosym, geometry, graphicx, color, setspace, sectsty, comment, footmisc, caption, pdflscape, array, hyperref, float, verbatim, listings, commath, enumitem, xfrac, pgfplots, mathtools, tikz, multirow, diagbox, chngpage, booktabs, tabularx, dirtytalk, indentfirst, bbm, subcaption, dsfont, aeguill, amsthm, multicol, fancyhdr, stmaryrd, siunitx, longtable, adjustbox, natbib}
\usepackage{makecell}
\usepackage[utf8]{inputenc}
\usepackage[T1]{fontenc}
\usepackage[affil-it]{authblk} % For author/affiliation formatting
\usepackage{xcolor}
\usepackage{newunicodechar}
\newunicodechar{−}{\ensuremath{-}}
\usepackage{ulem}

\newcolumntype{R}[2]{%
  >{\adjustbox{angle=#1,lap=\width-(#2)}\bgroup}%
  l%
  <{\egroup}%
}
\newcommand*\rot{\multicolumn{1}{R{45}{1.2em}}}

\title{\textbf{Is Bitcoin A Hedge Against Central Banking? Evidence from AI-Driven Monetary Policy Expectations}\thanks{The opinions and analyses expressed in this paper are the responsibility of the authors and do not necessarily reflect the views of Deutsche Bank. The complete replication package for this study is publicly available at: \url{https://github.com/maximenc/Bitcoin-Monetary-Policy-Expectations}. We thank Sofia Buono, Can Dogan, Annabelle Ren Fang, and Francesca Russell for useful discussions. We thank Alexis Bogroff for helpful advice in the choice of models and code implementations. We are grateful to Théophile Legrand for his comments and suggestions. We thank Jean-Charles Bricongne and Thomas Renault and the participants of the conference of the Economic Data Science Society. We acknowledge the reviewers for their insightful comments, which significantly contributed to the improvement of this work.

\textsuperscript{\dag}Corresponding authors: \href{mailto:m.nicolas@ucl.ac.uk}{m.nicolas@ucl.ac.uk}; \href{mailto:f.sicard@ucl.ac.uk}{f.sicard@ucl.ac.uk}
}}

\author[1,2]{Maxime L. D. Nicolas\textsuperscript{\dag}}
\author[2,3]{François Sicard\textsuperscript{\dag}}
\author[2,4]{Marion Laboure}
\author[2]{Zixin Sun}
\author[5]{Anah\'{\i} Rodr\'{\i}guez-Mart\'{\i}nez}

\affil[1]{\small Institute of Finance \& Technology, University College London, UK}
\affil[2]{\small Department of Arts and Sciences, University College London, UK}
\affil[3]{\small Centre for Blockchain Technologies, University College London, UK}
\affil[4]{\small Deutsche Bank AG, London, UK}
\affil[5]{\small Department of Computer Science, University College London, UK}

\date{\today}

\begin{document}

\maketitle
\begin{abstract}
\linespread{1.1}\selectfont
\noindent

This study investigates the transmission of monetary policy narratives to Bitcoin prices, distinguishing policy expectations from realized policy implementation. We introduce a weekly Monetary Policy Expectations (MPE) index derived from the Large Language Model (LLM)-based classification of 118,000+ market messages, providing a granular measure of hawkish and dovish monetary policy discourse.
We demonstrate that changes in the MPE index provide evidence of significant linear predictive information for Bitcoin returns at short-to-medium horizons, with significant Granger causality at multiple lags. 
A Long Short-Term Memory (LSTM) framework combined with SHapley Additive exPlanations (SHAP) further identifies nonlinear and regime-dependent relationships between monetary-policy expectations and Bitcoin returns, indicating that Bitcoin functions as a sensitive barometer of central bank signaling. In particular, hawkish monetary-policy narratives are associated with negative price responses  that are not accounted for by contemporaneous Federal Funds Rate adjustments.
These findings highlight Bitcoin’s structural sensitivity to global monetary discourse, establishing LLM-derived monetary-policy sentiment as a high-frequency measure of central-bank communication and as an informative leading macroeconomic indicator for the digital asset landscape.\\

\noindent \textbf{Keywords}: Monetary Policy Expectations, Bitcoin, Cryptocurrency, Large Language Models, Machine Learning, Explainable AI.

\end{abstract}
%\tableofcontents

\newpage

\section{Introduction}

% 0. Background / ("The Hook")
% 1. Situation + Litterature Review
% 2. Problem 
% 3. Actions / Results 
% 4. Contributions
% 5. Implications
% Outline

% 0. Background / ("The Hook")
For over a decade, Bitcoin has occupied a paradoxical position in the global financial landscape, oscillating between the promise of an autonomous ``digital gold" and the reality of a hyper-volatile risk asset. While traditional literature has increasingly positioned cryptocurrencies as potential hedges against U.S. economic policy \citep{polyzos2025investigating} or safe-havens during systemic crises \citep{conlon2024enduring}, these characterizations often overlook the nuanced transmission of monetary narratives. If the mere linguistic tone of central bank communications can move traditional equity prices and credit spreads independently of realized policy actions \citep{schmeling2025does}, the extent to which digital assets are truly ``decoupled" from the central banking apparatus remains a critical, unresolved question. This paper challenges the narrative of independence by introducing a novel Large Language Model (LLM)-based classification designed to decode the latent monetary-policy sentiment within over 118,000 real-time investor messages. By measuring policy expectations separately from realized interest rate shifts, we demonstrate that Bitcoin exhibits a nonlinear structural sensitivity to hawkish discourse. Our findings suggest that far from being a detached alternative, Bitcoin is deeply embedded within the global macroeconomic narrative, reacting more to the projected ``tone" of central banks than to the fundamental mechanics of the monetary system itself.

% 1. Situation + Litterature Review
%%%%% Litterature Review

The fundamental economic nature of Bitcoin remains contested, with the academic literature divided on whether it functions as an ``inflation hedge" or a speculative risk asset in financial markets. 
Bitcoin has been characterized as a ``digital safe-haven" \citep{yae2024volatile} or an effective hedge against inflation expectations \citep{liu2024hedging}. For instance, \cite{rodriguez2025bitcoin} show that Bitcoin appreciates following inflationary shocks in a manner partially comparable to gold. Conversely, Bitcoin has been portrayed as a speculative, liquidity-sensitive risk asset whose price dynamics are tied to macrofinancial conditions and exchange-level shocks rather than monetary fundamentals alone \citep{lyocsa2020impact, corbet2020impact, nicholas2021bitcoin}. This characterization echoes earlier work by \cite{baur2018bitcoin} and \cite{de2019drivers}, which identified short-term trading as the primary driver of its price action.

Even where comparisons to gold are made, Bitcoin’s suitability as a safe-haven remains questionable. \cite{smales2019bitcoin} highlights that Bitcoin’s high volatility and lower liquidity undermine its effectiveness as a safe-haven in its current state of market maturity. The debate is further complicated by the characteristics of the market participants. \cite{auer2022distrust} find no evidence that US investors are driven by systemic distrust of fiat currencies. Instead, adoption appears driven by demographic factors, with ``digital natives" primarily seeking speculative growth rather than a stable store of value.

Beyond investor characteristics, recent empirical work highlights Bitcoin's systematic sensitivity to monetary policy. \cite{pietrzak2023can} and \cite{pyo2020fomc} demonstrate that Bitcoin reacts significantly to Fed information shocks and Federal Open Market Committee (FOMC) announcements. This reaction extends beyond macroeconomic data and policy decisions to the manner in which they are communicated. \cite{gorodnichenko2023voice} reveal that vocal emotions of Federal Reserve chairs during press conferences move financial markets, suggesting that the ``tone" of policy delivery is a critical determinant. 
Evidence across central banks jurisdictions further underscores this relationship. While \cite{karau2023monetary} finds that euro-area rate reductions affect Bitcoin prices, the results indicate that Bitcoin remains driven largely by global risk sentiment. Similarly, \cite{ma2022monetary} show that unexpected US monetary tightening leads to a persistent drop in Bitcoin prices comparable in magnitude to gold, while \cite{pyo2020fomc} identify specific price movements following FOMC announcements. Furthermore, \cite{elsayed2024international} show that while spillovers between international policy and cryptocurrencies are often muted, they sharpen significantly during periods of unconventional policy or negative shadow rates.

In line with the sensitivity to macrofinancial conditions, the effectiveness of Bitcoin as a hedge appears equally nuanced and regime-dependent. \cite{conlon2024enduring} find that while Bitcoin acts as a strong hedge for the US dollar in the short run, it may actually increase aggregate risk in the long run. Similarly, \cite{liu2024hedging} document that while cryptocurrency futures effectively hedged inflation expectations throughout early 2022, this property eroded following systemic shocks such as the Luna and FTX collapses. This mirrors findings by \cite{conlon2020safe}, who showed that Bitcoin failed to protect investors during the COVID-19 market crash. 
Evidence on inflation hedging remains equally mixed. \cite{smales2024cryptocurrency} argues that Bitcoin does not offer a viable alternative to gold, as it tends to respond negatively to CPI surprises. By contrast, \cite{selmi2018bitcoin} suggest that Bitcoin’s hedging and safe-haven properties are regime-dependent, particularly in relation to oil price turmoil, while \cite{cong2024inflation} document a positive association between inflation expectations and Bitcoin purchases at the household level in developing economies, rationalizing its adoption as a local hedge.

These findings point to a broader role for macroeconomic uncertainty in shaping Bitcoin’s price dynamics. \cite{matkovskyy2020effects} link Bitcoin volatility to Economic Policy Uncertainty (EPU), while \cite{corbet2020impact} show it reacts counter-intuitively to ``good" macro news. This complexity is further reinforced by \cite{wang2023factors} and \cite{manickavasagam2025bitcoin}, who identify the trade-weighted USD and consumer confidence as stronger drivers than traditional stress indices. 
Capturing these nonlinear and state-dependent relationships has motivated the use of machine learning approaches. \cite{carbo2024determinants}, \cite{kose2025deep}, and \cite{kim2025bitcoin} employed Long Short-Term Memory (LSTM) and Convolutional Neural Network (CNN) architectures to identify the US dollar and stock indices as primary determinants of Bitcoin price movements, while \cite{goodell2023explainable} used explainable AI (XAI) techniques to forecast regime transitions during periods of geopolitical crises. 

% 2. Problem 
Despite this progress, a critical gap remains. Existing studies primarily focus on realized policy shocks or general sentiment, offering limited insight into the role of narrative-driven expectations. Yet, as evidenced by the impact of ``vocal cues" in central bank communication \citep{gorodnichenko2023voice}, narrative-driven expectations may exert a disproportionate influence on price formation prior to actual policy implementation. While \cite{picault2022media} show that media-based sentiment contains forward-looking information, there is currently no high-frequency measure that isolates the specific impact of monetary policy expectations on Bitcoin returns. This limitation reflects the difficulty traditional models face in capturing complex, nonlinear financial narratives \citep{goodell2023explainable}. \\

% 3. Actions / Results
To address this gap, we introduce a novel, high-frequency monetary policy expectations (MPE) index. Unlike traditional approaches, which rely on realized policy measures or news-based sentiment proxies constructed using lexicon methods \citep{picault2022media}, our index leverages an LLM to classify over 118,000 investor messages from social media, yielding a granular measure of dovish versus hawkish sentiment. This measure is incorporated into an LSTM forecasting framework, with SHapley Additive exPlanations (SHAP) used to interpret the resulting dynamics. We measure the role of MPE separately from realized policy actions, by conditioning on the realized rate path rather than by assuming the two coincide.

% 4. Contributions
The contribution of this study is threefold. First, it provides a comprehensive analysis of Bitcoin’s key price drivers, integrating signals from macroeconomics, financial markets, and, crucially, investor sentiment. Second, it demonstrates the application of XAI not only to forecast Bitcoin prices but also to economically interpret the model’s predictions, thereby addressing the ``black box'' limitation of traditional machine learning models. Third, it validates the use of an LLM in econometrics by producing a novel, high-frequency indicator for monetary policy expectations that improves upon static dictionary-based approaches.

% 5. Implications
These findings challenge the view of Bitcoin as an asset detached from macroeconomic fundamentals. Instead, we establish that Bitcoin acts as a sophisticated barometer of global liquidity, reacting rationally to the forward-looking pulse of central bank rhetoric. By showing that narrative-driven expectations meaningfully influence Bitcoin's price returns, we highlight the utility of LLM-derived sentiment as a leading macroeconomic indicator for both investors and policymakers.

The remainder of the paper is organized as follows. Section \ref{sec:data-method} outlines the methodology, including the construction of the MPE Index using Qwen2.5-7b (\ref{sec:mpexp}) and the econometric framework (\ref{sec:granger}). The predictive modeling strategy, employing an LSTM network (\ref{sec:lstm}) and SHAP interpretability (\ref{sec:shap}), is described in Section \ref{sec:xai}. Section \ref{sec:results} presents the empirical results, covering summary statistics (\ref{sec:sum-stats}), causal analysis (\ref{sec:granger-vmd}), out-of-sample forecasting (\ref{sec:forecast}), and regime-dependent interactions (\ref{sec:interactions}). Finally, Section \ref{sec:discuss} discusses Bitcoin's policy sensitivity, and Section \ref{sec:conclusion} concludes.

\section{Data and Methodology} \label{sec:data-method}

\subsection{Monetary Policy Expectations Index} \label{sec:mpexp}

We introduce a novel measure of policy expectations, constructed using an LLM, to classify investor discourse on social media. The index is designed to bridge the temporal gap between low-frequency official policy announcements and the high-frequency dynamics of financial markets.
Our dataset is sourced from StockTwits, the most popular financial social media platform with a user base exceeding 10 million as of early 2026.\footnote{StockTwits messages have been shown to reflect investor sentiment and to carry information for returns \citep{renault2017intraday}.} We capture the comprehensive historical record of messages associated with the cashtags ``\$FED" and ``\$MACRO" from September 2014 to March 2025 (N=118,562).\footnote{Investors use ``cashtags,'' a portmanteau of cash and hashtag, which consist of a dollar sign followed by a symbol to reference specific assets or topics.} These tags represent the primary channels for digital investor deliberations concerning U.S. monetary policy, Federal Reserve interventions, and broader macroeconomic conditions.

Unlike traditional dictionary-based approaches \citep{picault2017words, hubert2021signaling}, we employ \texttt{Qwen2.5-7b}, a 7-billion parameter model released by Alibaba Cloud \citep{bai2023qwen}, to parse financial context. This architecture was selected not only for its favorable balance of computational efficiency and high performance, which is critical for efficiently processing hundreds of thousands of messages, but also for its proven capacity to excel in specialized financial text classification \citep{yang2024evaluating}. We assess the reliability of this choice in Appendix \ref{sec:reliability-models-temps}, comparing \texttt{Qwen2.5-7b} against \texttt{mistral-nemo:12b}, a 12-billion parameter model developed by Mistral AI in collaboration with NVIDIA \citep{mistral_nemo_2024}, at two temperature settings each, for four configurations in total. \texttt{Qwen2.5-7b} is retained as the baseline classifier for its higher internal consistency and greater stability across temperature settings. We further justify this choice against a manually labeled sample of 1,095 messages, roughly 1\% of the corpus, in Appendix \ref{sec:validity-manual}.

The classification pipeline is implemented using the \texttt{ollama} Python library to facilitate high-throughput inference across the large-scale dataset. Each message is mapped to a symmetric five-point scale, i.e., Very Hawkish (-2), Hawkish (-1), Neutral (0), Dovish (+1), and Very Dovish (+2), as summarized in Table \ref{tab:sentiment_definitions}.\footnote{Note that the index is oriented so that higher values indicate a more dovish stance; conventions differ across the literature.} This scoring framework aligns with established literature in central bank communication analysis \citep{blinder2008central}, where weights serve as directional multipliers to ensure that extreme policy stances exert a proportionally greater influence on the aggregate signal. Detailed system prompts and model instructions are provided in Appendix \ref{sec:mistral-prompt}. Because the \$FED and \$MACRO streams are macro-oriented rather than policy-specific, the majority of collected messages do not concern monetary policy and are absorbed into the Neutral class. We therefore submit the neutral messages to a second, strictly binary LLM pass that asks only whether a message concerns monetary policy, retaining genuine no-change expectations at zero and discarding off-topic chatter; the procedure is detailed in Appendix \ref{sec:reliability-neutral}. This correction has a limited impact: the corrected and uncorrected weekly series correlate at $\rho = 0.945$ (Spearman $0.986$) and agree on the sign of the index in every one of the \num{545} weeks, so no turning point is created or removed.

\begin{table}[H]
\centering
\caption{Monetary Policy Sentiment Classification: Categories and Definitions}
\label{tab:sentiment_definitions}
\caption*{\small \textbf{Notes:} This table defines the scale used for the LLM-driven classification. The Model Definition summarizes the semantic instructions provided to the \texttt{Qwen2.5-7b} model. Symmetric weights ranging from -2 to +2 capture both the direction and intensity of monetary sentiment.}
\begin{tabular}{l c p{10cm}}
\toprule
Category & Score & Definition (Instructional Prompt) \\
\midrule
Very Hawkish & -2 & Messages expressing strong conviction for aggressive tightening, multiple rate hikes, or urgent inflation containment. \\
Hawkish & -1 & Messages suggesting a preference for tighter policy, concerns about rising inflation, or a `higher-for-longer' stance. \\
Neutral & 0 & Messages that are purely descriptive, balanced, or express no clear bias toward future policy directions. \\
Dovish & +1 & Messages suggesting a preference for easing, concern about economic slowing, or a pause in rate hikes. \\
Very Dovish & +2 & Messages expressing strong conviction for aggressive easing, rate cuts, or urgent economic stimulus. \\
\bottomrule
\end{tabular}
\end{table}

The MPE index is subsequently computed as the weighted weekly average of these scores, providing a proxy for market-implied policy stances. A weighted average based on engagement (likes and reshares) was used to better reflect concerns shared by many individuals. Table \ref{tab:mpe_summary_stats} presents the descriptive statistics of the classified corpus.

\begin{table}[H]
\centering
\caption{Descriptive Statistics of the MPE Index}
\label{tab:mpe_summary_stats}
\begin{small}
\caption*{\small \textbf{Notes:} This table summarizes the 118,562 messages classified from the \$FED and \$MACRO StockTwits messages (2014--2025). Categories represent the stance toward monetary policy: Very Hawkish ($-2$), Hawkish ($-1$), Dovish ($+1$) and Very Dovish ($+2$). Messages that concern monetary policy without expressing a direction are reported as \textit{Neutral ($0$)}, whereas messages that do not concern monetary policy are reported as \textit{Not MPE-related}.  \textit{Share (\%)} represents the percentage of total messages in each category; \textit{Avg. Likes} and \textit{Avg. Reshares} denote the mean number of likes and reshares received per message, respectively; and \textit{Avg. Engag.} (Average Engagement) is defined as the sum of mean likes and reshares per message.}
\end{small}
\resizebox{0.99\textwidth}{!}{%
\begin{tabular}{l r r c c c}
\toprule
& Obs. ($N$) & Share (\%) & Avg. Likes & Avg. Reshares & Avg. Engag. \\
\midrule
\textit{Very Hawkish ($-2$)} & 1,486 & 1.25 & 0.91 & 0.05 & 0.95 \\
\textit{Hawkish ($-1$)} & 5,086 & 4.29 & 0.72 & 0.06 & 0.78 \\
\textit{Neutral ($0$)} & 41,716 & 35.18 & 0.32 & 0.03 & 0.35 \\
\textit{Dovish ($+1$)} & 5,863 & 4.95 & 0.56 & 0.04 & 0.60 \\
\textit{Very Dovish ($+2$)} & 1,915 & 1.62 & 0.79 & 0.04 & 0.84 \\
\textit{Not MPE-related} & 62,496 & 52.71 & 0.22 & 0.02 & 0.23 \\
\midrule
\textit{TOTAL / Average} & 118,562 & 100.00 & 0.31 & 0.02 & 0.33 \\
\bottomrule
\end{tabular}}
\end{table}

The distribution exhibits a slight dovish tilt, with 6.57\% of messages reflecting easing expectations compared to 5.54\% favoring tightening. The dataset is predominantly neutral (87.89\%), of which the larger part (52.71\% of the corpus) is not about monetary policy, indicating that the vast majority of messages do not articulate a specific directional stance on monetary policy. However, polarized sentiments elicit disproportionately higher levels of engagement; specifically, messages expressing ``Very Hawkish" views generate an average engagement score of 0.95—nearly three times higher than the baseline for policy-related neutral messages (0.35), and over four times the level of off-topic ones (0.23). This disparity suggests that while explicit policy expectations are statistically rarer, they possess significantly higher ``signal strength" and peer validation within the investor community. 

We investigate the relation between the MPE index and market-based measures of policy expectations in Appendix \ref{sec:correlation-market}, where it correlates at $-0.374$ with the 6M$-$1M Treasury bill spread and $-0.345$ with the 3M$-$1M spread, both with the sign that convergent validity predicts. Further quantitative details regarding the lexical distribution, including bigram frequency analysis and extended keyword lists, are provided in Appendix \ref{sec:topic-analysis}.

\begin{figure}[H]
\centering
\caption{Bitcoin Prices, Returns, and Monetary Policy Indicators} \label{fig:btc-monetary-ffr-evol}
\resizebox{\textwidth}{!}{%
\begin{tabular}{p{\textwidth}}
\small{ \textbf{Notes:} This figure presents weekly Bitcoin price levels (scaled in thousands of U.S. dollars), Bitcoin returns, the monetary policy expectations index, and the effective federal funds rate over the period from September 2014 to March 2025.}
\end{tabular}}
\includegraphics[width=0.8\linewidth]{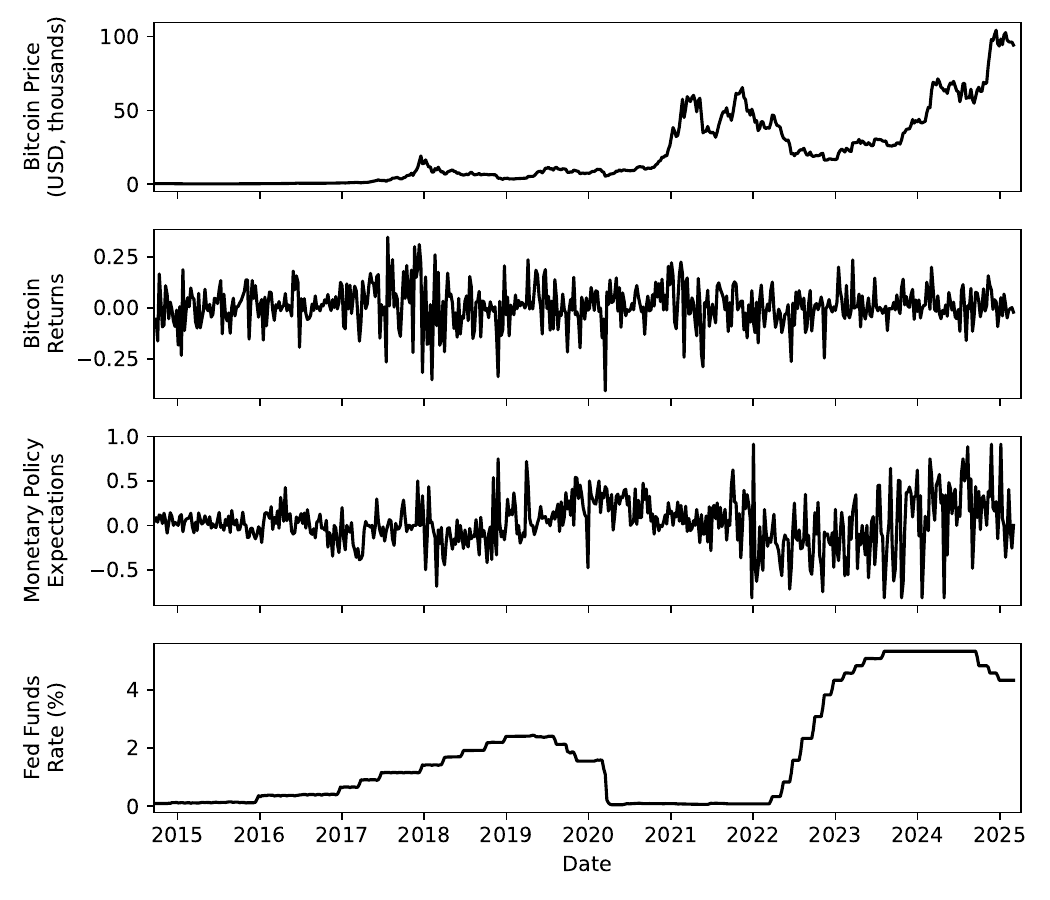}
\end{figure}

Figure \ref{fig:btc-monetary-ffr-evol} illustrates the temporal co-evolution of Bitcoin market dynamics and U.S. monetary policy expectations over the period from September 2014 to March 2025. The panels present Bitcoin's price trajectory and returns against our novel MPE index and the realized Federal Funds Rate (EFFR). Visual inspection reveals distinct regimes where spikes in hawkish sentiment (falling index values) precede or coincide with periods of effective rate adjustments.

\begin{figure}[H]
    \caption{Weekly Event Study: MPE Reaction by Pre-Announcement Regime.}
    \label{fig:event_study_fomc}
    \centering
    \resizebox{\textwidth}{!}{%
    \begin{tabular}{p{\textwidth}}
    \small{ \textbf{Notes:} This figure illustrates the aggregate behavior of the MPE index in a $\pm 6$-week window surrounding FOMC meetings. Events are partitioned into \textit{Hawkish} (\textit{MPE} $<- 0.01$), \textit{Neutral}, and \textit{Dovish} (\textit{MPE} $> 0.01$) regimes based on the average index value during the pre-announcement period ($t-5$ to $t-1$). Solid lines represent the mean path for each group, while shaded areas denote the $\pm 1$ standard deviation band. The vertical dashed line at $t=0$ marks the week of the FOMC announcement. Only two meetings fall in the \textit{Neutral} regime, so that group is omitted from the figure.}
    \end{tabular}}
    \includegraphics[width=0.85\textwidth]{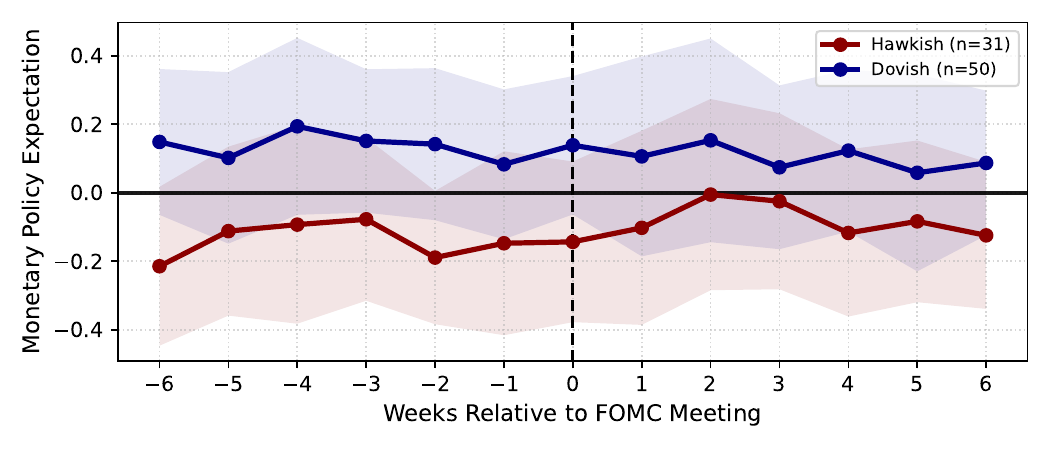}
\end{figure}

To distinguish our measure from existing studies that focus solely on the immediate reactions to policy announcements \citep{corbet2020cryptocurrency, pyo2020fomc}, the index is constructed to capture deeply anchored, persistent market narratives rather than merely serving as a redundant restatement of official Federal Reserve communications. A weekly event study is conducted around FOMC announcement dates. Figure \ref{fig:event_study_fomc} illustrates the aggregate temporal dynamics of the index within a ±6-week window, with observations partitioned into \textit{Dovish} ($> 0.01$), \textit{Neutral} ($[-0.01, 0.01]$, to avoid small noisy classifications around 0), and \textit{Hawkish} ($< -0.01$) states based on their pre-announcement mean values. Only two of the $83$ meetings fall in the \textit{Neutral} band, so the figure reports the \textit{Hawkish} ($n=31$) and \textit{Dovish} ($n=50$) regimes alone. The analysis reveals that the identified sentiment regimes exhibit remarkable persistence, maintaining their respective trajectories well after the $t=0$ announcement. This continuity indicates that the index captures market expectations that are typically continued, rather than substantially adjusted, by official policy communications.

\begin{figure}[H]
\centering
\caption{Bitcoin Weekly Returns Across Monetary Policy and Interest Rate Regimes.} \label{fig:btc-monetary-ffr}
\resizebox{\textwidth}{!}{%
\begin{tabular}{p{\textwidth}}
\small{ \textbf{Notes:} This figure illustrates the distribution of weekly Bitcoin returns categorized by monetary policy expectations (Panel a) and changes in the Effective Federal Funds Rate (Panel b). The horizontal boxplots represent the interquartile range (IQR), with the red vertical line indicating the median and the white circle ($\circ$) denoting the arithmetic mean $(\mu)$ with extreme values plotted individually. Regimes are defined as follows: Dovish (\textit{MPE} $>0.01$), Neutral, and Hawkish (\textit{MPE} $<-0.01$); and Falling ($\Delta F F R<0$), Flat, and Rising ($\Delta F F R>0$). Sample sizes ($n$) and mean returns ($\mu$) for each sub-group are provided in the left margin.}
\end{tabular}}
% Panel A
    \begin{subfigure}[b]{0.9\textwidth}
        \centering
        \caption{Monetary Policy Expectations Regimes}
        \includegraphics[width=\linewidth]{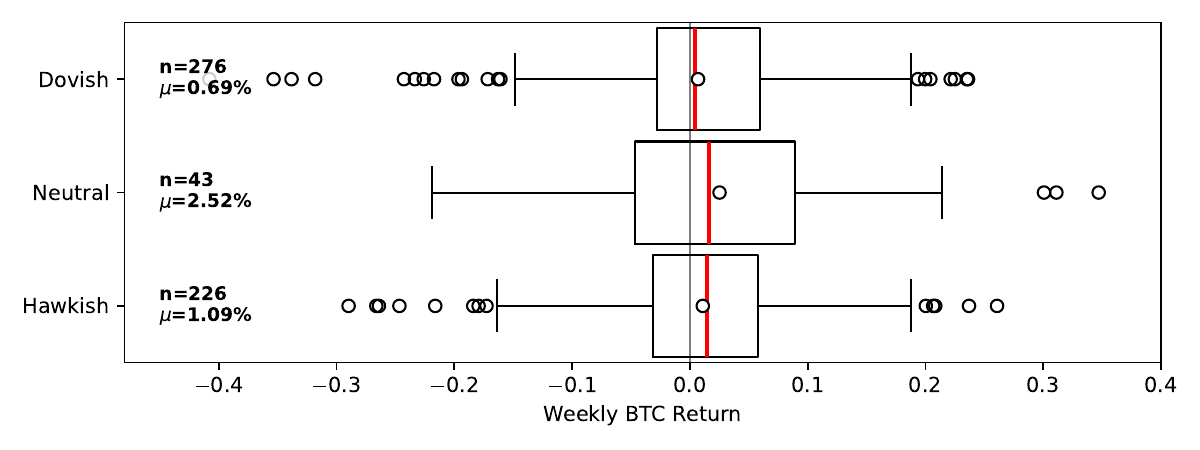}
        \label{fig:btc-monetary-a}
    \end{subfigure}
    \vspace{0em} % Add space between the two panels
    % Panel B
    \begin{subfigure}[b]{0.9\textwidth}
        \centering
        \caption{Fed Funds Rate Direction}
        \includegraphics[width=\linewidth]{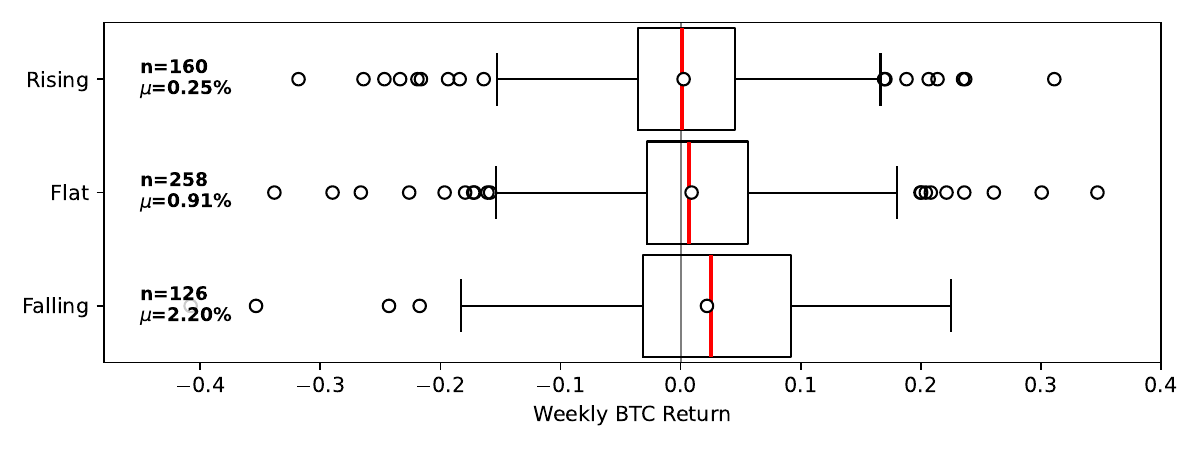}
        \label{fig:btc-ffr-b}
    \end{subfigure}
\end{figure}

To investigate the conditional distribution of Bitcoin returns, we partition the sample into distinct monetary policy regimes based on both policy expectations and realized rate adjustments. As before, we categorize the MPE index into \textit{Dovish}, \textit{Neutral}, and \textit{Hawkish} states, and similarly segment the Effective Federal Funds Rate (EFFR) cycle into \textit{Falling}, \textit{Flat}, and \textit{Rising} periods. Figure \ref{fig:btc-monetary-ffr} illustrates the resulting distributional shifts. 
In panel (a), the distribution shows that the \textit{Neutral} regime exhibits a higher mean return ($2.52\%$) than either polarized state ($1.09\%$ under \textit{Hawkish} and $0.69\%$ under \textit{Dovish} expectations), while the medians are close ($1.64\%$ against $1.43\%$ and $0.46\%$). The gap is driven by the tails, and it rests on only $43$ weeks with the widest dispersion of the three regimes (an interquartile range of $13.5$ percentage points against roughly $8.8$): quiet policy periods bring occasional large moves rather than reliably higher returns. In panel (b), realized rate cuts (\textit{Falling} regime) coincide with the highest average returns, suggesting that liquidity injections function as a primary valuation driver. However, the substantial interquartile ranges across all categories indicate that while monetary policy exerts directional pressure, idiosyncratic factors specific to the crypto-asset class continue to drive significant volatility independent of the macro regime.

\subsection {Main Macroeconomic Variables}

We construct a weekly dataset consisting of 18 macrofinancial and macroeconomic indicators, spanning the period from September 2014 to March 2025, to evaluate their predictive efficacy for Bitcoin returns (\textit{Btc}). Recent studies emphasize that aggregate macroeconomic conditions are essential for capturing the complex dynamics of cryptocurrency markets \citep{wang2023factors}. We categorize our predictors into three theoretical partitions, namely, Macro Conditions, Monetary Policy \& Financial Conditions, and Global Risk \& Market Sentiment, as detailed in Appendix \ref{sec:def-variables} (Table \ref{tab:var-def}), along with the sources of collection and applied transformations. 

The first partition, Macro Conditions, considers broad proxies for financial market performance, labor market health, and public attention to macroeconomic developments. Specifically, it incorporates the S\&P 500 index (\textit{SP500}) and weekly jobless claims (\textit{JoblessClaim}) to capture the real-time pulse of the U.S. economy, alongside internet search metrics as direct proxies for retail attention and inflation expectations. Google Search  Trends for inflation (\textit{GgleInfl}) is used as a real-time measure of public attention to price stability \citep{guzman2011internet}, while Google Search Trends for recession  (\textit{GgleReces}) serves as a proxy for economic anxiety, as demonstrated by \cite{woloszko2018tracking}. This partition also includes a news-based sentiment index (\textit{NewsSent}) to capture the broader information environment \citep{shapiro2022measuring}.

The second partition, Monetary Policy \& Financial Conditions, moves beyond the traditional reliance on a single interest rate. By integrating the Effective Federal Funds Rate (\textit{FFR}) with our novel Monetary Policy Expectations (\textit{MPE}) index and 5-year inflation expectations (\textit{5yInflExp}), we model policy as a multidimensional, forward-looking construct that accounts for both realized actions and market-priced trajectories. This partition further incorporates input costs via Brent crude (\textit{Brent}), safe-haven demand via Gold (\textit{Gold}), credit market stress via high-yield spreads (\textit{HighYield}), and the effective exchange rate (\textit{ExchRate}), collectively reflecting the cost--risk trade-offs that characterize prevailing financial conditions.

Finally, the Global Risk \& Market Sentiment partition aggregates systemic risk indicators, including the VIX index (\textit{VIX}), Policy Uncertainty (\textit{PolUncert}), and Geopolitical Risk (\textit{GeopolRisk}). In this context, Google Search Trends for climate change (\textit{GgleClimate}) is employed as a proxy for salient attention to global warming \citep{choi2020attention}, capturing the shifting investor beliefs and environmental narratives that increasingly influence asset valuations, alongside an equity market volatility index based on infectious-disease news (\textit{Infect}). These variables collectively account for Bitcoin's regime-dependent nature, in which shifts in global risk appetite typically trigger portfolio rebalancing. By synthesizing these three dimensions, the model accounts for the diverse fundamental and narrative-driven forces that shape the cryptocurrency ecosystem.

Data harmonization was performed to align all series to a consistent ``week-ending Friday'' frequency. We retrieved series at their native frequencies and resampled them to this common grid. For daily price and sentiment series, we considered the Friday closing value, such as Bitcoin (\textit{Btc}) and the S\&P 500 index (\textit{SP500}), or the weekly average for variables like yields, news sentiment (\textit{NewsSent}), and Google Search Trends to capture end-of-week investor sentiment while smoothing noise.

\subsection{Granger Causality and Variational Mode Decomposition} \label{sec:granger}

Having defined the broader macroeconomic variable set, we now focus specifically on the MPE index and subject it to a more rigorous causal investigation, decomposing its relationship with Bitcoin returns across both time and frequency dimensions. To this end, we employ the Granger causality test \citep{granger1969investigating} to rigorously assess this relationship. We test the null hypothesis that the MPE index  does not Granger-cause Bitcoin returns against the alternative that it does. 
The test is specified using the following bivariate Vector Autoregression (VAR) framework:

\begin{equation}
    R_t = \alpha_1 + \sum_{i=1}^{p} \beta_{1,i} R_{t-i} + \sum_{j=1}^{p} \gamma_{1,j} MPE_{t-j} + \epsilon_{1,t}~,
\end{equation}

%\begin{equation}
%    MPE_t = \alpha_2 + \sum_{i=1}^{p} \beta_{2,i} MPE_{t-i} + \sum_{j=1}^{p} \gamma_{2,j} R_{t-j} + \epsilon_{2,t}
%\end{equation}

\noindent where $R_t$ denotes Bitcoin returns at time $t$, and $MPE_t$ represents the Monetary Policy Expectations index. The parameters $\alpha$ represent the intercept terms, $\beta$ and $\gamma$ are the coefficients on lagged endogenous variables, $p$ denotes the optimal lag length selected using the Akaike Information Criterion (AIC), and $\epsilon_t$ are white noise error terms. 
Granger causality from $MPE_t$ to $R_t$ is tested via the joint null hypothesis
\begin{equation}
    H_0: \gamma_{1,1} = \gamma_{1,2} = \dots = \gamma_{1,p} = 0 ~,
\end{equation}
against the alternative that at least one $\gamma_{1,j} \neq 0$.
Rejection of $H_0$ implies that past values of monetary policy expectations contain statistically significant predictive information for Bitcoin returns.

Considering the non-stationary and multi-scale characteristics inherent in financial time series, standard time-domain analysis may obscure relationships that manifest at specific frequencies. To address this, we apply Variational Mode Decomposition (VMD), a non-recursive and adaptive signal processing technique introduced by \cite{dragomiretskiy2013variational}, to decompose Bitcoin returns into a finite set of intrinsic mode functions (IMFs), each capturing dynamics at distinct time scales. The decomposition separates the series into a slow-moving trend, a medium-term cycle and a high-frequency component, so that causality can be tested at each horizon separately rather than on their sum.
Unlike Empirical Mode Decomposition (EMD), VMD mitigates mode mixing by decomposing the Bitcoin return series $f(t)$ into a discrete number of band-limited IMFs, $u_k(t)$, each centered around a specific frequency $\omega_k$. 
To decompose our time-series data, we employ the \texttt{vmdpy} library \citep{carvalho2020evaluating}. The decomposition is obtained by solving the following constrained variational problem:

%\begin{equation}
%    \min_{\{u_k\}, \{\omega_k\}} \left\{ \sum_{k} \left\| \partial_t \left[ (\delta(t) + \frac{j}{\pi t}) * u_k(t) \right] e^{-j\omega_k t} \right\|_2^2 \right\} \quad \text{s.t.} \quad \sum_{k} u_k = f
%\end{equation}

\begin{equation}
    \min_{\{u_k\}, \{\omega_k\}} \left\{ \sum_{k} \left\| \partial_t \left[ (\delta(t) + \frac{j}{\pi t}) * u_k(t) \right] e^{-j\omega_k t} \right\|^2 \right\} \quad \text{s.t.} \quad \sum_{k} u_k = f ~,
\end{equation}

\noindent where $*$ denotes convolution, $\delta(t)$ is the Dirac distribution, and $\|\cdot\|$ represents the $L^2$ norm. 
The resulting IMFs represent short-term (high-frequency), medium-term, and long-term (low-frequency) components of Bitcoin returns. 
Prior to conducting the Granger causality analysis, the stationarity of each variable is verified using the Augmented Dickey–Fuller (ADF) test \citep{dickey1979distribution}. The causality tests are then re-estimated for each decomposed mode to assess whether the influence of MPE on Bitcoin returns exhibits frequency-dependent behavior.

\subsection{Explainable AI Framework} \label{sec:xai}

While the frequency-decomposed Granger analysis identifies \textit{when} and \textit{at what} scale monetary policy expectations influence Bitcoin returns, it does not fully characterize the mechanisms through which these effects materialize. To address this, we adopt a nonlinear modeling framework capable of capturing the complex, nonlinear dependencies between monetary policy expectations and Bitcoin returns. We employ an LSTM network, an architecture designed to overcome the limitations of traditional autoregressive models (e.g. ARIMA) and standard Recurrent Neural Networks (RNNs), which are prone to the vanishing gradient problem. 

To ensure the robustness of our results and prevent look-ahead bias, we implement a strict temporal walk-forward evaluation strategy, allowing for consistent out-of-sample validation across evolving macroeconomic regimes. The LSTM is employed primarily as an explanatory framework rather than as a forecasting model intended to outperform conventional benchmarks. Beyond predictive performance, our objective is to interpret these nonlinear relationships in economically meaningful terms. We therefore complement the LSTM framework with SHAP analysis, enabling a transparent decomposition of model predictions and a systematic evaluation of the relative importance and interaction effects of monetary policy expectations vis-à-vis other macrofinancial drivers of Bitcoin returns.
The complete replication package for this study, including all code used to implement the LSTM models and the SHAP analysis described in the following subsections, is provided in the supplemental information.

\begin{figure}[H]
\centering
\caption{Cross-Validation Framework} \label{fig:train-val-test}
\resizebox{\textwidth}{!}{%
\begin{tabular}{p{\textwidth}}
\small{ \textbf{Notes:} Schematic representation of the recursive partitioning of the dataset into four sequential folds. The Training set (dark shading) is anchored in September 2014 and expands by 12-month increments across folds, ranging from 5.5 years ($N \approx 288$ weeks) in Fold 1 to 7.5 years ($N \approx 391$ weeks) in Fold 4. The Validation set (diagonal hatching) maintains a fixed 24-month duration for hyperparameter optimization. The Test set (dotted shading) covers a non-overlapping 12-month out-of-sample period.}
\end{tabular}}
\includegraphics[width=0.8\linewidth]{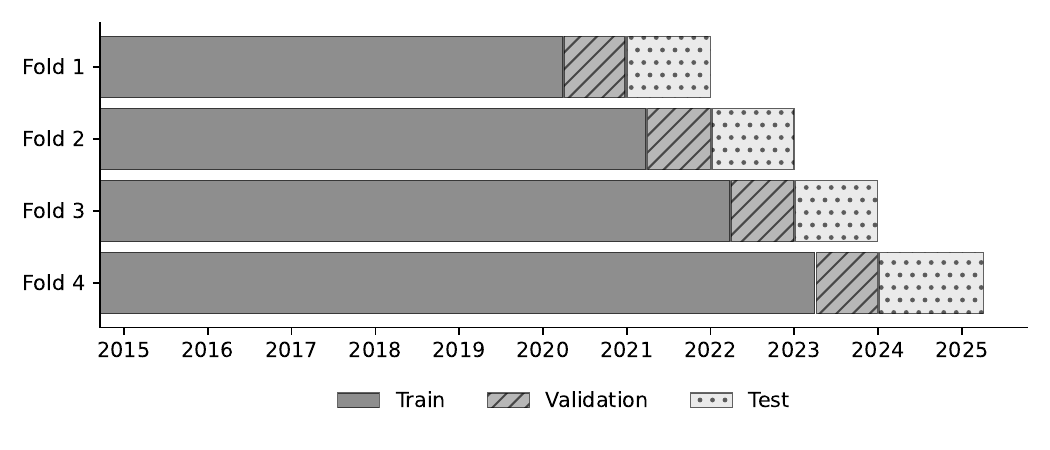}
\end{figure}

\subsubsection{LSTM Forecasting Architecture} \label{sec:lstm}

We implement an LSTM network \citep{hochreiter1997long} to capture the long-range temporal dependencies inherent in weekly Bitcoin returns time series. The LSTM architecture is designed to map a sequence of input features $X = \{x_{t-L+1}, ..., x_t\}$ to a one-week ahead future return $y_{t+1}$, where the input features comprise $18$ macrofinancial covariates and the lagged Bitcoin return. 

As illustrated in Figure \ref{fig:train-val-test}, we partition the dataset into four sequential folds using an expanding window approach. Each fold consists of a training set anchored in September $2014$, with subsequent folds expanding to include the previous fold’s validation period and additional data, resulting in training windows ranging from $288$ to $391$ weeks.
Validation sets, ranging from $36$ to $40$ weeks, are used for hyperparameter optimization; the shorter validation periods reduce the temporal gap between training and testing, improving the model’s adaptation to recent macroeconomic regimes. 
Non-overlapping test sets, spanning $52$ to $64$ weeks, are kept strictly out-of-sample and are not used for hyperparameter optimization, model fitting, or primary model selection. They provide an independent assessment of the resulting LSTM framework while being sufficiently long to support statistical comparisons with ARIMA benchmarks.
This four-fold recursive partitioning ensures the model is exposed to a diverse range of market and macroeconomic regimes, including the pre-COVID expansion, the $2020$ COVID shock and subsequent recovery, the $2022$ monetary tightening shock, and the $2024$–$2025$ ETF and policy pivot phase, while maintaining a sufficiently large training sample in each fold to support robust nonlinear estimation.

The LSTM architecture consists of an input layer with 
$N$ features, a single LSTM layer with hyperparameter-tuned units 
$u \in \{8, 16, 32\}$ and hyperbolic tangent (tanh) activation to capture nonlinear dependencies, dropout regularization ($p \in [0.1, 0.45]$) to mitigate overfitting, and a fully connected dense layer with linear activation to predict continuous returns. Hyperparameters -- including the lookback window ($L \in \{4,8,13\}$), learning rate ($1\times 10^{−4} \leq lr \leq 1 \times 10^{−3}$), optimizer choice (Adam or RMSprop), batch size ($8$ or $16$) -- are optimized via Optuna using a Tree-structured Parzen Estimator (TPE) search strategy \citep{akiba2019optuna} over $75$ trials per fold. 
To assess the stability of hyperparameter optimization in the high-dimensional and stochastic LSTM optimization landscape, we perform $20$ independent Optuna searches within each temporal fold, each using a distinct random seed. Each search selects its optimal configuration exclusively according to the validation MSE, without access to the corresponding test period. Gradient clipping (clipnorm $= 1.0$) is applied to stabilize training and prevent the model from overfitting extreme weekly shocks.

For each fold, the optimal configuration identified by each of the $20$ independent Optuna searches is subsequently retrained $10$ times using different random seeds on the combined training and validation set. This produces $200$ final LSTM realizations per fold. The repeated retraining isolates the effect of stochastic weight initialization and optimization variability, while the multiple independent searches assess whether the resulting nonlinear relationships are robust to alternative regions of the hyperparameter landscape. For forecasting evaluation, the $10$ retrainings associated with each Optuna search are treated as stochastic ensemble members and their predictions are averaged to obtain an ensemble forecast. The test-period observations are only introduced at this final evaluation stage. The resulting out-of-sample performance is used to assess the predictive adequacy of the LSTM framework relative to the ARIMA benchmark, rather than to determine the hyperparameters used during model development. The resulting models and their predictions are subsequently used as the basis for the SHAP analysis, which is the primary objective of the nonlinear modeling framework.
The model is trained to minimize the Mean Squared Error (MSE) loss function:
\begin{equation}
    L(\theta) = \frac{1}{T} \sum_{t=1}^{T} (y_t - \hat{y}_t)^2 ~,
\end{equation}
\noindent where $y_t$ is the actual return, $\hat{y}_t$ is the predicted return, and $\theta$ represents the network weights. 
Early stopping with adaptive learning rates is employed during hyperparameter tuning, while final training on the combined training+validation set uses a fixed number of epochs determined by the convergence of the selected Optuna configuration. 

\subsubsection{Model interpretability via SHAP} \label{sec:shap}

The LSTM architecture described above produces forecasts of Bitcoin returns but does not directly provide explanations for its predictions.
To address this, we complement the LSTM model with a model-agnostic interpretability framework that attributes each forecast to the contribution of individual input features.
We employ SHAP, a unified framework grounded in cooperative game theory \citep{lundberg2017unified}. Unlike permutation-based feature importance measures, which assign a single global score to each feature across the entire dataset, SHAP decomposes each individual model prediction into feature-level contributions, thereby enabling interpretability at the level of individual forecasts.
This allows us to assess not only the absolute importance of our MPE index, but also its relative contribution compared with a wide range of macroeconomic covariates, thereby situating the role of narrative-driven sentiment within the wider macroeconomic information set that drives Bitcoin returns. 

Formally, for a model $f$ and an input $x$, the SHAP value $\phi_i$ is assigned to the feature $i$, defined as the weighted average of its marginal contribution over all possible feature coalitions:

\begin{equation}
\label{eq:SHAP_phi}
    \phi_i(f, x) = \sum_{z' \subseteq x'} \frac{|z'|! (M - |z'| - 1)!}{M!} [f_x(z') - f_x(z' \setminus i)] ~,
\end{equation}

\noindent where $x'$ is the set of all input features, $z'$ is a subset of features, $M$ is the total number of input features, and $f_x(z')$ is the model output conditional on features $z'$. By construction, the SHAP values satisfy the \textit{additive feature attribution} property:

\begin{equation}
    f(x) = \phi_0 + \sum_{i=1}^{M} \phi_i ~,
\end{equation}

\noindent where $\phi_0$ is the baseline model output, set to the mean Bitcoin return in the training set.

In economic terms, a positive value of $\phi_i$ indicates that, given the macrofinancial environment represented in the input sequence, feature $i$ pushes the model's forecast towards a higher Bitcoin return relative to its baseline prediction, whereas a negative value indicates that it pushes the forecast towards a lower return. The magnitude of $\phi_i$, measured in the units of the predicted return, reflects the strength of this contribution to that particular forecast. Because this decomposition is performed at the level of individual predictions, the contribution of a given variable may vary across weeks and macroeconomic regimes. Importantly, $\phi_i$ should not be interpreted as a causal or structural effect of variable $i$ on the realised Bitcoin return; rather, it quantifies how the trained LSTM uses that variable, in conjunction with the other inputs, to generate a particular prediction.

In our workflow, SHAP values are computed for each fold and multiple training runs to account for model stochasticity. We store per-run, per-fold values for both \textit{global} and \textit{lag-specific} analyses, allowing a structured analysis along three dimensions:

\begin{enumerate}
\item \textbf{Global Importance:} SHAP values are averaged across samples and temporal lags to produce an overall mean absolute contribution for each feature. These aggregated values are further averaged across runs and folds, yielding a robust ranking of predictors by overall influence.

\item \textbf{Disaggregated Temporal Importance:} For each feature, SHAP values are retained at the level of individual lags to capture temporal dynamics. Mean and standard deviation across runs and folds are computed, allowing assessment of which past observations (lags) drive model predictions.

\item \textbf{Directional \& Interaction Effects:} We visualize the distribution of $\phi_i$ values against feature magnitude to determine if, for example, high hawkish sentiment (low $MPE$ index) drives negative returns (negative $\phi_{MPE}$), thereby testing our hypothesis of regime-dependent sensitivity.
\end{enumerate}

This dual-level approach allows comprehensive interpretability: global aggregation facilitates cross-feature comparison, whereas disaggregated lag-level SHAP highlights the temporal structure of predictive signals. In addition, directional and interaction analyses allow examination of regime-dependent relationships between MPE and Bitcoin price returns. All computations, including fold-wise storage, aggregation, and temporal profiling, are implemented in Python using TensorFlow and the SHAP library.

\section{Empirical Results} \label{sec:results}

\subsection{Summary Statistics} \label{sec:sum-stats}

Table \ref{tab:stat-des} presents the descriptive statistics for Bitcoin returns and the set of predictors. The MPE index is centred close to zero, with a mean of 0.03 and median of 0.02, indicating no strong average directional tendency over the sample. Its near-zero skewness (-0.05) suggests that positive and negative monetary policy expectation signals occur with broadly comparable magnitudes. However, its kurtosis of 4.21 indicates heavier tails than a normal distribution. Although half of the observations fall between -0.10 and 0.16, the 5th and 95th percentiles extend to -0.45 and 0.48. The index therefore remains relatively stable in most weeks but occasionally records substantial shifts in expectations. The federal funds rate, by contrast, is positively skewed (0.87) and platykurtic (2.28), reflecting a policy rate that moves infrequently and in discrete increments while expectations reprice immediately in response to policy announcements and macroeconomic news \citep{gurkaynak2005sensitivity}.

The properties of the remaining variables further support the chosen modeling approach. Bitcoin returns exhibit substantial weekly volatility ($SD = 0.10$) and negative skewness ($-0.36$), indicating prevalent downside risk and non-Gaussian fat tails ($Kurt = 4.79$). The extreme kurtosis in Jobless Claims (\textit{JoblessClaim}) ($300.59$) and Google searches for climate (\textit{GgleClimate}) ($237.38$) highlight the structural breaks and unprecedented volatility characteristic of the sample period. Additionally, the positive skewness in Policy Uncertainty (\textit{PolUncert}) ($1.75$) and \textit{Geopolitical Risk} ($2.20$) confirms their role as ``jump'' predictors that spike during systemic crises.

\begin{table}[H]
  \centering
  \small
  \renewcommand{\arraystretch}{1.1}
  \caption{Descriptive Statistics of the Macroeconomic and Financial Variables.}
  \label{tab:stat-des}
\caption*{\small \textbf{Notes:} This table reports descriptive statistics for the variables employed in the Bitcoin price forecasting framework. The sample consists of weekly observations from September 2014 to March 2025. Specifically, the reported statistics include the mean, standard deviation (SD), skewness, kurtosis, minimum value, 5th percentile (5\%), 25th percentile (25\%), median (Med), 75th percentile (75\%), 95th percentile (95\%), and maximum value (Max). The final column reports the number of available observations for each series. Detailed definitions, data sources, and transformation procedures for all variables are provided in Appendix~\ref{sec:def-variables}.
}
\resizebox{\linewidth}{!}{%
\begin{tabular}{lcccccccccccc}
\toprule
 & Mean & SD & Skew & Kurt & Min & 5\% & 25\% & Med & 75\% & 95\% & Max & n \\
\midrule
\textit{Btc} & 0.01 & 0.10 & -0.36 & 4.79 & -0.41 & -0.15 & -0.03 & 0.01 & 0.06 & 0.16 & 0.35 & 545 \\
\textit{MPE} & 0.03 & 0.27 & -0.05 & 4.21 & -0.82 & -0.45 & -0.10 & 0.02 & 0.16 & 0.48 & 0.91 & 546 \\
\textit{NewsSent} & -0.00 & 0.18 & -1.20 & 5.09 & -0.66 & -0.34 & -0.09 & 0.03 & 0.11 & 0.24 & 0.32 & 546 \\
\textit{PolUncert} & 0.05 & 0.34 & 1.75 & 8.71 & -0.57 & -0.36 & -0.17 & -0.01 & 0.18 & 0.69 & 2.14 & 545 \\
\textit{SP500} & 0.00 & 0.02 & -0.89 & 10.40 & -0.16 & -0.03 & -0.01 & 0.00 & 0.02 & 0.03 & 0.11 & 545 \\
\textit{Brent} & 0.00 & 0.05 & 0.36 & 9.28 & -0.25 & -0.08 & -0.03 & 0.00 & 0.03 & 0.08 & 0.37 & 545 \\
\textit{Gold} & 0.00 & 0.02 & -0.06 & 4.65 & -0.09 & -0.03 & -0.01 & 0.00 & 0.01 & 0.03 & 0.08 & 545 \\
\textit{HighYield} & 0.00 & 0.01 & -2.15 & 29.91 & -0.10 & -0.01 & -0.00 & 0.00 & 0.00 & 0.01 & 0.05 & 545 \\
\textit{GeopolRisk} & 0.03 & 0.27 & 2.20 & 15.02 & -0.50 & -0.31 & -0.14 & -0.01 & 0.16 & 0.46 & 2.08 & 545 \\
\textit{VIX} & 0.01 & 0.18 & 2.06 & 12.96 & -0.43 & -0.20 & -0.08 & -0.01 & 0.08 & 0.33 & 1.35 & 545 \\
\textit{USDollar} & 0.00 & 0.01 & -0.01 & 4.78 & -0.04 & -0.02 & -0.01 & 0.00 & 0.01 & 0.01 & 0.04 & 545 \\
\textit{Infect} & 6.36 & 8.66 & 2.37 & 11.66 & 0.00 & 0.09 & 0.38 & 1.52 & 9.89 & 22.32 & 63.05 & 546 \\
\textit{JoblessClaim} & -0.00 & 0.12 & 15.20 & 300.59 & -0.23 & -0.08 & -0.03 & -0.00 & 0.02 & 0.07 & 2.37 & 545 \\
\textit{ExchRate} & 0.00 & 0.01 & 0.71 & 6.66 & -0.02 & -0.01 & -0.00 & 0.00 & 0.00 & 0.01 & 0.04 & 545 \\
\textit{FFR} & 1.78 & 1.88 & 0.87 & 2.28 & 0.04 & 0.08 & 0.12 & 1.16 & 2.40 & 5.33 & 5.33 & 546 \\
\textit{5yInflExp} & 2.06 & 0.26 & -0.68 & 2.88 & 1.19 & 1.58 & 1.89 & 2.12 & 2.25 & 2.41 & 2.59 & 546 \\
\textit{GgleInfl} & 20.63 & 11.79 & 1.77 & 7.71 & 8.18 & 9.65 & 12.70 & 15.28 & 28.00 & 43.47 & 100.00 & 546 \\
\textit{GgleReces} & 8.75 & 9.32 & 5.47 & 43.93 & 2.23 & 3.23 & 5.00 & 6.30 & 8.45 & 22.00 & 100.00 & 546 \\
\textit{GgleClimate} & 2.55 & 5.51 & 14.74 & 237.38 & 0.46 & 0.93 & 1.47 & 2.00 & 3.00 & 3.93 & 100.00 & 546 \\
\bottomrule
\end{tabular}
  }
\end{table}

Table \ref{tab:corr} confirms the low redundancy of the MPE index. Its low correlation with the effective Federal Funds Rate ($\rho=0.05$) indicates that the measure captures forward-looking expectations rather than realized policy adjustments. The index is decoupled from general market sentiment, as shown by its negligible co-movement with the S\&P $500$ ($\rho=0.09$) and VIX ($\rho=-0.03$). Additionally, the negative relationship with Google Search trends for inflation ($\rho=-0.17$) suggests that the LLM-driven approach captures a specific dimension of professionalized policy discourse that is distinct from retail-level noise. Crucially, the negative correlation with 5-year inflation expectations ($\rho=-0.22$) provides fundamental validity for the index; since the index rises when the crowd turns dovish, it demonstrates that hawkish shifts in the MPE index are associated with higher long-term inflation expectations. These properties show that the index can provide a distinct, non-redundant signal for the prediction model.

\begin{landscape}
\begin{table}[h]
\centering
\caption{Analysis of Correlations}
\label{tab:corr}

% --- Notes block (unchanged, but no need to resizebox it) ---
\begin{minipage}{0.98\linewidth}
\small
\textbf{Notes}: This table reports pairwise Pearson correlation coefficients among the variables used in the Bitcoin price forecasting model. Correlations are computed using weekly data over the sample period from September 2014 to March 2025. Detailed definitions, data sources, and transformation procedures for all variables are provided in Appendix~\ref{sec:def-variables}.
\end{minipage}

\vspace{0.75em}

% --- Correlation matrix ---
\small
\setlength{\tabcolsep}{6.5pt}
\renewcommand{\arraystretch}{1.3}

\begin{tabular}{lcccccccccccccccccccccc}
\toprule
& \rot{\textit{Btc}} & \rot{\textit{MPE}} & \rot{\textit{NewsSent}} & \rot{\textit{PolUncert}} & \rot{\textit{SP500}} & \rot{\textit{Brent}} & \rot{\textit{Gold}} & \rot{\textit{HighYield}} & \rot{\textit{GeopolRisk}} & \rot{\textit{VIX}} & \rot{\textit{USDollar}} & \rot{\textit{Infect}} & \rot{\textit{JoblessClaim}} & \rot{\textit{ExchRate}} & \rot{\textit{FFR}} & \rot{\textit{5yInflExp}} & \rot{\textit{GgleInfl}} & \rot{\textit{GgleReces}}  \\
\midrule
\textit{MPE} & 0.02 &  &  &  &  &  &  &  &  &  &  &  &  &  &  &  &  &  \\
\textit{NewsSent} & 0.02 & -0.07 &  &  &  &  &  &  &  &  &  &  &  &  &  &  &  &  \\
\textit{PolUncert} & -0.07 & 0.02 & 0.02 &  &  &  &  &  &  &  &  &  &  &  &  &  &  &  \\
\textit{SP500} & 0.18 & 0.09 & -0.07 & -0.05 &  &  &  &  &  &  &  &  &  &  &  &  &  &  \\
\textit{Brent} & 0.12 & -0.04 & -0.06 & -0.07 & 0.30 &  &  &  &  &  &  &  &  &  &  &  &  &  \\
\textit{Gold} & 0.08 & 0.07 & -0.03 & 0.05 & 0.12 & 0.09 &  &  &  &  &  &  &  &  &  &  &  &  \\
\textit{HighYield} & 0.15 & 0.09 & -0.06 & -0.09 & 0.75 & 0.35 & 0.21 &  &  &  &  &  &  &  &  &  &  &  \\
\textit{GeopolRisk} & -0.00 & 0.03 & 0.02 & 0.01 & -0.01 & 0.02 & 0.10 & 0.00 &  &  &  &  &  &  &  &  &  &  \\
\textit{VIX} & -0.19 & -0.03 & 0.10 & 0.09 & -0.73 & -0.27 & 0.04 & -0.42 & 0.02 &  &  &  &  &  &  &  &  &  \\
\textit{USDollar} & -0.08 & -0.05 & 0.05 & 0.03 & -0.28 & -0.09 & -0.49 & -0.36 & 0.02 & 0.02 &  &  &  &  &  &  &  &  \\
\textit{Infect} & -0.01 & 0.11 & -0.63 & 0.06 & -0.06 & -0.02 & 0.01 & -0.12 & 0.02 & 0.03 & 0.01 &  &  &  &  &  &  &  \\
\textit{JoblessClaim} & -0.01 & 0.01 & -0.04 & 0.08 & -0.27 & -0.17 & -0.04 & -0.38 & 0.01 & 0.05 & 0.14 & 0.24 &  &  &  &  &  &  \\
\textit{ExchRate} & -0.05 & -0.07 & 0.06 & 0.02 & -0.33 & -0.24 & -0.26 & -0.46 & -0.07 & 0.16 & 0.47 & 0.05 & 0.29 &  &  &  &  &  \\
\textit{FFR} & -0.00 & 0.05 & 0.02 & 0.04 & 0.02 & -0.04 & 0.07 & 0.04 & -0.01 & -0.02 & -0.03 & -0.10 & 0.01 & -0.03 &  &  &  &  \\
\textit{5yInflExp} & -0.06 & -0.22 & 0.31 & -0.01 & 0.02 & 0.00 & -0.05 & -0.01 & -0.02 & -0.03 & 0.06 & -0.15 & -0.11 & 0.05 & 0.50 &  &  &  \\
\textit{GgleInfl} & -0.06 & -0.17 & -0.13 & 0.03 & -0.03 & -0.01 & 0.03 & -0.05 & -0.03 & -0.02 & 0.06 & 0.27 & -0.01 & 0.03 & 0.47 & 0.56 &  &  \\
\textit{GgleReces} & -0.14 & -0.10 & -0.34 & 0.04 & -0.16 & -0.12 & -0.02 & -0.19 & -0.06 & 0.04 & 0.06 & 0.34 & 0.18 & 0.10 & 0.13 & 0.06 & 0.48 &  \\
\textit{GgleClimate} & -0.03 & -0.01 & -0.02 & -0.04 & -0.07 & 0.03 & 0.04 & -0.03 & 0.03 & 0.01 & 0.02 & 0.03 & -0.00 & 0.06 & 0.09 & 0.12 & 0.16 & 0.07 \\

\bottomrule
\end{tabular}

\end{table}
\end{landscape}

Building on the visual evidence from Figure \ref{fig:btc-monetary-ffr}, we formally test for structural breaks in Bitcoin's return distribution across these identified regimes. Table \ref{tab:regime_tests} reports the results of the Kruskal-Wallis H-test for differences in central tendency, Levene's W-test for homogeneity of variance, and the $k$-sample Anderson--Darling test for equality of the full return distribution.

\begin{table}[H]
\centering
\caption{Non-parametric, Variance and Distributional Tests Across Regimes}
\label{tab:regime_tests}
\caption*{\small \textbf{Notes:} This table presents non-parametric and variance tests for Bitcoin weekly returns across different monetary policy regimes. \textit{MPE} represents the LLM-derived MPE index, and \textit{Fed Funds Rate} denotes the change in the Effective Federal Funds Rate. The Kruskal-Wallis H-test evaluates differences in median returns, while Levene's W-test assesses the homogeneity of variance (volatility) across regimes. The $k$-sample Anderson--Darling test assesses equality of the return distribution. Max Tail Risk (q5) represents the 5th percentile of returns in the most extreme regime. Significance is denoted $^{*}p<0.10$, $^{**}p<0.05$, $^{***}p<0.01$.}
\adjustbox{max width=\textwidth}{%
\begin{tabular}{lccccccc}
\toprule
Variable & Kruskal H & p-val (Med) & Levene W & p-val (Vol) & AD $k$-sample & p-val (Dist) & Max Tail Risk (q5) \\
\midrule
\textit{MPE} & 0.3420 & 0.8428 & 3.0270$^{**}$ & 0.0493 & -0.1437 & 0.2500 & -0.1568 \\
\textit{FFR} & 5.4540$^{*}$ & 0.0654 & 2.6779$^{*}$ & 0.0696 & 2.7000$^{**}$ & 0.0219 & -0.1649 \\
\bottomrule
\end{tabular}}
\end{table}

The empirical results reveal a notable distinction between the market's reaction to policy \textit{realization} versus policy \textit{expectations}. Realized changes in the Fed Funds Rate (\textit{FFR}) exert a statistically significant influence on both median returns ($H=5.45, p<0.10$) and volatility ($W=2.68, p<0.10$) at the 10\% level. This indicates that Bitcoin undergoes a measurable re-pricing and risk adjustment upon the actual implementation of policy shifts. The MPE index, in contrast, leaves median returns unaffected ($p=0.84$) but is the most significant of the two for volatility ($W=3.03$, $p=0.049$), so the distinction runs along the level of returns rather than their dispersion: realized policy moves the median, while expectations move the variance. The Anderson--Darling test sharpens the same contrast: the full return distribution differs across \textit{FFR} regimes ($2.700$, $p=0.022$) but not across \textit{MPE} regimes ($p=0.250$).

The persistence of high tail risk ($q5$ ranging from $-15.68\%$ to $-16.49\%$) across \textit{MPE} and \textit{FFR} further indicates that downside volatility remains a structural feature of the asset class regardless of the macro regime. The contrasting effects on the median and variance are consistent with different transmission mechanisms. Realized rate changes directly affect discount rates, funding costs, and prevailing liquidity conditions, and may consequently alter Bitcoin's central valuation. Policy narratives, by contrast, relate to an uncertain future trajectory and may lead investors to assess their effects on liquidity and risk appetite differently. As a result, conflicting investor responses may increase the dispersion of returns without causing a consistent shift in the median. This interpretation is suggestive of belief heterogeneity rather than constituting a direct test of investor disagreement.

\subsection{Granger Causality and VMD} \label{sec:granger-vmd}

To evaluate the predictive power of our novel index, we conduct standard Granger causality tests. To ensure stationarity as determined by the ADF test, we employ the first differences of the MPE index, 5-year inflation expectations (\textit{5yInflExp}), and Google Search Trends for inflation (\textit{GgleInfl}), alongside the second difference of the Effective Federal Funds Rate (\textit{FFR}).
Table \ref{tab:granger-simple} reports the $p$-values for lags 1 through 6. The results show that changes in the MPE index provide evidence of significant linear predictive information for Bitcoin returns (\textit{Btc}) at the second, third, fifth and sixth lags ($p<0.10$), and at the $5\%$ level at lags two and five. This suggests that while social media-derived policy expectations are a leading indicator, their impact on Bitcoin price formation is not immediate, likely reflecting a delayed transmission of narratives into market activity.

In contrast, several traditional indicators -- including the S\&P 500 (\textit{SP500}) and the \textit{FFR} -- fail to demonstrate significant Granger causality. This suggests that realized policy levels and equity-market movements provide no detectable incremental information for predicting Bitcoin within the linear time-domain specification. Gold (\textit{Gold}), however, exhibits significant predictive power across almost all lags (1–5), while Google Search Trends for recession (\textit{GgleReces}) shows evidence of significant linear predictive information at every tested horizon, at the $5\%$ level for lags 1–4 and at the $10\%$ level for lags 5 and 6. These findings are consistent with the liquidity-channel interpretation developed throughout this paper. Rising public attention to recession risk may signal forthcoming changes in risk appetite and expected monetary policy accommodation, thereby providing information about crypto-asset market conditions several weeks in advance. By contrast, Gold’s predictive power may reflect its shared exposure with Bitcoin to changes in real interest rates and the US dollar, rather than substitution between the two assets as competing stores of value.

\begin{table}[H]
  \centering
  \small
  \renewcommand{\arraystretch}{1.1}
  \caption{Standard Granger causality tests: p-values from SSR F-tests}
  \label{tab:granger-simple}
\captionsetup{justification=raggedright, singlelinecheck=false}
\caption*{\small This table reports $p$-values from standard Granger causality tests (SSR $F$-tests) for lags 1–6. Significant values ($^{*} p<0.10, ^{**} p<0.05$) indicate that the variable Granger-causes Bitcoin (\textit{Btc}) returns. Detailed variable definitions and data sources are provided in Appendix~\ref{sec:def-variables}.
}
\resizebox{0.7\linewidth}{!}{%
\begin{tabular}{lcccccc}
\toprule
 & 1 & 2 & 3 & 4 & 5 & 6 \\
\midrule
\textit{MPE} & 0.109 & 0.039$^{**}$ & 0.095$^{*}$ & 0.111 & 0.032$^{**}$ & 0.066$^{*}$ \\
\textit{NewsSent} & 0.996 & 0.942 & 0.897 & 0.905 & 0.750 & 0.855 \\
\textit{PolUncert} & 0.196 & 0.400 & 0.350 & 0.462 & 0.562 & 0.583 \\
\textit{SP500} & 0.236 & 0.468 & 0.407 & 0.576 & 0.734 & 0.704 \\
\textit{Brent} & 0.998 & 0.400 & 0.561 & 0.469 & 0.345 & 0.435 \\
\textit{Gold} & 0.077$^{*}$ & 0.042$^{**}$ & 0.080$^{*}$ & 0.085$^{*}$ & 0.076$^{*}$ & 0.124 \\
\textit{HighYield} & 0.140 & 0.421 & 0.557 & 0.730 & 0.848 & 0.892 \\
\textit{GeopolRisk} & 0.893 & 0.056$^{*}$ & 0.109 & 0.165 & 0.250 & 0.310 \\
\textit{VIX} & 0.464 & 0.581 & 0.611 & 0.768 & 0.899 & 0.923 \\
\textit{USDollar} & 0.548 & 0.585 & 0.801 & 0.854 & 0.940 & 0.863 \\
\textit{Infect} & 0.364 & 0.358 & 0.226 & 0.298 & 0.346 & 0.340 \\
\textit{JoblessClaim} & 0.866 & 0.590 & 0.698 & 0.735 & 0.606 & 0.363 \\
\textit{ExchRate} & 0.527 & 0.870 & 0.841 & 0.912 & 0.859 & 0.374 \\
\textit{FFR} & 0.936 & 0.795 & 0.944 & 0.940 & 0.320 & 0.451 \\
\textit{5yInflExp} & 0.587 & 0.772 & 0.914 & 0.951 & 0.648 & 0.829 \\
\textit{GgleInfl} & 0.925 & 0.280 & 0.445 & 0.606 & 0.710 & 0.796 \\
\textit{GgleReces} & 0.010$^{**}$ & 0.014$^{**}$ & 0.032$^{**}$ & 0.040$^{**}$ & 0.070$^{*}$ & 0.072$^{*}$ \\
\textit{GgleClimate} & 0.656 & 0.905 & 0.540 & 0.698 & 0.816 & 0.872 \\
\bottomrule
\end{tabular}}
\end{table}

To highlight meaningful relationships that are obscured by high-frequency noise or dominant long-term trends, we apply VMD, partitioning the series into three IMFs: the long-term trend ($I M F_1$), medium-term cycles ($I M F_2$), and high-frequency shocks ($I M F_3$). 

As shown in Table \ref{tab:granger-vmd}, the VMD-based decomposition sharpens the predictive content of the MPE index. The medium-term component of the index (\textit{MPE} $IMF_2$) shows evidence of significant linear predictive information for Bitcoin at all three scales and at nearly every lag. Traditional macrofinancial indicators, including the Effective Federal Funds Rate (\textit{FFR}), the S\&P 500 index (\textit{SP500}), and 5-year inflation expectations (\textit{5yInflExp}), similarly exhibit high statistical significance ($p < 0.01$) across specific frequency components, despite their previous insignificance in the raw series analysis. Specifically, the long-term component of the S\&P 500 (\textit{SP500} $IMF_1$) shows a dominant influence on Bitcoin's primary trajectory (\textit{Btc} $IMF_1$) with a peak F-statistic of $26.260$. This suggests that while Bitcoin may appear decoupled from the S\&P 500 in the aggregate, its underlying long-term trend is deeply anchored by global equity performance. Additionally, the long-term components of the \textit{FFR} (\textit{FFR} $IMF_1$) show a strong influence on Bitcoin's medium-term behavior (\textit{Btc} $IMF_2$) at lag 1, highlighted by a peak F-statistic of 13.574. Furthermore, high-frequency rate components (\textit{FFR} $IMF_3$) significantly drive short-term Bitcoin returns (\textit{Btc} $IMF_3$) at lag 1 (F-statistic = 9.467). This suggests that while raw interest rate levels may appear disconnected from Bitcoin in aggregate tests, structural shifts in monetary policy implementation influence Bitcoin's medium-term and short-term returns. Returning to the MPE index, its effect is by far strongest in the high-frequency domain (\textit{Btc} $IMF_3$), where it peaks at $F=23.196$ at lag 2 and $F=15.104$ at lag 3 -- the largest MPE statistics in the table. The same component also drives the long-term trend (\textit{Btc} $IMF_1$, peak $F=6.144$ at lag 3) and the medium-term cycle (\textit{Btc} $IMF_2$, peak $F=4.340$ at lag 4).

Overall, these cross-frequency results point to distinct economic transmission channels rather than a single aggregate relationship. The association between the long-term components of the S\&P 500 and Bitcoin is consistent with both assets sharing a common risk-appetite and discount-rate factor. The slow-moving component of the \textit{FFR}, by contrast, may influence Bitcoin's medium-term cycle through persistent changes in funding costs and liquidity conditions, whereas high-frequency rate movements generate more immediate repricing. Notably, the medium-term component of \textit{MPE} predicts Bitcoin at all three frequencies, with its strongest effect occurring at high frequency. This suggests that monetary-policy narratives develop at the frequency of the policy cycle but are transmitted most forcefully through short-run risk-asset repricing. Additionally, Geopolitical Risk (\textit{GeopolRisk}) and the VIX index (\textit{VIX}) exhibit pervasive causality at the medium and short-term scales, whereas News Sentiment (\textit{NewsSent}) primarily drives the long-term trend. This heterogeneity justifies the transition to a nonlinear LSTM framework, as traditional linear models fail to integrate these distinct multi-scale dynamics, effectively omitting critical frequency-specific information necessary for accurate forecasting.

% \resizebox{0.99\textwidth}{!}{%
\begingroup
\footnotesize % Adjusting font size to fit width without resizebox
\setlength{\tabcolsep}{4pt} % Tighten column spacing to fit page

\begin{longtable}{lrr|lrr|lrr}
\caption{VMD-Based Granger Causality Results ($p < 0.05$)} \label{tab:granger-vmd} \\
\multicolumn{9}{p{0.99\linewidth}}{\footnotesize \textbf{Note}: This table reports the $F$-statistics and $p$-values for Granger causality tests where the null hypothesis is that $IMF_j^{Predictor}$ does not Granger-cause $IMF_i^{BTC}$. The original series are decomposed into three scales: $IMF_1$ (long-term trend), $IMF_2$ (medium-term), and $IMF_3$ (short-term/high-frequency). The test considers a maximum lag of $p=6$, and only results significant at the 5\% level are reported. Predictor notation $(j, p)$ refers to the $j$-th IMF of the variable at lag $p$.} \\* 
\toprule
\multicolumn{3}{c}{$IMF_1^{B T C}(t)$} & \multicolumn{3}{c}{\textbf{\textit{$I M F_2^{B T C}(t)$}}} 
& \multicolumn{3}{c}{\textbf{\textit{$I M F_3^{B T C}(t)$}}} \\
Predictor(j,p) & F-stat & p-val & Predictor(j,p) & F-stat & p-val & Predictor(j,p) & F-stat & p-val \\
\midrule
\endfirsthead

\multicolumn{9}{c}%
{{\bfseries \tablename\ \thetable{} -- continued from previous page}} \\
\toprule
Predictor(j,p) & F-stat & p-val & Predictor(j,p) & F-stat & p-val & Predictor(j,p) & F-stat & p-val \\
\midrule
\endhead

\midrule
\multicolumn{9}{r}{{Continued on next page}} \\
\bottomrule
\endfoot

\bottomrule
\endlastfoot

$\textit{MPE}(2, 2)$ & 3.062 & 0.048 & $\textit{MPE}(2, 3)$ & 3.648 & 0.013 & $\textit{MPE}(2, 2)$ & 23.196 & 0.000 \\
$\textit{MPE}(2, 3)$ & 6.144 & 0.000 & $\textit{MPE}(2, 4)$ & 4.340 & 0.002 & $\textit{MPE}(2, 3)$ & 15.104 & 0.000 \\
$\textit{MPE}(2, 4)$ & 4.791 & 0.001 & $\textit{MPE}(2, 5)$ & 4.090 & 0.001 & $\textit{MPE}(2, 4)$ & 4.736 & 0.001 \\
$\textit{MPE}(2, 5)$ & 3.971 & 0.002 & $\textit{MPE}(2, 6)$ & 2.865 & 0.009 & $\textit{MPE}(2, 5)$ & 3.551 & 0.004 \\
$\textit{MPE}(2, 6)$ & 3.442 & 0.002 & $\textit{SP500}(1, 2)$ & 4.439 & 0.012 & $\textit{MPE}(2, 6)$ & 3.456 & 0.002 \\
$\textit{NewsSent}(1, 2)$ & 3.842 & 0.022 & $\textit{SP500}(1, 4)$ & 5.267 & 0.000 & $\textit{SP500}(1, 4)$ & 8.199 & 0.000 \\
$\textit{NewsSent}(3, 1)$ & 11.226 & 0.001 & $\textit{SP500}(1, 5)$ & 5.428 & 0.000 & $\textit{SP500}(1, 5)$ & 7.392 & 0.000 \\
$\textit{NewsSent}(3, 2)$ & 4.978 & 0.007 & $\textit{SP500}(1, 6)$ & 4.735 & 0.000 & $\textit{SP500}(1, 6)$ & 7.020 & 0.000 \\
$\textit{SP500}(1, 1)$ & 9.381 & 0.002 & $\textit{SP500}(2, 1)$ & 4.949 & 0.027 & $\textit{SP500}(3, 1)$ & 9.541 & 0.002 \\
$\textit{SP500}(1, 2)$ & 26.260 & 0.000 & $\textit{SP500}(2, 2)$ & 10.917 & 0.000 & $\textit{Brent}(3, 1)$ & 7.823 & 0.005 \\
$\textit{SP500}(1, 3)$ & 5.328 & 0.001 & $\textit{SP500}(2, 3)$ & 7.513 & 0.000 & $\textit{Brent}(3, 2)$ & 3.215 & 0.041 \\
$\textit{SP500}(1, 4)$ & 4.134 & 0.003 & $\textit{SP500}(2, 4)$ & 3.451 & 0.008 & $\textit{Brent}(3, 3)$ & 4.152 & 0.006 \\
$\textit{SP500}(1, 5)$ & 3.351 & 0.005 & $\textit{SP500}(2, 5)$ & 2.859 & 0.015 & $\textit{Gold}(3, 1)$ & 12.000 & 0.001 \\
$\textit{SP500}(1, 6)$ & 2.984 & 0.007 & $\textit{Gold}(1, 4)$ & 3.254 & 0.012 & $\textit{Gold}(3, 2)$ & 7.548 & 0.001 \\
$\textit{SP500}(3, 4)$ & 2.619 & 0.034 & $\textit{Gold}(1, 5)$ & 2.734 & 0.019 & $\textit{Gold}(3, 3)$ & 7.049 & 0.000 \\
$\textit{SP500}(3, 5)$ & 2.736 & 0.019 & $\textit{Gold}(1, 6)$ & 2.122 & 0.049 & $\textit{Gold}(3, 4)$ & 3.174 & 0.014 \\
$\textit{SP500}(3, 6)$ & 2.368 & 0.029 & $\textit{Gold}(2, 1)$ & 11.418 & 0.001 & $\textit{Gold}(3, 5)$ & 2.636 & 0.023 \\
$\textit{Brent}(1, 1)$ & 18.834 & 0.000 & $\textit{Gold}(3, 5)$ & 2.683 & 0.021 & $\textit{HighYield}(1, 4)$ & 4.005 & 0.003 \\
$\textit{Gold}(1, 1)$ & 4.275 & 0.039 & $\textit{Gold}(3, 6)$ & 2.278 & 0.035 & $\textit{HighYield}(1, 5)$ & 3.518 & 0.004 \\
$\textit{Gold}(3, 5)$ & 2.240 & 0.049 & $\textit{HighYield}(1, 4)$ & 2.993 & 0.018 & $\textit{HighYield}(1, 6)$ & 3.337 & 0.003 \\
$\textit{HighYield}(1, 2)$ & 3.977 & 0.019 & $\textit{HighYield}(1, 5)$ & 3.150 & 0.008 & $\textit{HighYield}(2, 4)$ & 2.426 & 0.047 \\
$\textit{HighYield}(2, 2)$ & 4.166 & 0.016 & $\textit{HighYield}(1, 6)$ & 2.675 & 0.014 & $\textit{HighYield}(2, 5)$ & 2.309 & 0.043 \\
$\textit{GeopolRisk}(2, 2)$ & 3.634 & 0.027 & $\textit{HighYield}(2, 2)$ & 5.608 & 0.004 & $\textit{HighYield}(2, 6)$ & 2.447 & 0.024 \\
$\textit{GeopolRisk}(2, 3)$ & 4.834 & 0.002 & $\textit{HighYield}(2, 3)$ & 3.149 & 0.025 & $\textit{GeopolRisk}(2, 2)$ & 7.476 & 0.001 \\
$\textit{GeopolRisk}(2, 4)$ & 4.096 & 0.003 & $\textit{GeopolRisk}(1, 1)$ & 8.173 & 0.004 & $\textit{GeopolRisk}(2, 3)$ & 9.080 & 0.000 \\
$\textit{GeopolRisk}(2, 5)$ & 4.398 & 0.001 & $\textit{GeopolRisk}(2, 2)$ & 5.816 & 0.003 & $\textit{GeopolRisk}(2, 4)$ & 4.507 & 0.001 \\
$\textit{GeopolRisk}(2, 6)$ & 3.545 & 0.002 & $\textit{GeopolRisk}(2, 3)$ & 5.343 & 0.001 & $\textit{GeopolRisk}(2, 5)$ & 2.977 & 0.012 \\
$\textit{VIX}(1, 2)$ & 12.960 & 0.000 & $\textit{GeopolRisk}(2, 4)$ & 2.930 & 0.020 & $\textit{GeopolRisk}(2, 6)$ & 2.378 & 0.028 \\
$\textit{VIX}(1, 3)$ & 3.617 & 0.013 & $\textit{GeopolRisk}(2, 5)$ & 3.248 & 0.007 & $\textit{GeopolRisk}(3, 3)$ & 2.767 & 0.041 \\
$\textit{VIX}(1, 4)$ & 2.776 & 0.026 & $\textit{GeopolRisk}(2, 6)$ & 2.779 & 0.011 & $\textit{VIX}(1, 4)$ & 3.249 & 0.012 \\
$\textit{VIX}(1, 5)$ & 2.509 & 0.029 & $\textit{VIX}(1, 2)$ & 6.499 & 0.002 & $\textit{VIX}(1, 5)$ & 3.451 & 0.004 \\
$\textit{VIX}(2, 3)$ & 3.157 & 0.024 & $\textit{VIX}(1, 3)$ & 3.980 & 0.008 & $\textit{VIX}(1, 6)$ & 3.719 & 0.001 \\
$\textit{VIX}(2, 4)$ & 2.940 & 0.020 & $\textit{VIX}(1, 4)$ & 5.799 & 0.000 & $\textit{VIX}(2, 2)$ & 3.280 & 0.038 \\
$\textit{VIX}(2, 5)$ & 2.457 & 0.032 & $\textit{VIX}(1, 5)$ & 4.789 & 0.000 & $\textit{VIX}(2, 3)$ & 3.407 & 0.017 \\
$\textit{USDollar}(3, 4)$ & 2.733 & 0.028 & $\textit{VIX}(1, 6)$ & 3.998 & 0.001 & $\textit{VIX}(2, 4)$ & 2.593 & 0.036 \\
$\textit{USDollar}(3, 5)$ & 2.300 & 0.044 & $\textit{VIX}(2, 2)$ & 3.712 & 0.025 & $\textit{USDollar}(3, 1)$ & 7.708 & 0.006 \\
$\textit{USDollar}(3, 6)$ & 2.120 & 0.050 & $\textit{VIX}(2, 3)$ & 3.039 & 0.029 & $\textit{USDollar}(3, 2)$ & 3.553 & 0.029 \\
$\textit{Infect}(2, 1)$ & 16.609 & 0.000 & $\textit{VIX}(2, 5)$ & 3.251 & 0.007 & $\textit{USDollar}(3, 3)$ & 3.797 & 0.010 \\
$\textit{Infect}(2, 2)$ & 6.870 & 0.001 & $\textit{VIX}(2, 6)$ & 2.483 & 0.022 & $\textit{Infect}(1, 6)$ & 2.422 & 0.026 \\
$\textit{Infect}(3, 6)$ & 2.540 & 0.020 & $\textit{USDollar}(3, 4)$ & 2.655 & 0.032 & $\textit{Infect}(2, 4)$ & 2.576 & 0.037 \\
$\textit{JoblessClaim}(1, 1)$ & 26.299 & 0.000 & $\textit{USDollar}(3, 5)$ & 2.471 & 0.032 & $\textit{Infect}(2, 5)$ & 2.415 & 0.035 \\
$\textit{JoblessClaim}(1, 2)$ & 12.598 & 0.000 & $\textit{Infect}(1, 6)$ & 2.139 & 0.048 & $\textit{Infect}(2, 6)$ & 2.783 & 0.011 \\
$\textit{ExchRate}(1, 2)$ & 3.099 & 0.046 & $\textit{Infect}(2, 6)$ & 2.270 & 0.036 & $\textit{JoblessClaim}(1, 6)$ & 2.162 & 0.045 \\
$\textit{FFR}(2, 4)$ & 2.533 & 0.040 & $\textit{Infect}(3, 2)$ & 3.758 & 0.024 & $\textit{ExchRate}(3, 2)$ & 4.929 & 0.008 \\
$\textit{5yInflExp}(2, 2)$ & 4.058 & 0.018 & $\textit{Infect}(3, 3)$ & 3.398 & 0.018 & $\textit{ExchRate}(3, 3)$ & 3.319 & 0.020 \\
$\textit{5yInflExp}(2, 3)$ & 5.322 & 0.001 & $\textit{JoblessClaim}(3, 2)$ & 4.479 & 0.012 & $\textit{FFR}(1, 6)$ & 2.223 & 0.040 \\
$\textit{5yInflExp}(2, 4)$ & 5.347 & 0.000 & $\textit{JoblessClaim}(3, 3)$ & 3.337 & 0.019 & $\textit{FFR}(2, 2)$ & 4.525 & 0.011 \\
$\textit{5yInflExp}(2, 5)$ & 4.666 & 0.000 & $\textit{FFR}(1, 1)$ & 13.574 & 0.000 & $\textit{FFR}(2, 3)$ & 4.059 & 0.007 \\
$\textit{5yInflExp}(2, 6)$ & 4.389 & 0.000 & $\textit{FFR}(2, 5)$ & 2.537 & 0.028 & $\textit{FFR}(3, 1)$ & 9.467 & 0.002 \\
$\textit{GgleReces}(3, 4)$ & 3.011 & 0.018 & $\textit{5yInflExp}(1, 1)$ & 6.399 & 0.012 & $\textit{5yInflExp}(2, 2)$ & 3.730 & 0.025 \\
$\textit{GgleReces}(3, 5)$ & 2.638 & 0.023 & $\textit{5yInflExp}(1, 3)$ & 4.040 & 0.007 & $\textit{5yInflExp}(2, 3)$ & 7.251 & 0.000 \\
$\textit{GgleReces}(3, 6)$ & 2.326 & 0.032 & $\textit{5yInflExp}(1, 4)$ & 2.934 & 0.020 & $\textit{5yInflExp}(2, 4)$ & 4.564 & 0.001 \\
$\textit{GgleClimate}(1, 1)$ & 7.656 & 0.006 & $\textit{5yInflExp}(1, 5)$ & 2.655 & 0.022 & $\textit{5yInflExp}(2, 5)$ & 2.992 & 0.011 \\
 &  &  & $\textit{5yInflExp}(1, 6)$ & 2.358 & 0.029 & $\textit{5yInflExp}(2, 6)$ & 2.157 & 0.046 \\
 &  &  & $\textit{5yInflExp}(2, 2)$ & 9.442 & 0.000 & $\textit{GgleReces}(1, 5)$ & 2.872 & 0.014 \\
 &  &  & $\textit{5yInflExp}(2, 3)$ & 8.836 & 0.000 & $\textit{GgleReces}(1, 6)$ & 2.417 & 0.026 \\
 &  &  & $\textit{5yInflExp}(2, 4)$ & 2.545 & 0.039 & $\textit{GgleReces}(3, 2)$ & 4.707 & 0.009 \\
 &  &  & $\textit{5yInflExp}(2, 5)$ & 2.825 & 0.016 & $\textit{GgleReces}(3, 3)$ & 4.905 & 0.002 \\
 &  &  & $\textit{5yInflExp}(2, 6)$ & 2.222 & 0.040 & $\textit{GgleReces}(3, 6)$ & 2.361 & 0.029 \\
 &  &  & $\textit{GgleInfl}(2, 2)$ & 3.111 & 0.045 &  &  &  \\
 &  &  & $\textit{GgleReces}(3, 4)$ & 2.512 & 0.041 &  &  &  \\
\end{longtable}
\endgroup

\subsection{Out of Sample Model Evaluation} \label{sec:forecast}

The frequency-decomposed Granger analysis reveals a scale-dependent causal structure linking monetary policy expectations to Bitcoin returns. We now evaluate whether these multi-scale relationships translate into tangible out-of-sample predictive information when embedded within the nonlinear LSTM framework (Section \ref{sec:lstm}).
%
%To this end, we implement a walk-forward validation framework across four distinct macroeconomic folds, as presented in Figure \ref{fig:train-val-test}. 
The statistical significance of our model's performance is assessed against an optimized ARIMA baseline using the Diebold-Mariano (DM) test for equal predictive accuracy. This provides an out-of-sample benchmark for the LSTM framework rather than constituting the primary objective of the analysis.

Table \ref{tab:model_comparison} summarizes the 1-step-ahead out-of-sample performance of the LSTM ensemble compared to the ARIMA baseline. 
The results show that the LSTM achieves a lower mean RMSE than the ARIMA ($0.0850$ versus $0.0873$), corresponding to a $2.6\%$ reduction in average RMSE. The LSTM achieves a lower RMSE in two of the four folds, with the largest improvement occurring in Fold $1$, where RMSE falls from $0.1295$ for ARIMA to $0.1146$ for the LSTM. In Fold $3$, the LSTM similarly achieves a lower RMSE ($0.0582$ versus $0.0671$). The standard deviation of the LSTM RMSE across folds ($0.022$) is also comparable to that of ARIMA ($0.026$), indicating broadly stable performance across the different macroeconomic regimes considered, despite the high volatility and idiosyncratic noise characteristic of Bitcoin returns.

The DM test results reveal that 
%the null hypothesis of equal predictive accuracy cannot be rejected ($p > 0.05$), indicating that 
the differences in predictive accuracy are not statistically significant in the first three folds ($p > 0.05$). In Fold $4$, the loss differential is statistically significant ($p = 0.016$), although the ARIMA model achieves the lower RMSE in this period.\footnote{To isolate the contribution of the index itself from that of the nonlinear architecture, we further compare a univariate ARIMA with an ARIMAX augmented by the MPE in Appendix \ref{sec:arimax-mpe}: adding the index lowers the pooled RMSE from $0.0902$ to $0.0895$, so the MPE carries genuine predictive information for Bitcoin returns even within a purely linear model.} Thus, we do not interpret the results as demonstrating that the LSTM is uniformly or statistically superior to ARIMA as a forecasting model. Rather, the out-of-sample comparison serves as a predictive adequacy check: the LSTM provides a competitive nonlinear representation of Bitcoin returns and, on average, achieves a lower RMSE than the linear benchmark. This provides a basis for examining the economic structure learned by the nonlinear model without making forecasting superiority the central claim.

As a robustness analysis, we additionally report in Appendix \ref{sec:model-info} an alternative, strictly out-of-sample selection criterion based on the lowest test-set RMSE among the independently retrained LSTM models. This criterion is not used for model development or hyperparameter optimization and is therefore not part of the primary forecasting procedure. Instead, it provides a deliberately stringent sensitivity check on the interpretability results by examining whether the principal SHAP patterns persist when the LSTM realization with the strongest observed out-of-sample performance is considered. Under this criterion, the resulting LSTM ensemble achieves a mean RMSE of $0.0759$, compared with $0.0873$ for the ARIMA benchmark (a reduction of $\sim 13\%$), with lower RMSE in three of the four folds.
Detailed results for the full suite of LSTM searches and runs, including architectural parameters and validation performance,  are reported in Table \ref{tab:lstm_metrics} in Appendix \ref{sec:model-info}.

\begin{table}[H]
\centering
\caption{Predictive Performance Comparison: LSTM vs. ARIMA Baseline}
\label{tab:model_comparison}
\begin{small}
\caption*{\small \textbf{Notes:} This table reports the out-of-sample forecasting performance comparison between the LSTM Ensemble and the ARIMA baseline across four sequential walk-forward folds. Performance is measured by the Root Mean Square Error (RMSE) and Mean Absolute Error (MAE). The Diebold-Mariano (DM) $p$-value ($p$-val) indicates the statistical significance of the predictive accuracy between models. ARIMA $(p, d, q)$ orders are optimized per fold via the Akaike Information Criterion (AIC). Overall Mean values and Standard Deviations (Std Dev) are calculated across the four folds, with the latter provided in parentheses. Hyperparameter configuration is selected according to validation performance.}
\end{small}
\resizebox{0.9\textwidth}{!}{%
\begin{tabular}{@{}lcccccc@{}}
\toprule
 & \multicolumn{3}{c}{LSTM Ensemble} & \multicolumn{3}{c}{ARIMA (Baseline)} \\
\cmidrule(lr){2-4} \cmidrule(l){5-7}
Fold & RMSE & MAE & DM $p$-val & RMSE & MAE & Order $(p,d,q)$ \\ \midrule
 1 & 0.1146 & 0.0926 & 0.450 & 0.1295 & 0.1067 & (2, 0, 2) \\
 2 & 0.0965 & 0.0831 & 0.443 & 0.0876 & 0.0623 & (2, 0, 2) \\
 3 & 0.0582 & 0.0428 & 0.684 & 0.0671 & 0.0477 & (2, 0, 2) \\
 4 & 0.0705 & 0.0562 & 0.016 & 0.0651 & 0.0497 & (0, 0, 1) \\
\midrule
Overall (Mean) & 0.0850 & 0.0687 & --- & 0.0873 & 0.0666 & --- \\
Overall (Std Dev) & (0.022) & (0.020) & --- & (0.026) & (0.024) & --- \\
\bottomrule
\end{tabular}}
\end{table}

%\FS{[FS: I think, here, it is clear that FOLDS are different because they represent changes in market regime? However, at the moment, the diference in prediction between LSTM and ARIMA is still significative (like $35\%$). Could we try if considering 4 folds (or even 3 if necessary) reduces this difference in prediction? I know that we are interested in explainability with the SHAP, something that ARIMA does not provide, but it would be solid if we can reduce this difference in predition... If we convert the std into standard error, then we do not have overlap between the prediction error bars, so we cannot even use this argument to reconcialiate both approaches. Also, Best Val Loss for FOLD2 shows a std bigger than the mean, which is a bit weak.]}

\subsection{Explainability Analysis} \label{sec:explainability}

\begin{figure}[htbp]
    \caption{Mean Absolute SHAP Values: Feature Importance.}
    \label{fig:shap_grid}
    \centering
    \resizebox{\textwidth}{!}{%
    \begin{tabular}{p{\textwidth}}{
    \small \textbf{Notes:} Panels (a) through (d) illustrate the mean absolute SHAP values for each of the four cross-validation folds, while Panel (e) presents the cross-fold average. Error bars represent $\pm 1$ standard deviation around the mean. Features are ranked in descending order of their aggregate mean absolute contribution.}
    \end{tabular}}

    % Row 1
    \begin{subfigure}[b]{0.49\textwidth}
        \centering
        \caption{Fold 1}
        \vspace{-2pt} % Reduces space between title and plot
        \includegraphics[width=\linewidth]{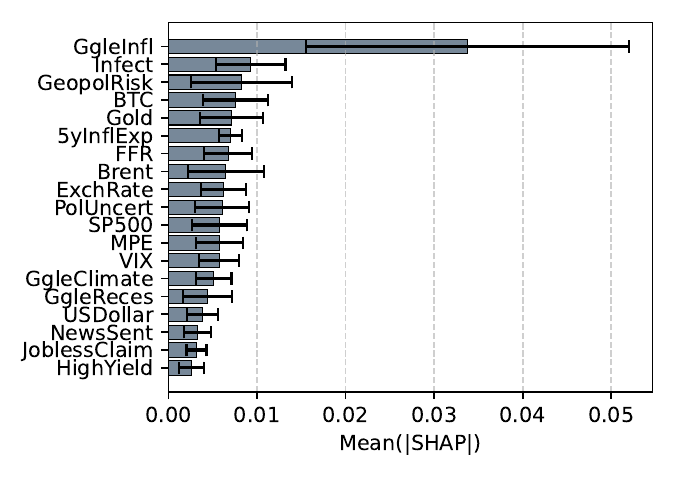}
        \label{fig:shap_f1}
    \end{subfigure}
    \begin{subfigure}[b]{0.49\textwidth}
        \centering
        \caption{Fold 2}
        \vspace{-2pt}
        \includegraphics[width=\linewidth]{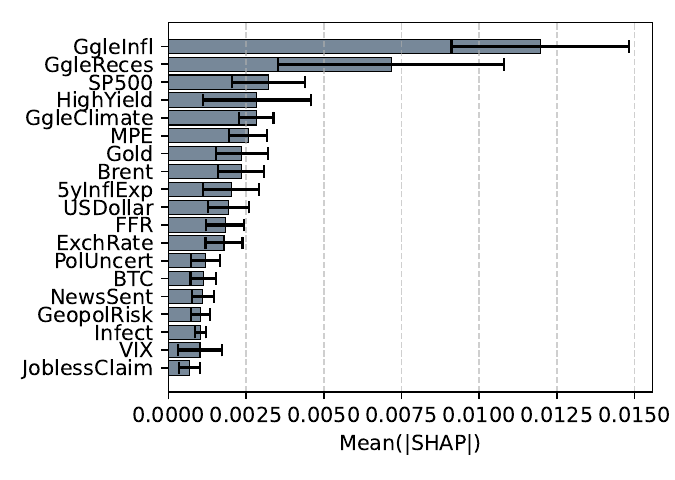}
        \label{fig:shap_f2}
    \end{subfigure}

    \vspace{-10pt} % Reduces vertical space between rows

    % Row 2
    \begin{subfigure}[b]{0.49\textwidth}
        \centering
        \caption{Fold 3}
        \vspace{-2pt}
        \includegraphics[width=\linewidth]{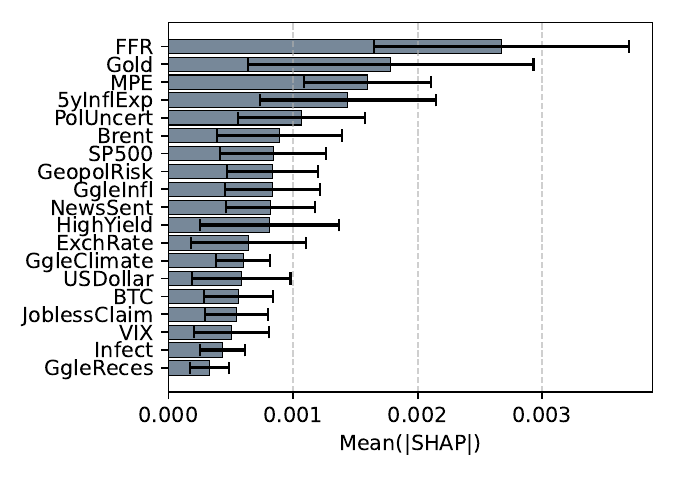}
        \label{fig:shap_f3}
    \end{subfigure}
    \begin{subfigure}[b]{0.49\textwidth}
        \centering
        \caption{Fold 4}
        \vspace{-2pt}
        \includegraphics[width=\linewidth]{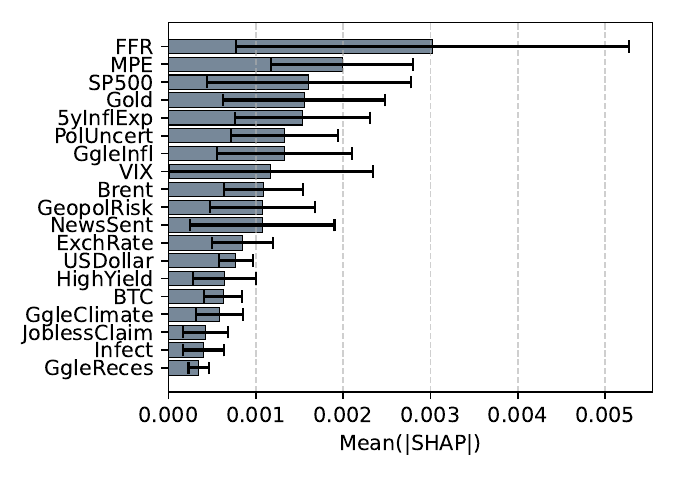}
        \label{fig:shap_f4}
    \end{subfigure}

    \vspace{-10pt}

    % Row 3
    \begin{subfigure}[b]{0.48\textwidth}
        \centering
        \caption{Average}
        \includegraphics[width=\linewidth]{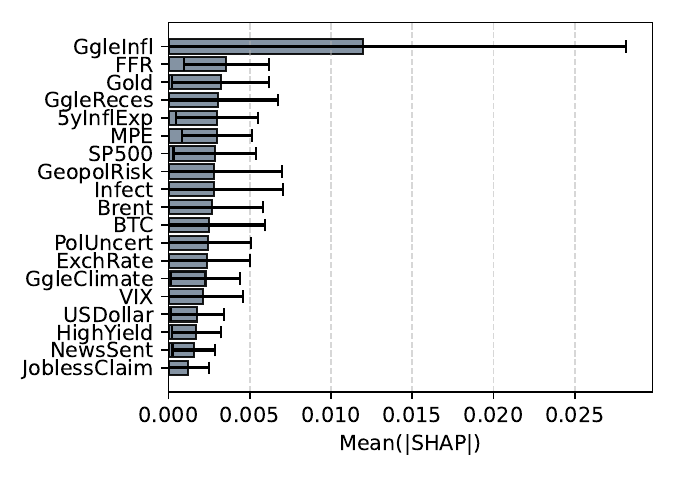}
        \label{fig:shap_avg}
    \end{subfigure}
    
\end{figure}

The relative contribution of the MPE index and each macroeconomic variable to the model's predictive power is illustrated in Figure~\ref{fig:shap_grid}. We employ SHAP values to quantify feature importance, providing a theoretically grounded approach to interpreting the nonlinear mappings of the LSTM network. Panels (a -- d) present the average importance rankings for each individual walk-forward fold, reflecting the sensitivity of the model to different market regimes. Panel (e) displays the aggregate mean absolute SHAP values, identifying the primary drivers of Bitcoin (\textit{Btc}) returns over the entire sample period.

On average across all folds, the MPE index ranks among the top predictors, confirming that monetary policy narratives carry significant weight in the model's decision-making process. Panel (e) reveals that Google Search Trends for inflation (\textit{GgleInfl}) emerge as the most important predictor, ahead of the \textit{FFR}, gold (\textit{Gold}), Google Search Trends for recession (\textit{GgleReces}), and 5-year inflation expectations (\textit{5yInflExp}), with the \textit{MPE} ranking sixth of nineteen. Retail investor attention to inflation captures the related forward-looking macro anxiety that our sentiment index is designed to measure; together, these variables suggest that Bitcoin pricing is predominantly shaped by the broader inflation-monetary policy context rather than by traditional financial market indicators.  Conversely, News Sentiment (\textit{NewsSent}) and Jobless Claims (\textit{JoblessClaim}) appear at the bottom across all folds, indicating that general news sentiment and labor market conditions are comparatively less informative for Bitcoin returns. 

The strong showing of the Google Trends variables deserves comment. \textit{GgleInfl} is the most important predictor on average and \textit{GgleReces} ranks fourth, placing retail search behavior alongside the \textit{FFR} and market-based inflation expectations. The fact that two attention measures outrank most financial variables suggests that Bitcoin responds to the salience of inflation and recession risk in public discourse, and not only to its realization in prices. The contribution of \textit{GgleReces} is, however, concentrated in the first two folds.

The prominence of the \textit{MPE} alongside the \textit{FFR}, rather than being subsumed by it, confirms that expectation-based narratives carry information not captured by realized policy actions, directly supporting our central hypothesis. However, the ranking of the \textit{MPE} varies across folds: it places twelfth in Fold 1 and sixth in Fold 2 --- covering the post-COVID liquidity surge and the initial inflation shock --- before rising to third in Fold 3 and second in Fold 4, precisely where the realized \textit{FFR} ranks first. This reordering suggests that Bitcoin's relevant information set shifts as inflation moves from an emerging narrative to realized policy action: Google searches for inflation (\textit{GgleInfl}) dominate the early folds, when public attention may have carried information about inflation risk before it was reflected in policy instruments, whereas the \textit{FFR} and \textit{MPE} become more central once policy is being implemented. The higher rank of the \textit{MPE} does not, however, reflect a larger absolute contribution: its mean absolute SHAP value declines from 0.006 in Fold 1 to 0.003 in Fold 2 and 0.002 in Folds 3 and 4, so the rise in rank comes from the sharper decline of competing predictors. What changes across folds is the relative composition of predictive information rather than the strength of the \textit{MPE}'s own effect. The wide error bars in Panel (e) point in the same direction, reinforcing the value of walk-forward validation for assessing the robustness of feature importance in volatile financial series.

Table \ref{tab:global_shap_val_loss} confirms the predictive significance of the \textit{MPE}, which ranks as the sixth most influential predictor in the LSTM model with a mean absolute SHAP value of $0.003$. While the realized \textit{FFR} holds a slightly higher average importance (Rank 2, $0.004$), the proximity of these values indicates that policy expectations carry nearly as much weight in Bitcoin return formation as actual policy implementation. Notably, in Fold 2 --- the period of greatest macro instability --- the \textit{MPE} ($0.003$) outweighs the \textit{FFR} ($0.002$), suggesting that forward-looking narratives become a more dominant price driver during regimes of unprecedented policy shifts. The MPE's contribution is also the most evenly distributed across folds of any leading predictor.
%The consistently high ranking of the index validates its role as a primary, non-redundant signal that captures market sensitivity to monetary discourse independently of realized rate adjustments.

\begin{table}[H]
\centering
\caption{Global Feature Importance: Aggregated Mean Absolute SHAP Values.}
\label{tab:global_shap_val_loss}
\begin{small}
\caption*{\small \textbf{Notes:} This table reports the global feature importance for the LSTM forecasting model, quantified using aggregated Mean Absolute SHAP (SHapley Additive exPlanations) values. The variables are ranked based on their average contribution to the model's predictions across four cross-validation folds; values in parentheses represent the standard deviation of these contributions. Detailed definitions, data sources, and transformation procedures for all variables are provided in Appendix~\ref{sec:def-variables}.}
\end{small}
\resizebox{0.99\textwidth}{!}{%
\begin{tabular}{l c c llll}
\toprule
 & Rank & Average & Fold 1 & Fold 2 & Fold 3 & Fold 4  \\
Variable &  &  &  &  &  &   \\
\midrule
\textit{GgleInfl} & 1 & 0.012 (0.016) & 0.034 (0.018) & 0.012 (0.003) & 0.001 (0.000) & 0.001 (0.001) \\
\textit{FFR} & 2 & 0.004 (0.003) & 0.007 (0.003) & 0.002 (0.001) & 0.003 (0.001) & 0.003 (0.002) \\
\textit{Gold} & 3 & 0.003 (0.003) & 0.007 (0.004) & 0.002 (0.001) & 0.002 (0.001) & 0.002 (0.001) \\
\textit{GgleReces} & 4 & 0.003 (0.004) & 0.004 (0.003) & 0.007 (0.004) & 0.000 (0.000) & 0.000 (0.000) \\
\textit{5yInflExp} & 5 & 0.003 (0.003) & 0.007 (0.001) & 0.002 (0.001) & 0.001 (0.001) & 0.002 (0.001) \\
\textit{MPE} & 6 & 0.003 (0.002) & 0.006 (0.003) & 0.003 (0.001) & 0.002 (0.001) & 0.002 (0.001) \\
\textit{SP500} & 7 & 0.003 (0.003) & 0.006 (0.003) & 0.003 (0.001) & 0.001 (0.000) & 0.002 (0.001) \\
\textit{GeopolRisk} & 8 & 0.003 (0.004) & 0.008 (0.006) & 0.001 (0.000) & 0.001 (0.000) & 0.001 (0.001) \\
\textit{Infect} & 9 & 0.003 (0.004) & 0.009 (0.004) & 0.001 (0.000) & 0.000 (0.000) & 0.000 (0.000) \\
\textit{Brent} & 10 & 0.003 (0.003) & 0.006 (0.004) & 0.002 (0.001) & 0.001 (0.001) & 0.001 (0.000) \\
\textit{BTC} & 11 & 0.002 (0.003) & 0.008 (0.004) & 0.001 (0.000) & 0.001 (0.000) & 0.001 (0.000) \\
\textit{PolUncert} & 12 & 0.002 (0.003) & 0.006 (0.003) & 0.001 (0.000) & 0.001 (0.001) & 0.001 (0.001) \\
\textit{ExchRate} & 13 & 0.002 (0.003) & 0.006 (0.003) & 0.002 (0.001) & 0.001 (0.000) & 0.001 (0.000) \\
\textit{GgleClimate} & 14 & 0.002 (0.002) & 0.005 (0.002) & 0.003 (0.001) & 0.001 (0.000) & 0.001 (0.000) \\
\textit{VIX} & 15 & 0.002 (0.002) & 0.006 (0.002) & 0.001 (0.001) & 0.001 (0.000) & 0.001 (0.001) \\
\textit{USDollar} & 16 & 0.002 (0.002) & 0.004 (0.002) & 0.002 (0.001) & 0.001 (0.000) & 0.001 (0.000) \\
\textit{HighYield} & 17 & 0.002 (0.002) & 0.003 (0.001) & 0.003 (0.002) & 0.001 (0.001) & 0.001 (0.000) \\
\textit{NewsSent} & 18 & 0.002 (0.001) & 0.003 (0.002) & 0.001 (0.000) & 0.001 (0.000) & 0.001 (0.001) \\
\textit{JoblessClaim} & 19 & 0.001 (0.001) & 0.003 (0.001) & 0.001 (0.000) & 0.001 (0.000) & 0.000 (0.000) \\
\bottomrule
\end{tabular}}
\end{table}

\begin{figure}[htbp]
    \caption{Mean Absolute SHAP values: Feature Importance and Lag ($t-n$).}
    \label{fig:shap_lag_grid}
    \centering
    \resizebox{\textwidth}{!}{%
    \begin{tabular}{p{\textwidth}}
    \small{\small \textbf{Notes:} This figure decomposes the mean absolute SHAP values into specific weekly lags. Panels (a) through (d) illustrate the variation in lag dynamics across the four cross-validation folds, while Panel (e) reports the aggregate cross-fold average profile.}
    \end{tabular}}

    % Row 1
    \begin{subfigure}[b]{0.49\textwidth}
        \centering
        \caption{Fold 1}
        \vspace{-2pt}
        \includegraphics[width=\linewidth]{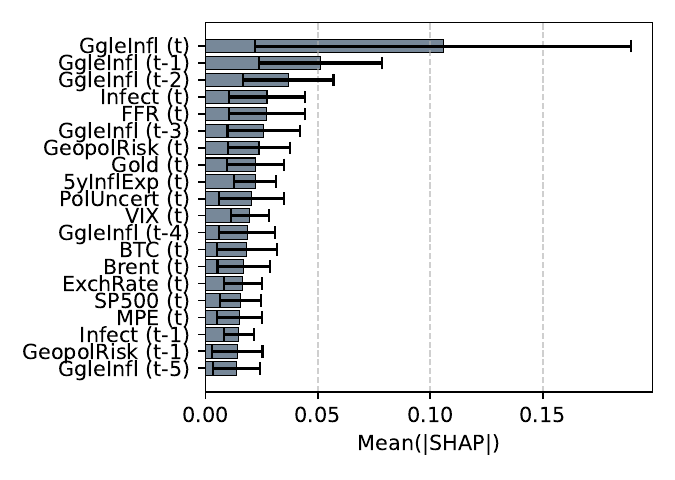}
        \label{fig:shap_lag_f1}
    \end{subfigure}
    \begin{subfigure}[b]{0.49\textwidth}
        \centering
        \caption{Fold 2}
        \vspace{-2pt}
        \includegraphics[width=\linewidth]{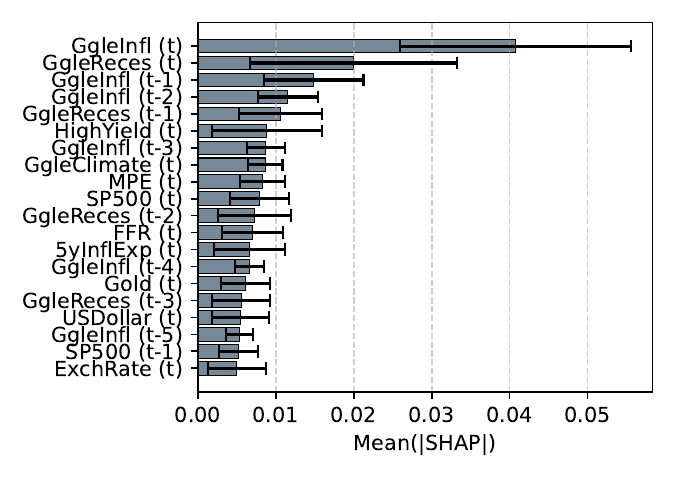}
        \label{fig:shap_lag_f2}
    \end{subfigure}

    \vspace{-10pt} % Reduces space between rows

    % Row 2
    \begin{subfigure}[b]{0.49\textwidth}
        \centering
        \caption{Fold 3}
        \vspace{-5pt}
        \includegraphics[width=\linewidth]{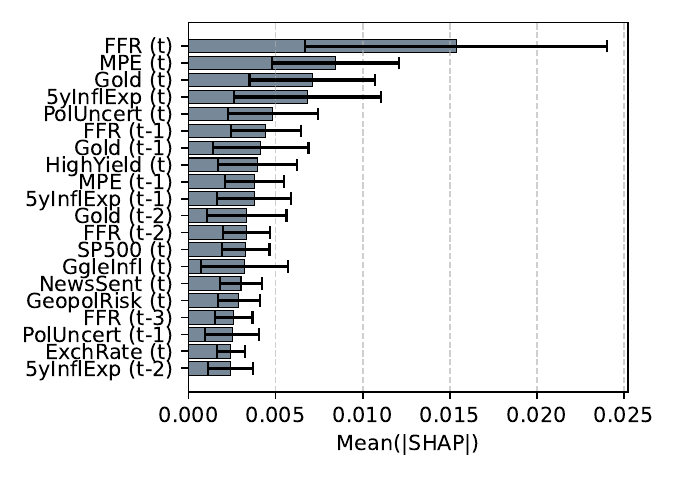}
        \label{fig:shap_lag_f3}
    \end{subfigure}
    \begin{subfigure}[b]{0.49\textwidth}
        \centering
        \caption{Fold 4}
        \vspace{-2pt}
        \includegraphics[width=\linewidth]{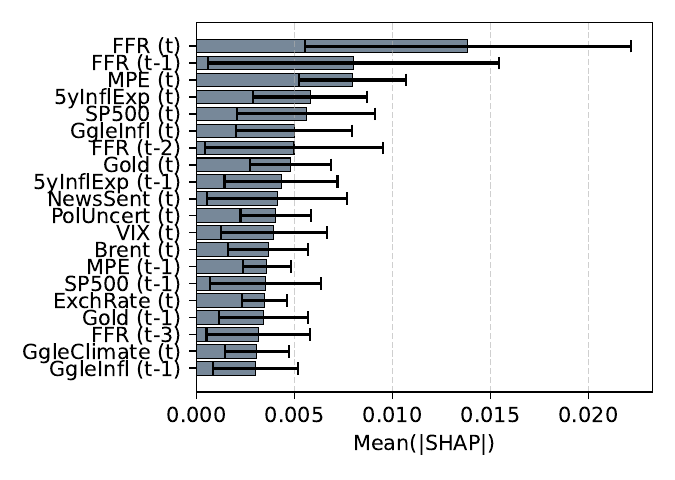}
        \label{fig:shap_lag_f4}
    \end{subfigure}

    \vspace{-10pt}

    % Row 3
    \begin{subfigure}[b]{0.49\textwidth}
        \centering
        \caption{Average}
        \vspace{-2pt}
        \includegraphics[width=\linewidth]{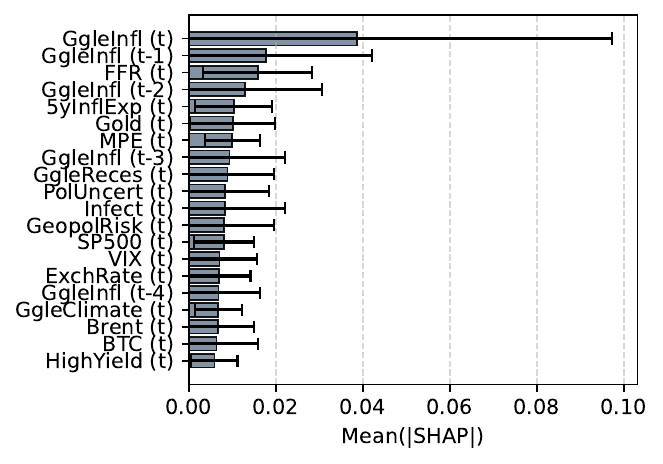}
        \label{fig:shap_lag_avg}
    \end{subfigure}

\end{figure}

Table~\ref{tab:disaggregated_shap} and Figure~\ref{fig:shap_lag_grid} disaggregate the SHAP values by weekly lag and show that predictive importance is heavily concentrated at lag $t$. Of the twenty highest-ranked feature--lag pairs, fifteen are dated at lag $t$, including the \textit{FFR} (Rank 3), \textit{5yInflExp} (Rank 5), \textit{Gold} (Rank 6) and the MPE index (Rank 7, 0.010). The only variable that retains importance further back in the lookback window is Google Search Trends for inflation (\textit{GgleInfl}), which appears at every lag from $t$ to $t-4$ (Ranks 1, 2, 4, 8 and 16). 

\begin{figure}[H]
    \centering
    \caption{Comparative Temporal Attribution Signatures of Top 10 Features.}\label{fig:persistence_comparative}
    \centering
    \resizebox{\textwidth}{!}{%
    \begin{tabular}{p{\textwidth}}
    \small{ \textbf{Notes:} The plot displays the Mean Absolute SHAP values across the lookback horizon ($t$ to $t-n$), aggregated over four walk-forward folds. Markers and line styles differentiate the top-ranked predictors, illustrating the heterogeneity in their predictive persistence.}
    \end{tabular}}
    \includegraphics[width=0.95\linewidth]{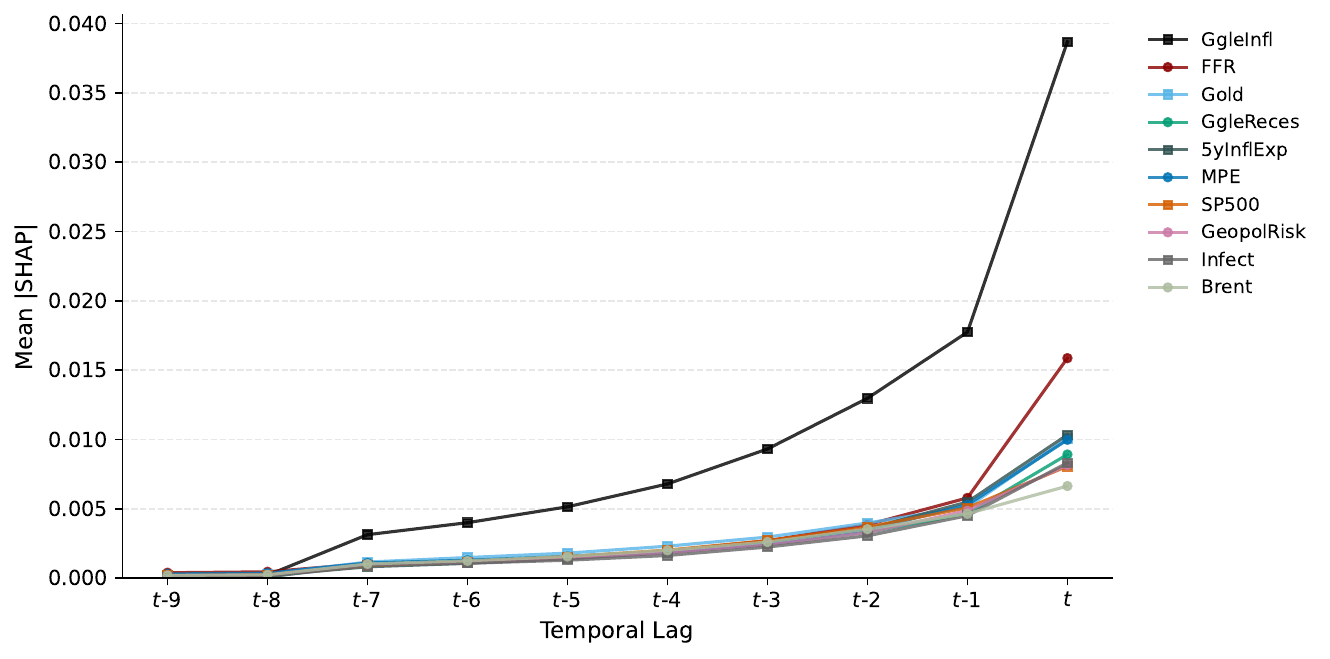}
\end{figure}

\begin{figure}[H]
    \centering
    \caption{Temporal Attribution Signatures of the Top 10 Covariates.}    \label{fig:temporal_stability}
    \centering
    \resizebox{\textwidth}{!}{%
    \begin{tabular}{p{\textwidth}}
    \small{ \textbf{Notes:} Each panel illustrates the evolution of predictive importance across the temporal lookback window ($t-n$ to $t$). The solid line represents the mean absolute SHAP value across all cross-validation folds, while the shaded area indicates the standard deviation.}
    \end{tabular}}
    \includegraphics[width=1.0\linewidth]{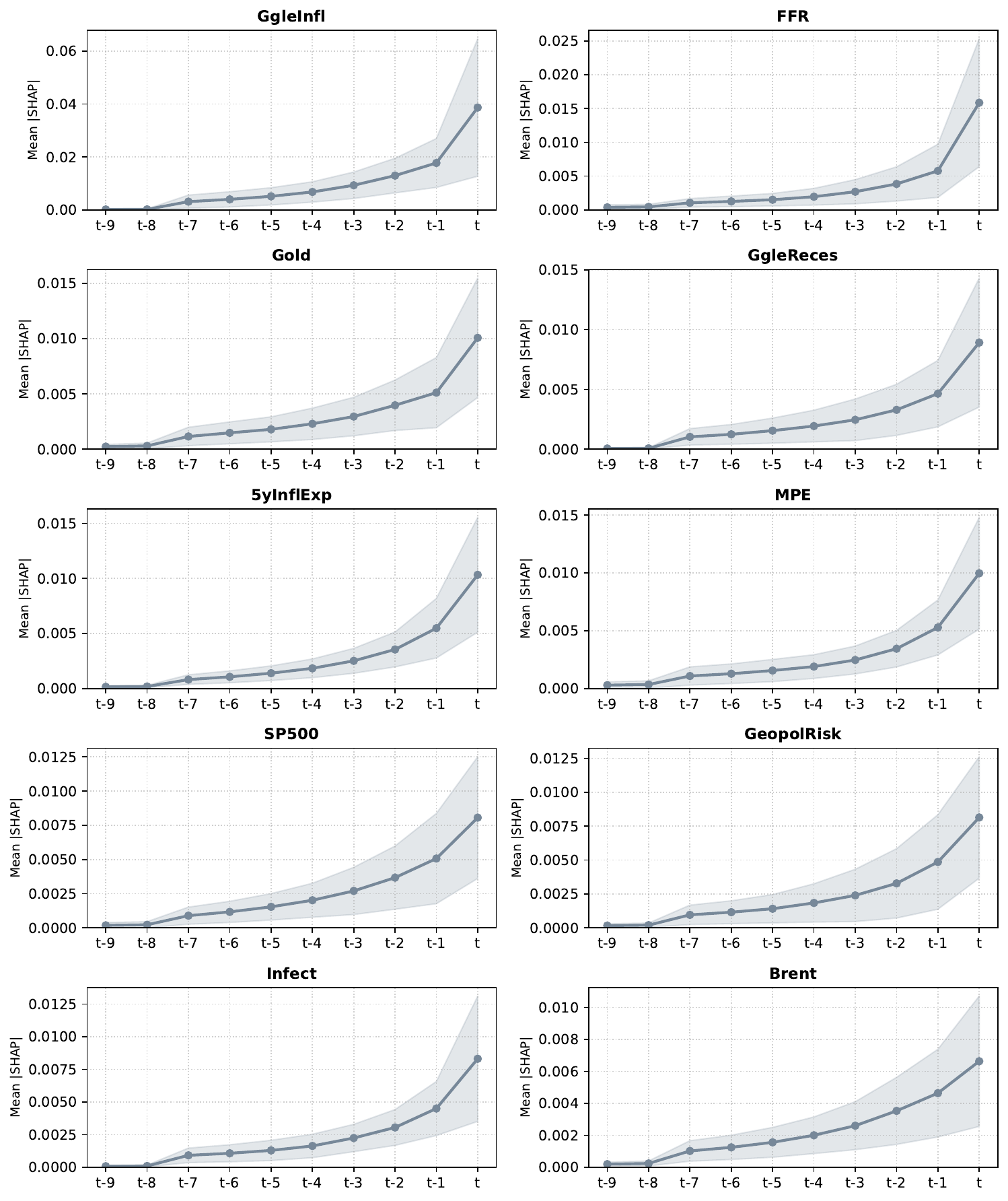}
\end{figure}

Figures~\ref{fig:persistence_comparative} and~\ref{fig:temporal_stability} display the same pattern across the full lookback horizon. For every one of the top ten covariates, mean absolute SHAP values rise monotonically as the lag shortens, followed by a pronounced step up at $t$. The predictors therefore differ in amplitude rather than in shape: \textit{GgleInfl} reaches $0.039$ at $t$ and separates visibly from the rest from $t-8$ onwards, the \textit{FFR} rises to $0.016$, and \textit{MPE}, \textit{Gold} and \textit{5yInflExp} converge around $0.010$, while the remaining series follow the same profile at lower levels. The concentration of predictive importance at lag $t$ indicates that the LSTM places greatest weight on the most recent information when forecasting returns at $(t+1)$, consistent with rapid weekly incorporation of macroeconomic signals and limited predictive memory. Google Search Trends for inflation (\textit{GgleInfl}) are the principal exception, retaining importance from $(t)$ to $(t-4)$. This persistence may reflect the gradual accumulation and diffusion of public attention, whereas the concentration of MPE importance at $(t)$ suggests that policy narratives are incorporated into Bitcoin prices more rapidly.

The concentration of SHAP importance at $t$ is not inconsistent with Granger-causality results,  which indicate MPE predictability across several longer lags. The two methods capture different dimensions of the relationship. On the one hand, the bivariate Granger tests identify distributed linear predictive effects, on the other hand, SHAP measures the conditional contribution of each lag within a multivariate nonlinear model that forecasts $t+1$. Taken together, the findings suggest that the most recent MPE observation provides a strong short-term nonlinear signal, while part of its effect is transmitted through slower linear relationships over the following weeks.

\begin{table}[H]
\centering
\caption{Disaggregated Feature-Lag Importance: Mean Absolute SHAP Values by Lag.}
\label{tab:disaggregated_shap}
\begin{small}
\caption*{\small \textbf{Notes:} This table provides a disaggregated view of feature importance by breaking down the SHAP values into specific time lags for each predictor. The variables are ranked based on their average contribution to the model's predictions across four cross-validation folds; values in parentheses represent the standard deviation of these contributions. Detailed definitions, data sources, and transformation procedures for all variables are provided in Appendix~\ref{sec:def-variables}.}
\end{small}
\resizebox{0.99\textwidth}{!}{%
\begin{tabular}{l c c llll}
\toprule
 & Rank & Average & Fold 1 & Fold 2 & Fold 3 & Fold 4  \\
Variable &  &  &  &  &  &   \\
\midrule
\textit{GgleInfl (t)} & 1 & 0.039 (0.059) & 0.106 (0.084) & 0.041 (0.015) & 0.003 (0.003) & 0.005 (0.003) \\
\textit{GgleInfl (t-1)} & 2 & 0.018 (0.024) & 0.051 (0.027) & 0.015 (0.006) & 0.002 (0.001) & 0.003 (0.002) \\
\textit{FFR (t)} & 3 & 0.016 (0.013) & 0.027 (0.017) & 0.007 (0.004) & 0.015 (0.009) & 0.014 (0.008) \\
\textit{GgleInfl (t-2)} & 4 & 0.013 (0.018) & 0.037 (0.020) & 0.012 (0.004) & 0.001 (0.001) & 0.002 (0.002) \\
\textit{5yInflExp (t)} & 5 & 0.010 (0.009) & 0.022 (0.009) & 0.007 (0.005) & 0.007 (0.004) & 0.006 (0.003) \\
\textit{Gold (t)} & 6 & 0.010 (0.010) & 0.022 (0.013) & 0.006 (0.003) & 0.007 (0.004) & 0.005 (0.002) \\
\textit{MPE (t)} & 7 & 0.010 (0.006) & 0.015 (0.010) & 0.008 (0.003) & 0.008 (0.004) & 0.008 (0.003) \\
\textit{GgleInfl (t-3)} & 8 & 0.009 (0.013) & 0.026 (0.016) & 0.009 (0.002) & 0.001 (0.000) & 0.002 (0.001) \\
\textit{GgleReces (t)} & 9 & 0.009 (0.011) & 0.013 (0.007) & 0.020 (0.013) & 0.001 (0.001) & 0.002 (0.001) \\
\textit{PolUncert (t)} & 10 & 0.008 (0.010) & 0.021 (0.014) & 0.004 (0.002) & 0.005 (0.003) & 0.004 (0.002) \\
\textit{Infect (t)} & 11 & 0.008 (0.014) & 0.027 (0.017) & 0.003 (0.001) & 0.002 (0.001) & 0.002 (0.001) \\
\textit{GeopolRisk (t)} & 12 & 0.008 (0.011) & 0.024 (0.014) & 0.003 (0.001) & 0.003 (0.001) & 0.003 (0.001) \\
\textit{SP500 (t)} & 13 & 0.008 (0.007) & 0.015 (0.009) & 0.008 (0.004) & 0.003 (0.001) & 0.006 (0.004) \\
\textit{VIX (t)} & 14 & 0.007 (0.009) & 0.020 (0.008) & 0.003 (0.002) & 0.002 (0.001) & 0.004 (0.003) \\
\textit{ExchRate (t)} & 15 & 0.007 (0.007) & 0.017 (0.008) & 0.005 (0.004) & 0.002 (0.001) & 0.003 (0.001) \\
\textit{GgleInfl (t-4)} & 16 & 0.007 (0.009) & 0.019 (0.012) & 0.007 (0.002) & 0.001 (0.000) & 0.001 (0.001) \\
\textit{GgleClimate (t)} & 17 & 0.007 (0.005) & 0.013 (0.005) & 0.009 (0.002) & 0.002 (0.001) & 0.003 (0.002) \\
\textit{Brent (t)} & 18 & 0.007 (0.008) & 0.017 (0.012) & 0.004 (0.002) & 0.002 (0.001) & 0.004 (0.002) \\
\textit{BTC (t)} & 19 & 0.006 (0.010) & 0.018 (0.013) & 0.003 (0.001) & 0.002 (0.001) & 0.002 (0.001) \\
\textit{HighYield (t)} & 20 & 0.006 (0.005) & 0.008 (0.006) & 0.009 (0.007) & 0.004 (0.002) & 0.002 (0.001) \\
\bottomrule
\end{tabular}}
\end{table}

%Other predictors show nonlinear patterns: \textit{Brent} crude displays an inverted U-shaped profile peaking at intermediate lags ($t-3$ to $t-2$), and the \textit{SP500} and \textit{PolUncert} exhibit hump-shaped trajectories. These findings confirm that Bitcoin's price discovery is a multi-scale process, requiring the LSTM's capacity to simultaneously track both persistent narrative-driven signals and recent macroeconomic shocks.

\subsection{Interaction Analysis} \label{sec:interactions}

To separate the predictive value of expectations from realized policy actions, we conduct a conditional interaction analysis. We examine the relationship between the MPE index and its corresponding SHAP contributions, partitioning the sample by the realized Effective Federal Funds Rate (\textit{FFR}) cycle: \textit{Rising} ($\Delta FFR > 0$), \textit{Flat} ($\Delta FFR = 0$), and \textit{Falling} ($\Delta FFR < 0$).

\begin{figure}[H]
    \caption{Conditional Marginal Effects: Interaction between Monetary Policy and FFR Regimes.}
    \label{fig:shap_interaction_grid_val_loss}
    \centering
    \resizebox{\textwidth}{!}{%
    \begin{tabular}{p{\textwidth}}
    
    \small{ \textbf{Notes:} This figure illustrates the conditional relationship between MPE scores ($x$-axis) and their corresponding SHAP values ($y$-axis), partitioned by Fed Funds Rate (FFR) regimes. The regimes are defined by the weekly change in the FFR: \textit{Rising} ($\Delta FFR > 0$), \textit{Flat} ($\Delta FFR = 0$), and \textit{Falling} ($\Delta FFR < 0$). Panels (a)-(d) present individual folds. Vertical lines represent the standard deviations of the SHAP values across runs for a corresponding MPE point. Regression lines represent linear fits with shaded 95\% confidence intervals; lines are only displayed for regimes reaching statistical significance at the 1\% level ($p < 0.01$).}
    \end{tabular}}

    \vspace{10pt}

    % Row 1
    \begin{subfigure}[b]{0.49\textwidth}
        \centering
        \caption{Fold 1}
        \vspace{-2pt}
        \includegraphics[width=\linewidth]{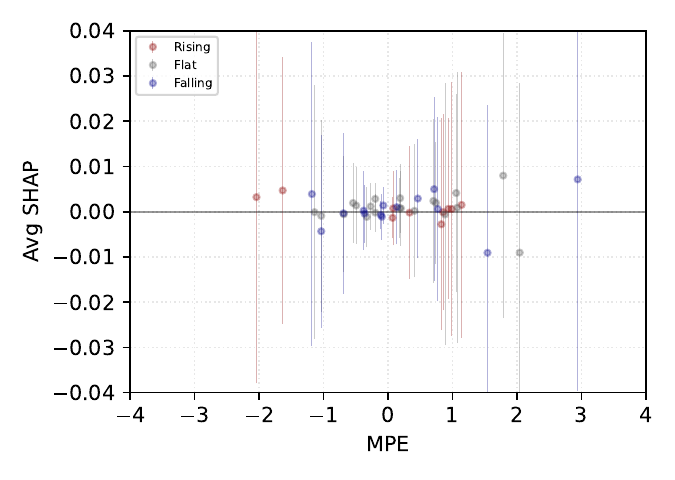}
        \label{fig:int_f1}
    \end{subfigure}
    \begin{subfigure}[b]{0.49\textwidth}
        \centering
        \caption{Fold 2}
        \vspace{-2pt}
        \includegraphics[width=\linewidth]{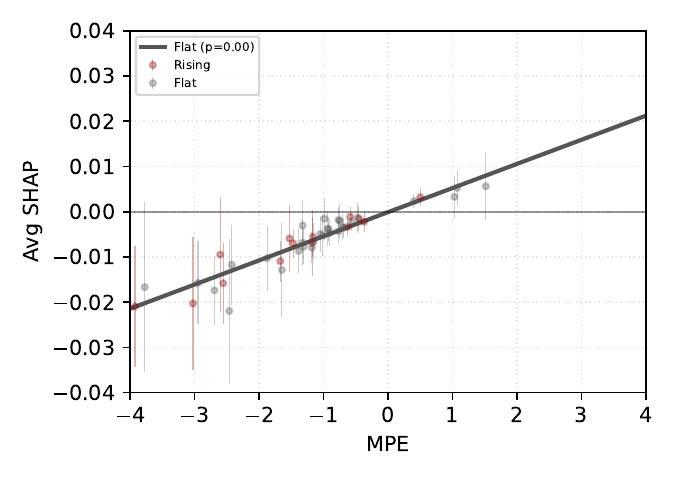}
        \label{fig:int_f2}
    \end{subfigure}

    \vspace{5pt} 

    % Row 2
    \begin{subfigure}[b]{0.49\textwidth}
        \centering
        \caption{Fold 3}
        \vspace{-2pt}
        \includegraphics[width=\linewidth]{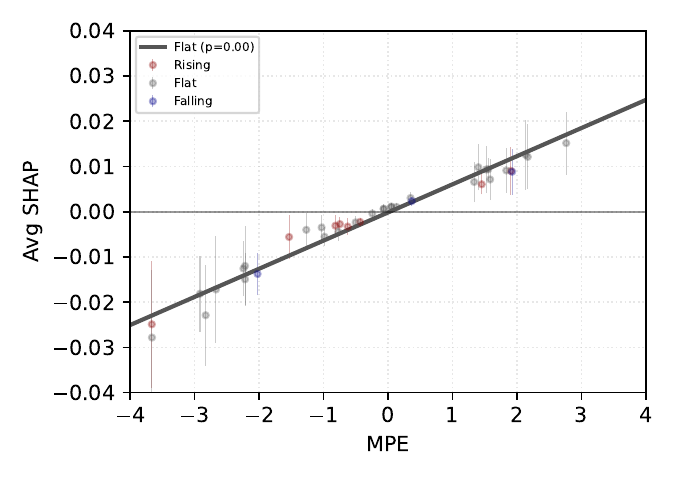}
        \label{fig:int_f3}
    \end{subfigure}
    \begin{subfigure}[b]{0.49\textwidth}
        \centering
        \caption{Fold 4}
        \vspace{-2pt}
        \includegraphics[width=\linewidth]{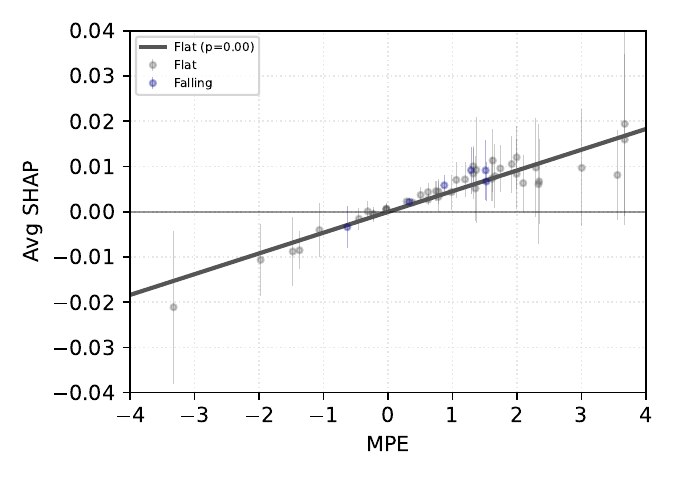}
        \label{fig:int_f4}
    \end{subfigure}

\end{figure}

Figure \ref{fig:shap_interaction_grid_val_loss} presents these marginal effects. Each data point represents the average SHAP value across ten independent runs for a given \textit{MPE} score, with vertical bars indicating the standard deviation across those runs. To identify robust trends, regression lines are displayed when the linear relationship between the stratified \textit{MPE} and SHAP values reaches a high level of statistical significance ($p < 0.01$).

In Folds 2 to 4 the relationship is upward sloping: as sentiment becomes more Hawkish, the SHAP value decreases, exerting negative pressure on predicted \textit{Btc} returns, while dovish sentiment is associated with positive SHAP contributions. The slope is estimated with high precision ($p < 0.01$) in the ``Flat" partition, where the realized rate is constant, so the relationship cannot be an artefact of contemporaneous rate movements. This constitutes direct evidence of the informational autonomy of the MPE index, confirming that narrative-driven sentiment extracts forward-looking signals that are largely independent of the contemporaneous level of realized policy.

The analysis further exposes temporal heterogeneity in market sensitivity across the four walk-forward folds. Fold 1 stands apart: its estimated relationship does not meet the 1\% significance threshold under the primary validation-based selection procedure (p-value: 0.03).\footnote{The retrospectively selected lowest-test-RMSE specification produces a significant slope of the opposite sign (see Appendix C, Figure~\ref{fig:shap_interaction_grid_testRMSE}). The Fold 1 relationship is therefore sensitive to model selection, whereas the conventional positive slope in Folds 2–4 is stable across both procedures.} This fold covers the earliest part of the sample, before the spot ETF approvals and the entry of institutional allocators, when Bitcoin traded largely as a speculative retail asset detached from the monetary cycle and its price responded to crypto-specific flows rather than to the policy outlook. The conventional sign appears only in Folds 2 to 4, once the asset had become integrated into mainstream portfolios and began to be priced against interest rates like other risk assets. Notably, Fold 2 corresponds to a period of significant macro instability, characterized by a transition to aggressive monetary tightening, with interest rates increased at an unprecedented rate to combat rising inflation. This variation confirms that Bitcoin's responsiveness to policy expectations is state-dependent.

Ultimately, the significance of the predictive relationship shows only within the ``Flat" regime --- where official rates remain unchanged --- which indicates that Bitcoin price discovery is shaped by the forward-looking pulse of policy discourse.

\section{Bitcoin as a Policy-Sensitive Asset and Policy Implications} \label{sec:discuss}

The empirical evidence presented in this study offers insights into the debate regarding Bitcoin's macroeconomic identity. By conditioning the impact of policy expectations on realized rate adjustments, our findings challenge the ``digital safe-haven'' narrative and characterize Bitcoin as a hyper-sensitive barometer of global liquidity conditions. As demonstrated in the Interaction Analysis (Section \ref{sec:interactions}), the negative effect of hawkish sentiment on Bitcoin returns is statistically significant ($p < 0.01$) from Fold 2 onwards within the ``Flat'' Federal Funds Rate regime. This implies that the cryptocurrency market prices in the threat of tightening long before liquidity is physically withdrawn from the system. This finding suggests that Bitcoin markets are sophisticated enough to incorporate soft information --- social media discourse and narrative shifts --- into price formation immediately. The low correlation between our index and the effective FFR ($\rho = 0.05$) confirms that this is not a redundancy; rather, the MPE index captures the ``shadow'' of the Fed --- the forward guidance and sentiment that bridges the gap between official FOMC meetings. This is consistent with the broader literature on the information content of central bank communication \citep{gorodnichenko2023voice, schmeling2025does}.

Contrary to the ``inflation hedge'' hypothesis \citep{liu2024hedging}, the VMD decomposition (Table \ref{tab:granger-vmd}) suggests that inflation expectations reach Bitcoin through the policy cycle rather than through a store-of-value channel. The long-term component of five-year inflation expectations (\textit{5yInflExp}) does not predict Bitcoin's own long-term component ($IMF_{1}$), the horizon at which a hedge against the erosion of purchasing power would be expected to operate. Its medium-term component ($IMF_{2}$), by contrast, predicts all three Bitcoin components at lags 2--6. The \textit{MPE} index behaves the same way: only its medium-term component transmits predictive information, and most forcefully at high frequency ($F = 23.196$ at lag 2). Both series therefore act at the frequency of the policy cycle rather than that of a shared long-run inflation trend.

These findings suggest that investors treat Bitcoin less as protection against inflation itself and more as a leveraged bet on the availability of cheap capital: what matters is not the expected price level, but the monetary-policy response it is expected to provoke. When expectations shift towards tightening, Bitcoin is repriced as a broader risk asset. This interpretation is consistent with \citet{ma2022monetary}, who document persistent Bitcoin price declines following unexpected monetary tightening, and with \citet{conlon2020safe}, who show that Bitcoin failed to provide safe-haven protection during the liquidity stress of the COVID-19 market crash.

This regime-dependency is most acute during the unprecedented 2022 tightening cycle --- an exogenous shock in which the Federal Funds Rate rose from near-zero to approximately 4.5\% within a single year, a pace unprecedented over the prior two decades. During this interval, as shown in Fold~2 of our walk-forward analysis, the predictive relationship between the MPE index and Bitcoin returns becomes nonlinear, consistent with a structural break driven by conditions outside the model's training distribution. \\

% \textcolor{blue}{The emergence of this relationship in Fold 2 coincides with the rapid 2022 monetary-tightening cycle, during which the federal funds rate rose from near zero to approximately (4.5\%) within a single year. This represented a large and rapid policy-regime shift, largely undertaken in response to rising inflation, rather than an exogenous monetary-policy shock. The statistically significant MPE-SHAP relationship within the ``Flat"-\textit{FFR} partition is consistent with Bitcoin becoming more sensitive to policy narratives during major monetary transitions. Since Figure \ref{fig:shap_interaction_grid_val_loss} estimates linear relationships within each partition and does not formally compare slopes across folds or regimes, the result should be interpreted as suggestive evidence of changing sensitivity rather than proof of nonlinearity or a structural break.}\\

%This further reinforces the view that Bitcoin is not a passive inflation hedge but a liquidity-sensitive asset whose sensitivity to monetary narratives is itself regime-dependent.\\

% POLICY IMPLICATIONS
% 1. Central bank communication as an independent policy instrument
These findings carry several important policy implications for both policymakers and market participants. For central banks, our results highlight the growing importance of communication as an independent policy instrument. The finding that hawkish narrative sentiment drives Bitcoin returns independently of realized rate changes implies that the mere anticipation of tightening, as reflected in investor communications, is sufficient to trigger significant repricing of Bitcoin returns. If retail investors systematically overreact to hawkish rhetoric, premature narrative tightening could induce procyclical volatility in crypto markets that spills over into broader financial conditions through wealth effects and sentiment contagion. Policymakers should therefore consider the crypto-asset ecosystem as an additional transmission channel of monetary communication, alongside the more traditional equity and bond market channels.

% 2. LLM-derived sentiment indices for regulators
For market participants and risk managers, our results suggest that monitoring LLM-derived sentiment indices constructed from investor communications represents a viable and informative complement to traditional macroeconomic indicators. The leading properties of our MPE index, documented in the temporal attribution analysis of Section \ref{sec:explainability}, imply that shifts in narrative sentiment precede realized price movements by a sufficient horizon to be actionable, offering potential utility for risk management and tactical asset allocation in portfolios with cryptocurrency exposure.

% 3. Real-time market surveillance tool for central banks
Additionally, the MPE index could serve as a real-time market surveillance instrument for central banks, allowing them to gauge market expectations before official macro data or FOMC releases — essentially a high-frequency barometer of how forward guidance is being received by investors.

% 3. Implications for retirement fund managers
The growing integration of Bitcoin into U.S. retirement vehicles --- underscored by the Department of Labor's May 2025 rescission of its 2022 guidance urging fiduciaries to exercise extreme caution toward cryptocurrency in 401(k) plans, and the subsequent August 2025 executive order directing federal agencies to expand access to digital assets in defined-contribution plans \citep{dol2025compliance, eo2025democratizing} --- makes these findings particularly timely. Our results suggest that Bitcoin's correlation with broader risk assets intensifies precisely during tightening cycles, the periods when pension funds are most in need of stable, inflation-hedging instruments. Institutional investors should therefore treat Bitcoin as a procyclical exposure to global liquidity conditions rather than a structural hedge, adjusting allocations dynamically in response to shifts in monetary policy narratives rather than relying on it as a passive store of value.

\section{Conclusion} \label{sec:conclusion}

This study contributes to the digital asset literature by introducing a novel, high-frequency MPE index, constructed using the \texttt{Qwen2.5-7b} LLM. By deploying generative AI to classify over 118,000 investor messages, we provide a methodological contribution that measures policy sentiment separately from realized policy actions—a distinction that traditional dictionary-based models fail to capture.

Our empirical results establish two critical dynamics. First, we demonstrate, using Granger causality, that Bitcoin markets are structurally tethered to forward-looking policy narratives at short- to medium-term horizons. Second, through our LSTM-SHAP analysis, we uncover a robust, regime-dependent asymmetry: hawkish sentiment exerts a pronounced negative effect on Bitcoin returns independent of the effective Federal Funds Rate. This shows that market prices reflect the \textit{threat} of tightening long before liquidity is officially withdrawn.

These findings have significant implications for policymakers and financial stability monitors. First, they confirm that Bitcoin is a sensitive barometer of global liquidity, often reacting to the ``shadow" of policy communication before official interventions occur. This suggests that central bank forward guidance has immediate, tangible spillover effects on the crypto-asset ecosystem that must be accounted for in financial risk assessments. Second, the value of our generative AI methodology underscores the need to use LLMs to capture the subtle, conditional nature of investors' conversations. As cryptocurrencies become increasingly integrated into traditional finance, LLM-derived sentiment indices will be essential tools for regulators seeking to measure the transmission of monetary policy in real-time. Future research should extend this framework to alternative cryptocurrency markets to assess the applicability of these findings across different investor communities.

\setlength\bibsep{0pt}
\bibliographystyle{apalike}

\clearpage

\appendix

\renewcommand{\theequation}{A.\arabic{equation}}
\setcounter{equation}{0}

\begin{comment}
    
1. Macro Conditions (replacing "Economic Activity" + parts of others)
Captures the broad macroeconomic and financial environment:

S\&P 500 (SP500)
Google Trends: Inflation (GgleInfl)
Google Trends: Recession (GgleReces)
Jobless Claims (JoblessClaim)
News Sentiment (NewsSent)

2. Monetary Policy \& Financial Conditions (replacing "Monetary Policy Decisions" + "Firm Investment Decisions")
Captures policy stance, interest rate dynamics, and financial market stress:

Federal Funds Rate (FFR)
Monetary Policy Expectations (MPE)
5-Year Inflation Expectations (5yInflExp)
High Yield Spread (HighYield)
Exchange Rate (ExchRate)
Brent Crude (Brent)
Gold (Gold)

3. Global Risk \& Market Sentiment (replacing "Forward-Looking Market Expectations" + "Household Consumption")
Captures systemic risk, uncertainty, and investor sentiment:

VIX
Policy Uncertainty (PolUncert)
Geopolitical Risk (GeopolRisk)
U.S. Dollar index (USDollar)
Google Trends: Climate (GgleClimate)
Infectious Disease index (Infect)

\end{comment}

\section{Definitions and Data Sources of Main Variables} \label{sec:def-variables}

\small
\renewcommand{\arraystretch}{1}
\begin{longtable}{@{} >{\itshape\raggedright\arraybackslash}p{4cm} p{11cm}@{}}
\caption{Definitions of the variables, by macroeconomic partition.}
\label{tab:var-def} \\
\captionsetup{ singlelinecheck=false}
\caption*{\small \textbf{Notes: }This table provides an overview of the variables used in the forecasting model. It details the specific data sources and the categorization of predictors across three key themes: macro conditions, monetary policy and financial conditions, and global risk and market sentiment. The table documents the specific transformation procedures applied to each series to ensure methodological consistency.} \\
\toprule
\textbf{Name} & \textbf{Definition, construction, and source (transformation)} \\
\midrule
\endfirsthead
\multicolumn{2}{l}{\small\textit{Table \thetable\ (continued)}} \\
\toprule
\textbf{Name} & \textbf{Definition, construction, and source (transformation)} \\
\midrule
\endhead
\midrule
\multicolumn{2}{r}{\small\textit{Continued on next page}} \\
\endfoot
\bottomrule
\endlastfoot
\multicolumn{2}{l}{\textbf{\textit{i) Macro Conditions}}} \\
S\&P 500 &
Weekly return of the S\&P 500 equity index (last price on Friday; Bloomberg Finance LP). \emph{Natural log return}. \\
Jobless Claims &
New unemployment insurance claims (last value on Friday; Bloomberg Finance LP). \emph{log difference}. \\
Ggle Inflation &
Search intensity for ``inflation'' in the U.S.; extracted and normalized via \texttt{pytrends}. \emph{Absolute level}. \\
Ggle Recession &
Search intensity for ``recession'' in the U.S. (Google Trends; extracted and normalized via \texttt{pytrends}). \emph{Absolute level}. \\
NewsSent &
News-based sentiment index \citep{shapiro2022measuring}. \emph{Absolute level}. \\
\addlinespace
\multicolumn{2}{l}{\textbf{\textit{ii) Monetary Policy \& Financial Conditions}}} \\
FFR &
Policy-controlled federal funds rate (last value on Friday; Bloomberg Finance LP). \emph{Absolute level}. \\
MPE &
Externally constructed monetary policy expectation measure based on StockTwits sentiment. \emph{Absolute level}. \\
5yInflExp &
Five-year inflation expectation series (last value on Friday; Bloomberg Finance LP). \emph{Absolute level}. \\
HighYield &
Bloomberg U.S. High Yield total return index, unhedged (last value on Friday; Bloomberg Finance LP). \emph{Growth rate}. \\
ExchRate &
BIS effective exchange rate for the U.S. dollar (weekly average; BIS). \emph{log difference}. \\
Brent &
Brent crude oil price (last price on Friday; Bloomberg Finance LP). \emph{Growth rate}. \\
Gold &
Gold price (last price on Friday; Bloomberg Finance LP). \emph{Natural log return}. \\
Btc &
Bitcoin price (last price on Friday; Haver). \emph{Natural log return}. \\
\addlinespace
\multicolumn{2}{l}{\textbf{\textit{iii) Global Risk \& Market Sentiment}}} \\
VIX &
CBOE implied volatility index (last value on Friday; Bloomberg Finance LP). \emph{Growth rate}. \\
PolUncert &
U.S. economic policy risk / uncertainty index (\texttt{policyuncertainty.com}). \emph{Growth rate}. \\
GeopolRisk &
Caldara--Iacoviello Geopolitical Risk index (weekly average; \texttt{policyuncertainty.com}). \emph{Growth rate}. \\
USDollar &
U.S. Dollar index (last value on Friday; Bloomberg Finance LP). \emph{Growth rate}. \\
GgleClimate &
Search intensity for ``climate change'' in the U.S. (Google Trends; extracted and normalized via \texttt{pytrends}). \emph{Absolute level}. \\
Infect &
Equity market volatility index based on infectious-disease news (weekly average; \texttt{policyuncertainty.com}). \emph{Absolute level}. \\
\end{longtable}

\section{LLM: Analysis \& Reliability Checks} \label{sec:mistral-prompt}

\subsection{Prompt Architecture} \label{sec:prompt-architecture}

\begin{table}[H]
\scriptsize
\centering
\caption{Prompt: Stocktwits classification}
\label{tab:prompt}
\captionsetup{justification=raggedright, singlelinecheck=false}
\caption*{\small \textbf{Notes:} This table displays the ``System-User" prompt architecture used to instruct \texttt{Qwen2.5-7b} and \texttt{mistral-nemo:12b} models. The prompt defines a five-point hawkish-to-dovish scale and enforces a strict classification output to ensure objective, scalable sentiment extraction from social media text.}
\vspace{0.3cm}
\begin{minipage}{\linewidth}
\lstset{
  language=Python,
  basicstyle=\scriptsize\ttfamily,
  frame=single,
  numbers=left,
  numberstyle=\tiny,
  breaklines=true,
  breakatwhitespace=true,
  keepspaces=true,
  columns=fullflexible,
  showstringspaces=false,
  tabsize=2
}
\begin{lstlisting}
system_prompt = """
    Analyze the tweet provided below to determine the stance on monetary policy expressed by the user. messages often feature concise language, hashtags, and abbreviations typical of social media communication. Consider the context and the specific wording to classify the stance into one of the following categories: "Very Hawkish", "Hawkish", "Neutral", "Dovish", and "Very Dovish".

    - "Very Hawkish": This category should be selected if the tweet indicates an aggressive stance towards fighting inflation, likely advocating for significant interest rate hikes or other contractionary measures. Look for strong language supporting drastic monetary tightening.
    - "Hawkish": Choose this if the tweet suggests a tendency towards tightening monetary policy to control inflation. It may not indicate extreme measures but shows a clear preference for policy tightening.
    - "Neutral": This applies if the tweet implies a balanced approach, without clear leanings towards tightening or easing monetary policy. It might suggest contentment with maintaining current policy levels or express views that neither strongly support nor oppose policy changes.
    - "Dovish": Select this category if the tweet points towards a preference for easing monetary policy, likely aiming to stimulate economic growth by advocating for lower interest rates or through quantitative easing.
    - "Very Dovish": This should be chosen if the tweet reflects a strong inclination towards stimulating the economy, suggesting substantial measures for easing monetary policy beyond standard adjustments, like a significant push for quantitative easing or deep cuts in interest rates.

    Consider not only the explicit content but also the tone, hashtags, and any economic indicators or events mentioned to understand the user's stance fully. Provide your classification along with a brief explanation of the key factors that influenced your decision.
    
    Remember, your response should only be a category, nothing more. This task is about classification accuracy, not explanation or detail.
    """

user_prompt = """ 
    Instructions:
    - Analyze the tweet: "{}"
    - Classify into categories: "Very hawkish," "Hawkish," "Neutral," "Dovish," or "Very dovish.".
    - Do not include any explanations, elaborations, or additional information in your response. Your role is solely to classify, not to explain or discuss, return only the category.
    """
\end{lstlisting}
\end{minipage}
\end{table}

\subsection{Reliability of the MPE Index Across Models and Temperatures} \label{sec:reliability-models-temps}

We run the entire classification pipeline under four configurations: two models, \texttt{Qwen2.5-7b} and \texttt{mistral-nemo:12b}, each at two decoding temperatures, $\tau=0.8$ and $\tau=0.3$. Prompt, five-point scale, parser, sample and the weekly engagement-weighted aggregation are held fixed across the four runs, so any difference between them is attributable to the classifier and its decoding temperature alone.

%%%%%
Three complementary measures are reported in Table~\ref{tab:mpe_run_agreement}. Agreement (\textit{Agree}) is the share of messages assigned the same direction by both runs; it is intuitive but inflated by the large neutral mass, on which agreement is easy. $r$ is the Pearson correlation of the two −1/0/+1 series, and measures whether the two runs move together across the scale rather than matching exactly. $\kappa_w$ is Cohen's kappa with quadratic weights over $\{−1, 0, +1\}$: it corrects the agreement rate by penalising a hawkish/dovish reversal more heavily than a miss into neutral. Values of $\kappa_w$ near 1 indicate agreement well beyond chance, and near 0 agreement no better than chance.

\begin{figure}[H]
    \centering
    \caption{Effect of decoding temperature on the weekly MPE index.}\label{fig:temp-mistral-qwen-temp}
    \centering
    \resizebox{\textwidth}{!}{%
    \begin{tabular}{p{\textwidth}}
    \small{ \textbf{Notes:} Weekly engagement-weighted MPE index, 2014--2025, for each model at both decoding temperatures ($\tau=0.8$ in blue, $\tau=0.3$ in red). Each panel holds the model fixed and varies only the temperature; $\rho$ is the correlation between the two series within the panel.}
    \end{tabular}}
    \includegraphics[width=0.95\linewidth]{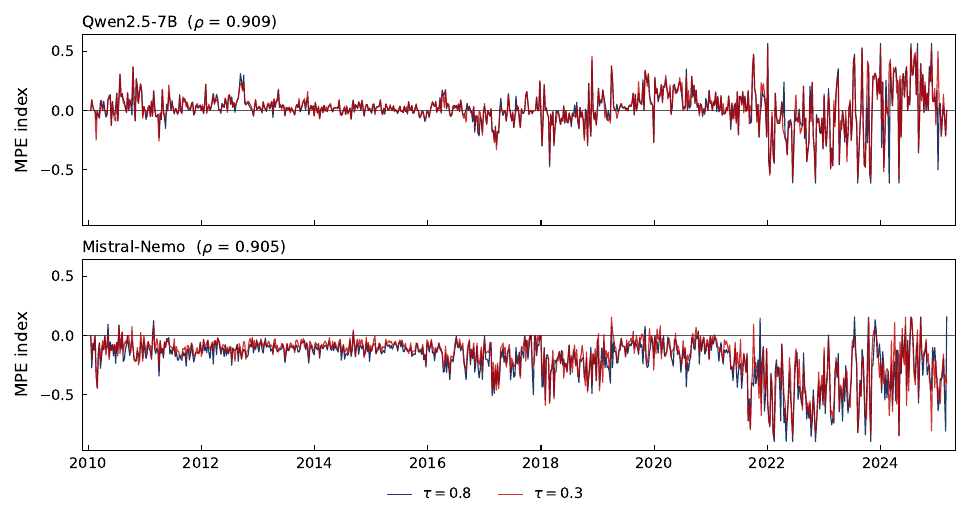}
\end{figure}

Temperature has little impact on either model. The two same-model pairs are by far the closest of the six: lowering $\tau$ from $0.8$ to $0.3$ leaves 99.0\% of Qwen's classifications unchanged ($\kappa_w=0.935$) and 91.8\% of Mistral-Nemo's ($\kappa_w=0.714$). Qwen is thus the more self-consistent of the two by a clear margin, but both are stable. The same holds for the weekly indices, which correlate at $0.909$ (Qwen) and $0.905$ (Mistral-Nemo) across temperatures (Table~\ref{tab:mpe_weekly_corr}) and are near-indistinguishable in Figure~\ref{fig:temp-mistral-qwen-temp}, tracking the same peaks and troughs throughout 2014--2025. The choice of temperature therefore has no material effect on the index. None of the variation in the index is attributable to decoding randomness.

\begin{table}[H]
\centering
\caption{Per-Message Agreement Between Classification Runs}
\label{tab:mpe_run_agreement}
\begin{small}
\caption*{\small \textbf{Notes:} Each pair of classification runs is compared on the collapsed three-category scale, in which the two hawkish classes ($-2,-1$) are merged into a single Hawkish category and the two dovish classes ($+1,+2$) into a single Dovish category, so that only the direction of the stance is compared. Agree.\ is the share of messages assigned the same direction by both runs, $r$ the Pearson correlation of the two $-1/0/+1$ series, and $\kappa_w$ Cohen's kappa with quadratic weights over $\{-1,0,+1\}$. $\tau$ denotes the decoding temperature.}
\end{small}
\resizebox{0.8\textwidth}{!}{%
\begin{tabular}{l ccc}
\toprule
Pair of runs & Agree. & $r$ & $\kappa_w$ \\
\midrule
\textit{Qwen2.5-7b ($\tau$=0.8) vs. Mistral-Nemo ($\tau$=0.8)} & 0.855 & 0.363 & 0.336 \\
\textit{Qwen2.5-7b ($\tau$=0.8) vs. Qwen2.5-7b ($\tau$=0.3)} & 0.990 & 0.935 & 0.935 \\
\textit{Qwen2.5-7b ($\tau$=0.8) vs. Mistral-Nemo ($\tau$=0.3)} & 0.881 & 0.394 & 0.373 \\
\textit{Mistral-Nemo ($\tau$=0.8) vs. Qwen2.5-7b ($\tau$=0.3)} & 0.855 & 0.360 & 0.334 \\
\textit{Mistral-Nemo ($\tau$=0.8) vs. Mistral-Nemo ($\tau$=0.3)} & 0.918 & 0.717 & 0.714 \\
\textit{Qwen2.5-7b ($\tau$=0.3) vs. Mistral-Nemo ($\tau$=0.3)} & 0.881 & 0.393 & 0.373 \\
\bottomrule
\end{tabular}}
\end{table}

\begin{figure}[H]
    \centering
    \caption{Qwen vs Mistral-Nemo: per-message confusion matrices.}\label{fig:temp-mistral-qwen}
    \centering
    \resizebox{\textwidth}{!}{%
    \begin{tabular}{p{\textwidth}}
    \small{ \textbf{Notes:} Joint classification of the same $N=118{,}562$ messages by the two models, both at decoding temperature $\tau=0.3$. The left panel uses the full five-point scale ($-2$ Very Hawkish to $+2$ Very Dovish); the right panel collapses it to the three directional categories by merging $(-2,-1)$ into Hawkish and $(+1,+2)$ into Dovish. Rows are the Qwen class, columns the Mistral-Nemo class, so diagonal cells are agreements and off-diagonal cells disagreements.}
    \end{tabular}}
    \includegraphics[width=0.95\linewidth]{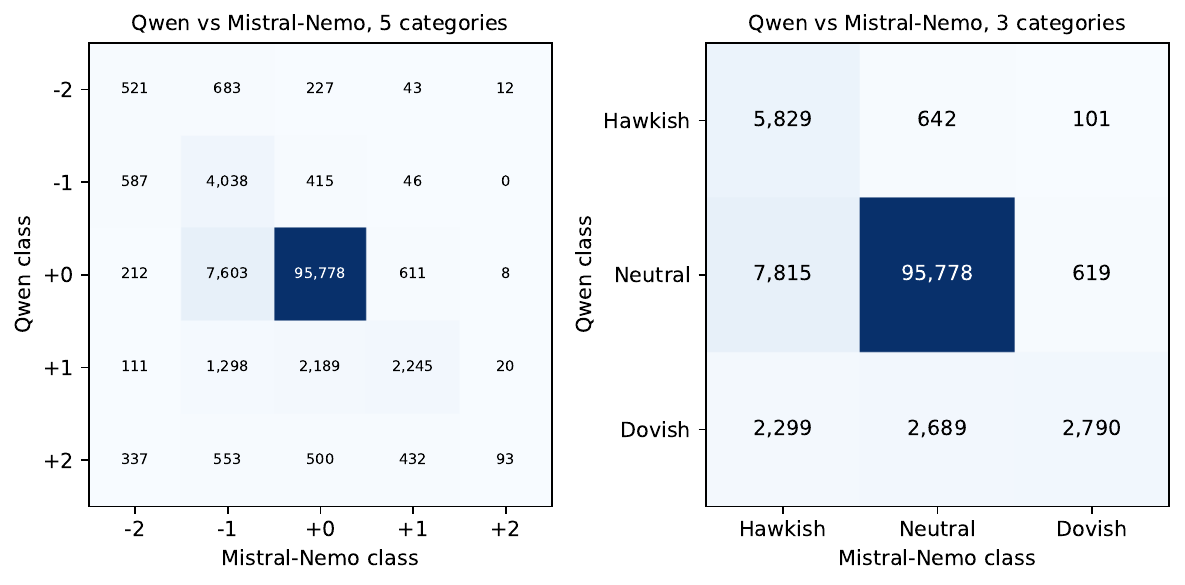}
\end{figure}

\begin{table}[H]
\centering
\caption{Correlation of the weekly engagement-weighted MPE indices across the four classification runs.}
\label{tab:mpe_weekly_corr}
\begin{tabular}{l cccc}
\toprule
 & (1) & (2) & (3) & (4) \\
\midrule
(1) Qwen2.5-7b ($\tau$=0.8) & 1.000 &  &  &  \\
(2) Mistral-Nemo ($\tau$=0.8) & 0.482 & 1.000 &  &  \\
(3) Qwen2.5-7b ($\tau$=0.3) & 0.909 & 0.466 & 1.000 &  \\
(4) Mistral-Nemo ($\tau$=0.3) & 0.485 & 0.905 & 0.447 & 1.000 \\
\bottomrule
\end{tabular}
\end{table}

Across models, per-message agreement remains high in absolute terms, at 85.5\% to 88.1\% depending on the pair, but the associated correlations and weighted kappas are much lower, at $r\approx0.36$--$0.39$ and $\kappa_w\approx0.33$--$0.37$. The confusion matrices in Figure~\ref{fig:temp-mistral-qwen} confirm this reading. The neutral $\times$ neutral cell alone accounts for 95,778 of the 118,562 jointly classified messages (80.8\%), and the off-diagonal mass is heavily concentrated in the neutral row and column rather than in the hawkish/dovish corners: only 101 messages are called hawkish by Qwen and dovish by Mistral-Nemo, and 2,299 the reverse. Outright sign reversals are therefore rare. Conditional on both models judging a message directional, they agree on its direction in 78.2\% of cases.

Mistral-Nemo calls 13.5\% of messages hawkish and 3.0\% dovish; Qwen calls 5.5\% hawkish and 6.6\% dovish. Mistral-Nemo is thus markedly more inclined to read a message as hawkish, while Qwen is close to balanced, and the asymmetry is sharpest at the extremes: Mistral-Nemo assigns Very Dovish ($+2$) to 0.1\% of messages against 1.6\% for Qwen. This calibration gap carries into the weekly index, which over the 545 weeks sits about $0.20$ lower for Mistral-Nemo ($-0.18$ versus $+0.02$ at $\tau=0.3$) and is hawkish in 94\% of weeks against 36\% for Qwen; the level shift is visible between the two panels of Figure~\ref{fig:temp-mistral-qwen-temp}.

We retain \texttt{Qwen2.5-7b} as the baseline classifier for its higher internal consistency ($\kappa_w=0.935$ against $0.714$) and more balanced calibration. All results below are re-estimated with the Mistral-Nemo index as a robustness check.

\subsection{Validity Against Manually Labeled Sample} \label{sec:validity-manual}

The comparison above shows that the two classifiers agree with each other, not that either agrees with a human reader. To check this, two of the authors manually labeled 1,100 messages on the same five-point scale. The exercise was blind: annotators saw only the text, never the model's output, and worked independently, recording a confidence score and a short justification for each label. The two sets of labels were then reconciled into a single consensus label, using the scores and notes to resolve disagreements, leaving a validation set of 1,095 messages, roughly 1\% of the corpus.

Sampling was deliberately not random. Directional messages are scarce, so a random draw would have been dominated by neutral messages and would have measured little beyond the models' ability to discard irrelevant text. We therefore required the sample to include at least 25\% of messages judged directional by Qwen and at least 25\% judged directional by Mistral-Nemo, drawn at random within each group. Directional messages accordingly make up 44\% of the validation set (615 neutral, 174 hawkish, 211 dovish, 14 very hawkish, 81 very dovish) against roughly 15\% of the corpus.

Qwen is the most accurate classifier on both margins (Table~\ref{tab:mpe_manual_3cat}). It matches the human label on 75.5\% of messages ($\kappa_w=0.530$) against 72.6\% for Mistral-Nemo ($\kappa_w=0.432$). Among messages that both the annotators and the model called directional, where agreeing on the absence of a signal earns no credit, the gap widens to 84.3\% ($\kappa_w=0.668$) against 76.5\% ($\kappa_w=0.541$).

\begin{figure}[H]
    \centering
    \caption{Model classifications against the reconciled manual labels.}\label{fig:manual-confusion}
    \centering
    \resizebox{\textwidth}{!}{%
    \begin{tabular}{p{\textwidth}}
    \small{ \textbf{Notes:} Confusion matrices for the 1,095 manually labeled messages, with the reconciled human label in rows and the model class in columns, so diagonal cells are agreements. The top panels use the full five-point scale ($-2$ Very Hawkish to $+2$ Very Dovish); the bottom panels collapse it to Hawkish ($-2,-1$), Neutral ($0$) and Dovish ($+1,+2$). Left: \texttt{Qwen2.5-7b}. Right: \texttt{mistral-nemo:12b}. Both models are run at $\tau=0.3$. }
    \end{tabular}}
    \includegraphics[width=0.95\linewidth]{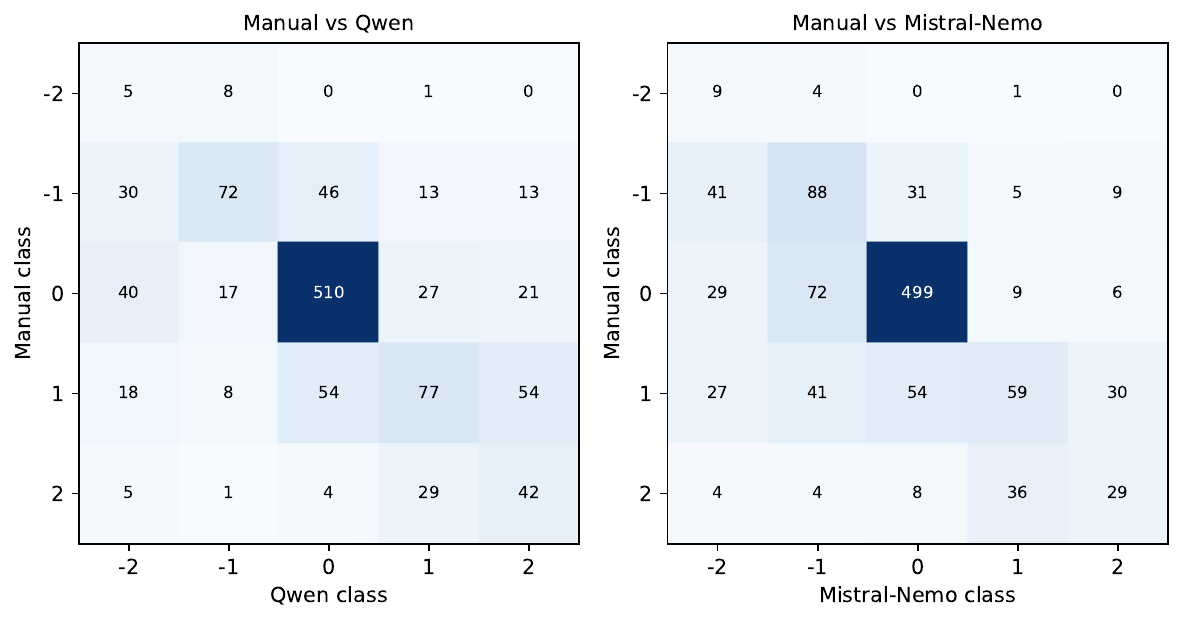}
    \includegraphics[width=0.9\linewidth]{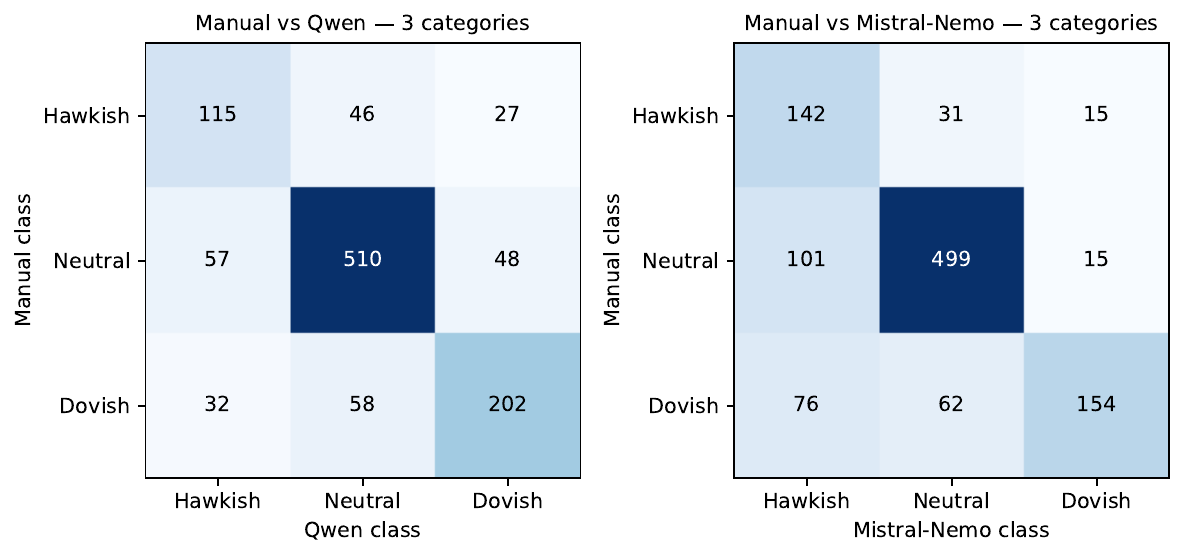}
\end{figure}

Figure~\ref{fig:manual-confusion} shows where the errors fall. Both models identify neutral messages reliably (83\% for Qwen, 81\% for Mistral-Nemo). The dominant error is missing direction altogether: 104 directional messages are called neutral by Qwen and 93 by Mistral-Nemo, so the index is attenuated toward zero rather than biased in sign.

The manual labels also confirm Mistral-Nemo's hawkish tilt, as it recovers 75.5\% of human hawkish messages but only 52.7\% of dovish ones and labels 319 messages hawkish where the annotators found 188. Qwen is more even (61.2\% and 69.2\%), with marginal totals close to the human ones, which supports its use as the baseline classifier.

\begin{table}[H]
\centering
\caption{Agreement with the Reconciled Manual Labels, Three-Category Scale}
\label{tab:mpe_manual_3cat}
\begin{small}
\caption*{\small \textbf{Notes:} Model classes and human labels are collapsed to Hawkish ($-2,-1$), Neutral ($0$) and Dovish ($+1,+2$). Panel (A) uses every labeled message ($N = 1,095$); panel (B) keeps only the messages that both the annotator and the model called directional, so that agreeing on ``no direction'' contributes nothing to the score ($N$ = Qwen2.5-7b 376, Mistral-Nemo 387). Agree.\ is the share of exactly matching classes, $r$ the Pearson correlation, and $\kappa_w$ Cohen's kappa with quadratic weights (over $\{-1,0,+1\}$ in panel A and $\{-1,+1\}$ in panel B).}
\end{small}
\resizebox{0.99\textwidth}{!}{%
\begin{tabular}{l lll lll}
\toprule
& \multicolumn{3}{c}{(A) All messages: Hawkish / Neutral / Dovish}
& \multicolumn{3}{c}{(B) Directional only: Hawkish / Dovish} \\
\cmidrule(lr){2-4} \cmidrule(lr){5-7}
Model &  Agree. & $r$ & $\kappa_w$ & Agree. & $r$ & $\kappa_w$ \\
\midrule
\textit{Qwen2.5-7b} & \textbf{0.755} & \textbf{0.531} & \textbf{0.530} & \textbf{0.843} & \textbf{0.669} & \textbf{0.668} \\
\textit{Mistral-Nemo} & 0.726 & 0.456 & 0.432 & 0.765 & 0.568 & 0.541 \\
\bottomrule
\end{tabular}}
\end{table}

\subsection{Reliability of Neutral Labels} \label{sec:reliability-neutral}

The neutral category is very large: 104,212 of the 118,562 classified messages (\SI{87.9}{\percent}) are labeled Neutral by \texttt{Qwen2.5-7b} at $\tau = 0.3$. That label conflates two distinct objects. A message may be about monetary policy while expressing no directional view (e.g. \textit{``FOMC meets Wednesday, consensus is a hold''}), in which case it is a genuine no-change expectation and belongs in the index at zero; or it may not concern monetary policy at all (e.g. \textit{``\$MACRO earnings season is a coin flip this quarter''}), in which case it only dilutes the weekly average.

We separate the two with a second pass over the neutral messages alone. The prompt, reproduced in Table~\ref{tab:prompt-neutral}, follows the same system-user architecture as the first pass but asks a strictly binary, topic-only question: is the message about monetary policy at all? It states explicitly that the directional question has already been settled and is not being asked again, so the model judges topic rather than opinion, and it returns one of two labels, ``Related'' or ``Unrelated''. This pass discards \SI{60.0}{\percent} of neutral messages (\SI{52.7}{\percent} of the corpus) as off-topic, so the corrected index rests on 56,066 messages rather than 118,562, with the retained real neutrals entering the weekly average at zero.

\begin{table}[H]
\scriptsize
\centering
\caption{Prompt: Neutral-class relevance triage}
\label{tab:prompt-neutral}
\captionsetup{justification=raggedright, singlelinecheck=false}
\caption*{\small \textbf{Notes:} This table displays the ``System-User" prompt architecture used to re-examine the Neutral class. The prompt asks a strictly binary, topic-only question (``is the message about monetary policy at all?") separating real neutrals, kept in the index at 0, from off-topic messages that only dilute it.
.}
\vspace{0.3cm}
\begin{minipage}{\linewidth}
\lstset{
  language=Python,
  basicstyle=\scriptsize\ttfamily,
  frame=single,
  numbers=left,
  numberstyle=\tiny,
  breaklines=true,
  breakatwhitespace=true,
  keepspaces=true,
  columns=fullflexible,
  showstringspaces=false,
  tabsize=2
}
\begin{lstlisting}

relevance_system_prompt = """
    You are given a message posted on a financial social media platform. The message has already been judged not to express any directional view on monetary policy - it is neither hawkish nor dovish. Your only task is to decide whether the message is ABOUT monetary policy at all. Classify it into exactly one of the following two categories:

    - "Related": The message concerns monetary policy or the institutions and instruments that set it.
      This includes the Federal Reserve, the FOMC, its members and their communications, interest
      rates and rate decisions, quantitative easing or tightening, the money supply, the policy
      response to inflation, or the future path of policy. The message may simply report, question,
      or comment on these topics without taking any side - that still counts as "Related".

    - "Unrelated": The message does not concern monetary policy. This includes messages about
      individual stocks or crypto prices, company earnings, commodities, geopolitics, fiscal or tax
      policy, general market moves, promotional content, jokes, insults, or content with no economic
      meaning at all. Mentioning an economic figure in passing, with no link to policy, is "Unrelated".

    Judge only the topic, not the opinion. Do not consider whether the message is hawkish or dovish;
    that question has already been answered and is not being asked again.
    Remember, your response must be only one of the two category names, nothing more.
    """

relevance_user_prompt = """
    Instructions:
    - Analyze the message: "{}"
    - Classify into exactly one category: "Related" or "Unrelated".
    - "Related" means the message is about monetary policy, the Federal Reserve, or interest rates,
      even if it expresses no opinion on their direction.
    - "Unrelated" means the message is about something else entirely.
    - Do not include any explanations, elaborations, or additional information in your response.
      Your role is solely to classify, not to explain or discuss, return only the category.
    """
\end{lstlisting}
\end{minipage}
\end{table}

\begin{figure}[H]
    \centering
    \caption{The weekly MPE index before and after the relevance correction.}\label{fig:relevance-correction}
    \centering
    \resizebox{\textwidth}{!}{%
    \begin{tabular}{p{\textwidth}}
    \small{ \textbf{Notes:} Weekly engagement-weighted MPE index from \texttt{Qwen2.5-7b} at $\tau = 0.3$, over the \num{545} weeks from September 2014 to March 2025. The original series (red) is built on all classified messages, so every neutral enters at zero; the corrected series (blue) keeps directional messages and the real neutrals identified by the relevance triage, and drops the off-topic ones. $\rho$ is the correlation between the two series. The correction leaves the timing of the movements unchanged while widening their amplitude.}
    \end{tabular}}
    \includegraphics[width=0.95\linewidth]{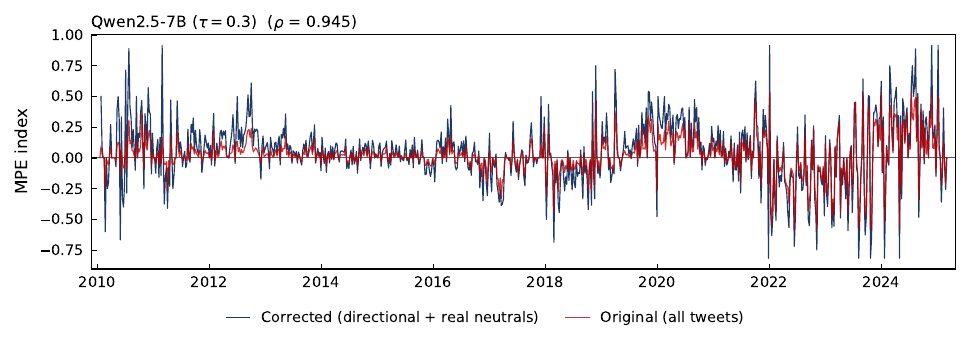}
\end{figure}

Panel~(A) of Table~\ref{tab:neutral_triage} gives the resulting composition of the corpus, and panel~(B) compares the two weekly series. They correlate at $\rho = 0.945$ (Spearman $0.986$) and agree on the sign of the index in every one of the \num{545} weeks, so no turning point is created or removed; what changes is scale, with the mean rising from $0.021$ to $0.049$ and the standard deviation from $0.157$ to $0.253$. Figure~\ref{fig:relevance-correction} shows the two series together: they move in step throughout, with the corrected index tracing the same peaks and troughs at consistently wider amplitude. 

\begin{table}[H]
\centering
\caption{Composition of the Neutral Class and the Effect of the Relevance Triage}
\label{tab:neutral_triage}
\begin{small}
\caption*{\small \textbf{Notes:} \texttt{Qwen2.5-7b} at $\tau = 0.3$, the configuration used for the final MPE index. Panel (A) splits the neutral class by a second, binary relevance prompt into real neutrals (about monetary policy, no direction) and off-topic messages; shares are relative to all classified messages, except the last line, which is relative to the neutral class. Panel (B) compares the weekly engagement-weighted index built on all classified messages (original) with the index built on directional messages plus real neutrals retained at zero (corrected), over the \num{545} weeks from September 2014 to March 2025.}
\end{small}
\resizebox{0.5\textwidth}{!}{%
\begin{tabular}{l r r}
\toprule
\multicolumn{3}{l}{\textbf{(A) Composition of the classified corpus}} \\
\midrule
& Messages & Share (\%) \\
\midrule
\textit{Classified messages}                          & 118,562 & 100.00 \\
\quad \textit{Directional (Hawkish or Dovish)}        &  14,350 &  12.10 \\
\quad \textit{Neutral}                                & 104,212 &  87.90 \\
\qquad \textit{real neutral (MPE-related)}            &  41,716 &  35.18 \\
\qquad \textit{off-topic (not MPE-related)}           &  62,496 &  52.71 \\
\textit{Retained in the corrected index}              &  56,066 &  47.29 \\
\textit{Off-topic share of the neutral class}         &         &  59.97 \\
\midrule
\multicolumn{3}{l}{\textbf{(B) Weekly index: original vs.\ corrected}} \\
\midrule
& Original & Corrected \\
\midrule
\textit{Mean}                                         &   0.021 &   0.049 \\
\textit{Standard deviation}                           &   0.157 &   0.253 \\
\textit{Correlation (Pearson / Spearman)}             & \multicolumn{2}{c}{0.945 / 0.986} \\
\textit{Sign agreement}                               & \multicolumn{2}{c}{1.000} \\
\textit{Directional-only benchmark ($\sigma$)}        & \multicolumn{2}{c}{0.729} \\
\bottomrule
\end{tabular}}
\end{table}

\subsection{Correlation with Market-Based Measures} \label{sec:correlation-market}

We compare the index to three market-based measures from September 2014 to March 2025. The fed funds futures implied rate is $100$ minus the front-month CME 30-Day Federal Funds futures price (\texttt{ZQ=F}, Yahoo Finance), and prices the average policy rate expected over the contract month. The OIS-equivalent expected change subtracts the effective federal funds rate (FRED \texttt{DFF}) from that implied rate, giving the net hike or cut priced in. The Treasury spreads are differences of constant-maturity yields (FRED \texttt{DGS1MO}, \texttt{DGS3MO}, \texttt{DGS6MO}), which widen when hikes are expected. All series are aggregated to weekly frequency.

\begin{figure}[H]
    \centering
    \caption{The MPE index and market-based measures of policy expectations.}\label{fig:mpe_vs_market_expectations}
    \centering
    \resizebox{\textwidth}{!}{%
    \begin{tabular}{p{\textwidth}}
    \small{ \textbf{Notes:} Weekly MPE index (blue, right axis) against each market-based
    measure (red, left axis) over the 545 weeks from September 2014 to March 2025, with $\rho$
    the correlation between the two series in each panel. The MPE index is oriented so that
    higher values are more dovish, whereas the market measures rise with expected tightening,
    so convergent validity implies series moving in opposite directions. The index is near-flat
    before 2022 and volatile thereafter, when attention to monetary policy on social media rises
    with the tightening cycle.}
    \end{tabular}}
    \includegraphics[width=0.95\linewidth]{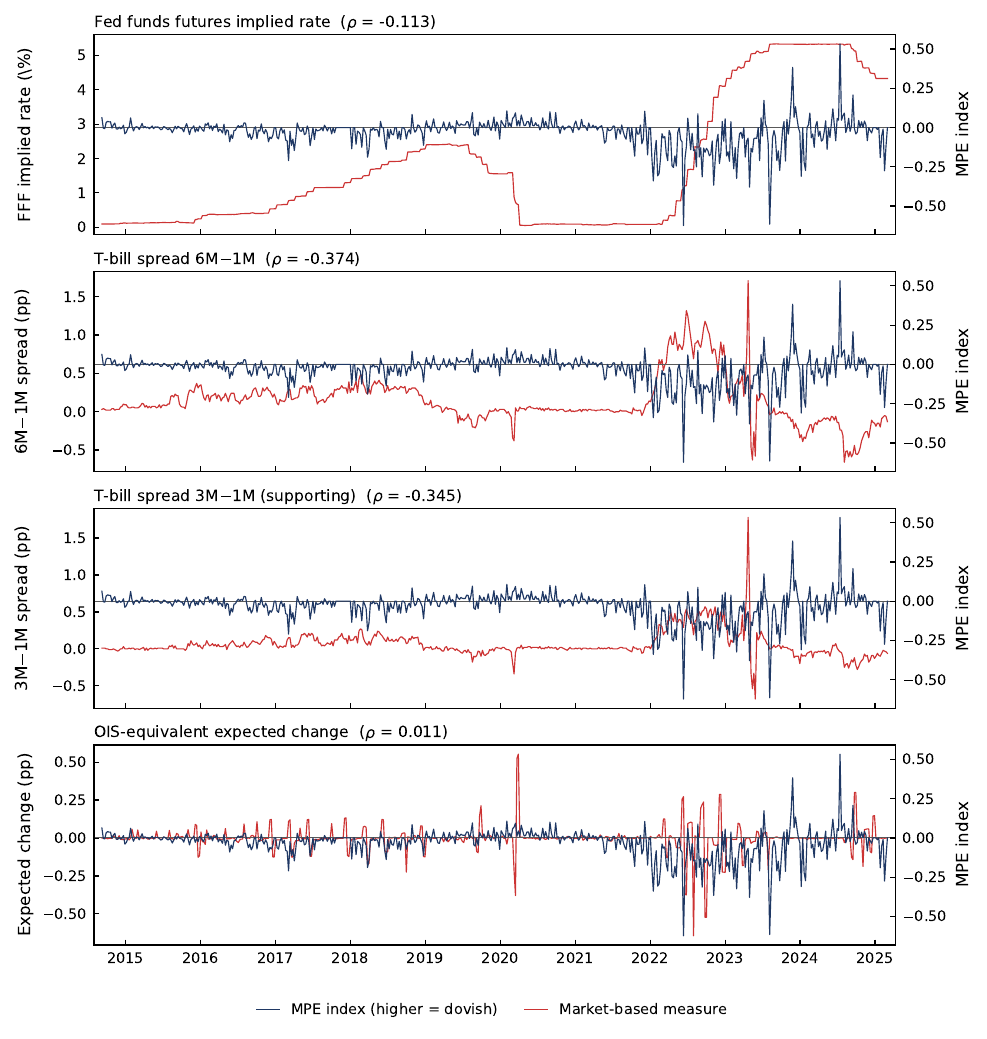}
\end{figure}

Table~\ref{tab:mpe_market_validity} shows that the predicted sign holds for the term-structure measures, and clearly so: the MPE index correlates at $-0.374$ with the 6M$-$1M bill spread and $-0.345$ with the 3M$-$1M spread. Both have the expected sign and economically meaningful. The correlation with the fed funds futures implied rate is also negative but weak ($-0.113$), and the OIS-equivalent expected change is essentially uncorrelated ($0.011$). Figure~\ref{fig:mpe_vs_market_expectations} shows this pattern graphically: the index co-moves with the bill spreads, which price the direction of policy, but not with the futures implied rate.

\begin{table}[H]
\centering
\caption{Correlation of the MPE Index with Market-Based Measures of Policy Expectations}
\label{tab:mpe_market_validity}
\begin{small}
\caption*{\small \textbf{Notes:} Pearson correlations in levels over 545 weeks from September 2014 to March 2025. The MPE index runs from Very Hawkish ($-2$) to Very Dovish ($+2$), so higher values are more dovish, whereas every measure rises when the market prices more tightening; convergent validity therefore predicts a \textit{negative} correlation with the MPE index (column 1). Rows (2)--(4) are the three market-based measures of the expected policy path: the fed funds futures implied rate is $100$ minus the CME \texttt{ZQ=F} front-month price, the OIS-equivalent expected change is that implied rate minus the effective federal funds rate, and the T-bill spreads are from FRED. Row (5) is a supporting variant of row~(4) and row (6), the effective federal funds rate (EFFR), is a realized anchor rather than an expectation. Lower triangle; the matrix is symmetric with a unit diagonal.}
\end{small}
\resizebox{0.8\textwidth}{!}{%
\begin{tabular}{l cccccc}
\toprule
& (1) & (2) & (3) & (4) & (5) & (6) \\
\midrule
\textit{(1) MPE index} & 1.000 &  &  &  &  &  \\
\textit{(2) FFF implied rate} & -0.113 & 1.000 &  &  &  &  \\
\textit{(3) OIS-equiv. exp. change} & 0.011 & -0.016 & 1.000 &  &  &  \\
\textit{(4) T-bill 6M$-$1M spread} & -0.374 & -0.231 & -0.091 & 1.000 &  &  \\
\textit{(5) T-bill 3M$-$1M spread} & -0.345 & -0.056 & -0.056 & 0.915 & 1.000 &  \\
\textit{(6) EFFR} & -0.114 & 0.999 & -0.061 & -0.227 & -0.054 & 1.000 \\
\bottomrule
\end{tabular}}
\end{table}

\subsection{Topic Analysis of MPE-Related Messages} \label{sec:topic-analysis}

\begin{figure}[H]
    \centering
    \caption{Thematic Divergence in Monetary Policy Narratives: Hawkish vs. Dovish Clusters}
    \label{fig:mpe_wordclouds}
    \begin{tabular}{p{0.99\textwidth}}
    \small{ \textbf{Notes:} This figure illustrates the lexical characteristics of the two primary policy regimes identified by our model. Panel (a) presents keywords from messages classified as Hawkish (-1, -2), while Panel (b) presents those from Dovish (+1, +2) classifications. Text data has been lemmatized and filtered for social media-specific noise. The size of each token represents its relative frequency within its respective corpus, with dark red denoting strong restrictive/tightening sentiment and dark blue denoting strong expansionary/easing sentiment.}
    \end{tabular}
    \includegraphics[width=\textwidth, trim=0cm 3cm 0cm 1.5cm, clip]{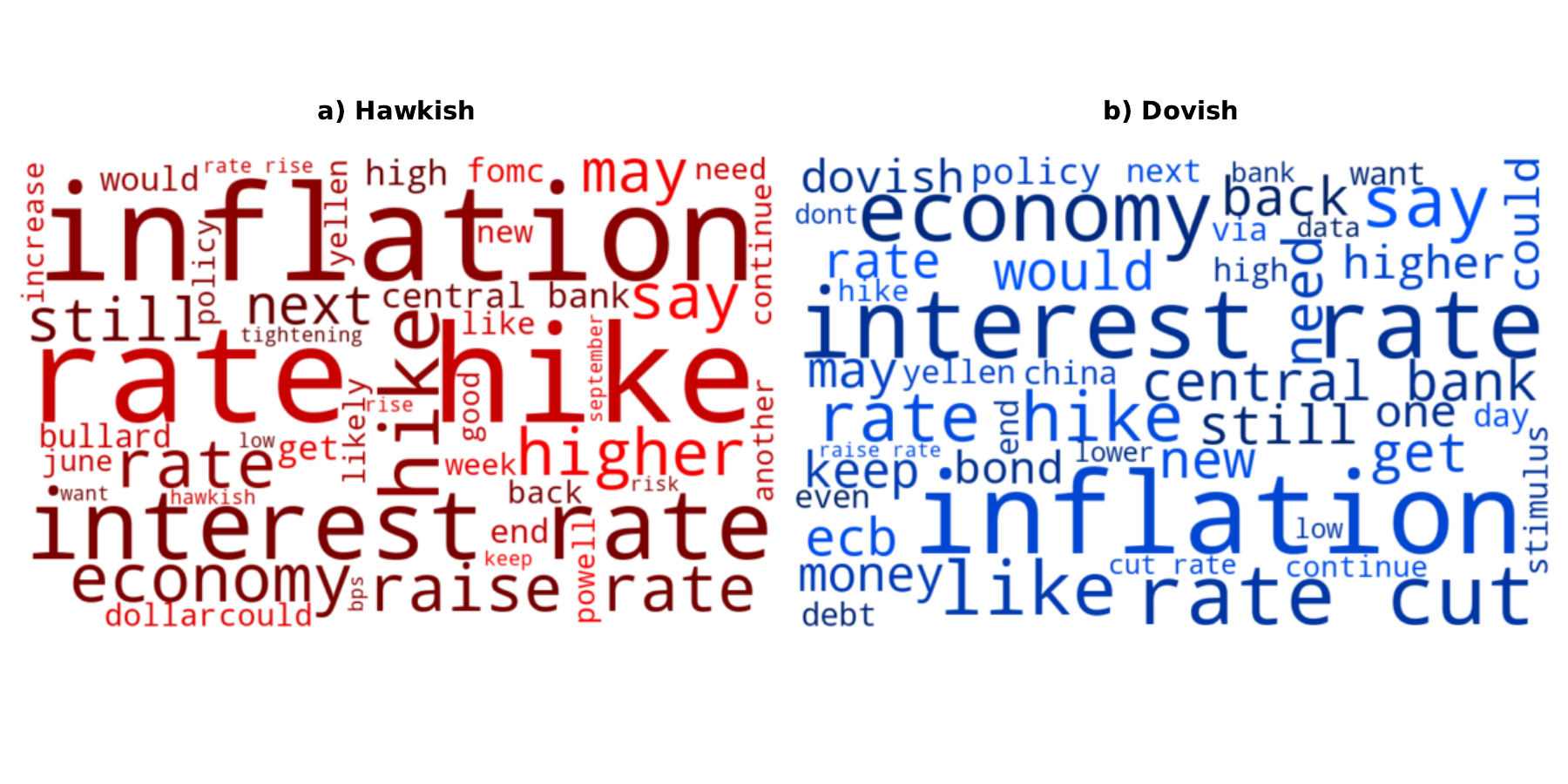}
\end{figure}

To validate the structural integrity of the newly developed MPE index, we conduct a qualitative lexical analysis of the underlying corpus. The methodology employs a Natural Language Processing (NLP) pipeline where raw social media text is lemmatized and filtered for social media-specific noise to isolate core economic concepts. As illustrated in Figure \ref{fig:mpe_wordclouds}, the thematic divergence between the two clusters confirms that the index captures distinct macroeconomic signals aligned with central bank narratives. The Hawkish regime is characterized by keywords and groups such as \textit{inflation}, \textit{raise rate}, \textit{rate hike}, and \textit{tightening}. Conversely, the Dovish regime features terms such as \textit{cut rate}, \textit{stimulus}, \textit{recession expect}, \textit{easing}, and \textit{dovish}.

\begin{figure}[H]
    \centering
    \caption{Narrative Validation: Inflation Discourse and the MPE Index}
    \label{fig:inflation_mpe_evolution}
    \begin{tabular}{p{0.95\textwidth}}
    \small{ \textbf{Notes:} This figure illustrates the co-evolution of the inflation-related narrative volume (left axis) and the aggregate MPE index (right axis) on a quarterly basis. The inflation theme is quantified as the percentage of messages containing specific economic keywords (e.g., `CPI', `PPI', and `PCE'). }
    \end{tabular}
    \includegraphics[width=0.95\textwidth]{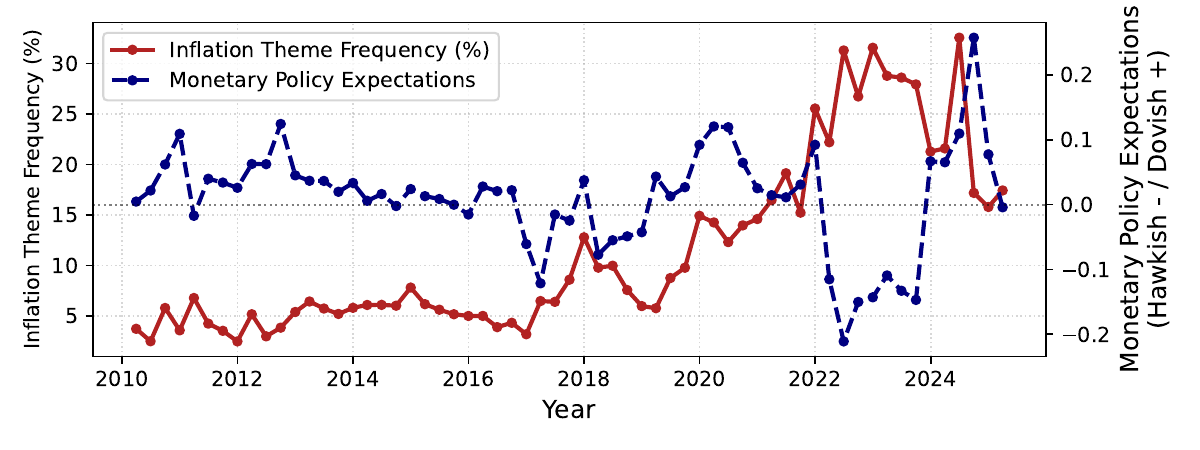}
\end{figure}

We further examine its co-movement with the dominant economic narratives of the sample period. The inflation narrative is isolated by tracking the relative frequency of messages containing a curated set of keywords, including \textit{`inflation'}, \textit{`cpi'}, \textit{`pce'}, \textit{`ppi'}, \textit{`prices'}, and \textit{`hike'}. As depicted in Figure \ref{fig:inflation_mpe_evolution}, there is a high degree of synchronization between the volume of inflation-related discourse and the aggregate policy stance. 

\newpage
\section{Additional Model Information, Validity, and Robustness Analysis} \label{sec:model-info}

\subsection{Individual Fold and Run Performance}

The LSTM is estimated using four consecutive temporal folds, each comprising a training, validation and strictly out-of-sample test window. Within each fold, $20$ independent Optuna searches are conducted, with each search evaluating $75$ candidate hyperparameter configurations. Hyperparameter selection is based on validation MSE, using the training period for model estimation and the subsequent validation period for model evaluation. For each of the 20 searches, the configuration achieving the lowest validation MSE is subsequently re-estimated ten times using different random initializations on the combined training and validation observations. The ten retrainings of a given configuration are treated as stochastic ensemble members, and their predictions are averaged to obtain the corresponding ensemble forecast. This procedure yields 200 final LSTM realizations per fold and allows the analysis to assess the stability of the learned nonlinear relationships with respect to the stochastic optimization landscape. Table~\ref{tab:lstm_metrics} reports the individual runs and their corresponding ensemble results, together with the selected optimizer, lookback window and number of hidden units.

The same hyperparameter search space was maintained across all folds to ensure methodological consistency and avoid ex post tailoring of the LSTM architecture to individual market regimes. Consequently, differences in out-of-sample performance across folds reflect, in part, the changing statistical and macroeconomic characteristics of the underlying periods rather than fold-specific adjustment of the model complexity. In particular, the LSTM is not expected to uniformly outperform the ARIMA benchmark across all regimes; rather, its role in this study is to provide a flexible nonlinear representation of the relationships between monetary-policy expectations and Bitcoin returns, which is subsequently interrogated using SHAP.

\begin{figure}[H]
    \caption{Conditional Marginal Effects for the Retrospectively Selected LSTM (lowest test-set RMSE)}
    \label{fig:shap_interaction_grid_testRMSE}
    \centering
    \resizebox{\textwidth}{!}{%
    \begin{tabular}{p{\textwidth}}
    
    \small{ \textbf{Notes:} This figure illustrates the conditional relationship between MPE scores ($x$-axis) and their corresponding SHAP values ($y$-axis), partitioned by Fed Funds Rate (FFR) regimes. The regimes are defined by the weekly change in the FFR: \textit{Rising} ($\Delta FFR > 0$), \textit{Flat} ($\Delta FFR = 0$), and \textit{Falling} ($\Delta FFR < 0$). Panels (a)-(d) present individual folds. Vertical lines represent the standard deviations of the SHAP values across runs for a corresponding MPE point. Regression lines represent linear fits with shaded 95\% confidence intervals; lines are only displayed for regimes reaching statistical significance at the 1\% level ($p < 0.01$).}
    \end{tabular}}

    \vspace{10pt}

    % Row 1
    \begin{subfigure}[b]{0.49\textwidth}
        \centering
        \caption{Fold 1}
        \vspace{-2pt}
        \includegraphics[width=\linewidth]{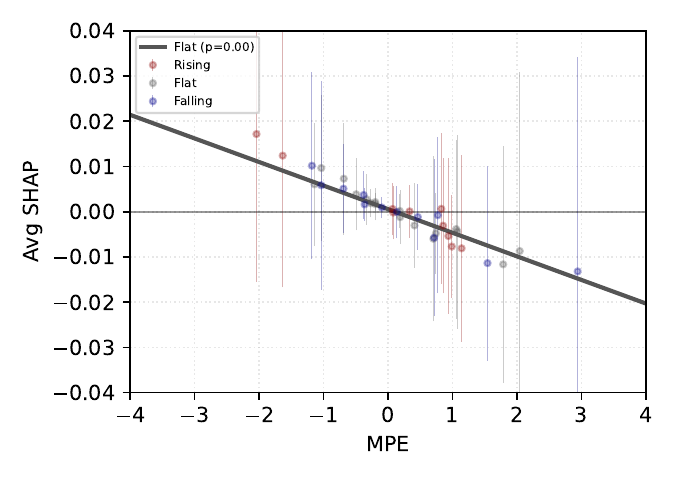}
        \label{fig:int_f1}
    \end{subfigure}
    \begin{subfigure}[b]{0.49\textwidth}
        \centering
        \caption{Fold 2}
        \vspace{-2pt}
        \includegraphics[width=\linewidth]{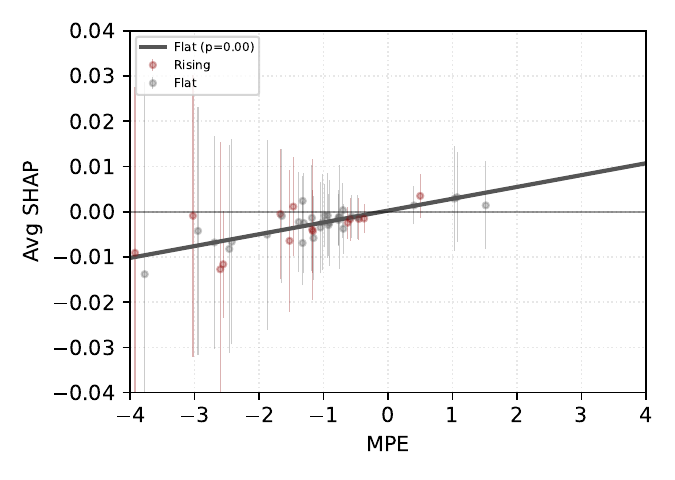}
        \label{fig:int_f2}
    \end{subfigure}

    \vspace{5pt} 

    % Row 2
    \begin{subfigure}[b]{0.49\textwidth}
        \centering
        \caption{Fold 3}
        \vspace{-2pt}
        \includegraphics[width=\linewidth]{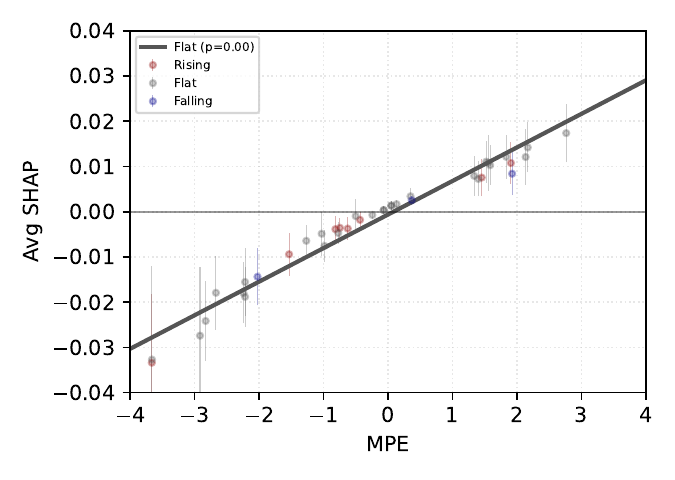}
        \label{fig:int_f3}
    \end{subfigure}
    \begin{subfigure}[b]{0.49\textwidth}
        \centering
        \caption{Fold 4}
        \vspace{-2pt}
        \includegraphics[width=\linewidth]{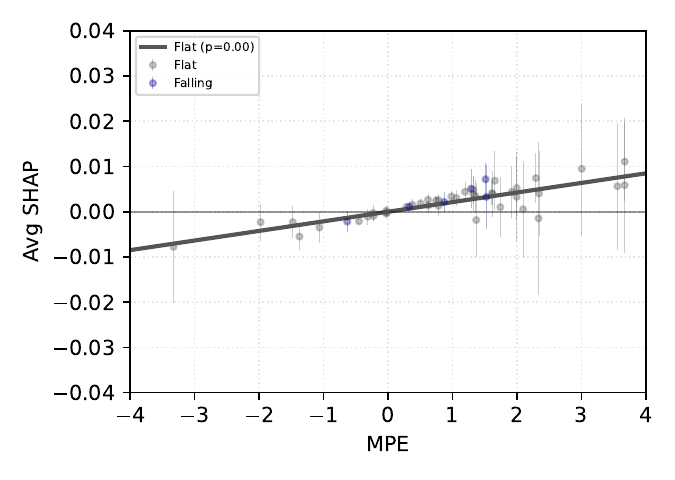}
        \label{fig:int_f4}
    \end{subfigure}

\end{figure}

As a complementary robustness analysis, we also examine the sensitivity of the results to retrospective selection based on out-of-sample test performance. Specifically, within each fold, we identify the LSTM search yielding the lowest ensemble test-set RMSE among the $20$ independently searched configurations. This procedure is used solely as a post-hoc robustness analysis and is not used for hyperparameter optimization, model development, or selection of the primary model specification. The corresponding forecasting results are reported in Table~\ref{tab:model_comparison_testRMSE}, while the associated feature-importance and interaction results are reported in Table~\ref{tab:global_shap} and Figure \ref{fig:shap_interaction_grid_testRMSE}. These results provide a deliberately stringent sensitivity check: they assess whether the substantive feature rankings and nonlinear relationships identified by SHAP remain evident when attention is restricted to the LSTM configuration that achieved the strongest realized performance on the held-out test observations.

\begin{table}[H]
\centering
\caption{Predictive Performance Comparison: LSTM (lowest test-set RMSE) vs. ARIMA Baseline}
\label{tab:model_comparison_testRMSE}
\begin{small}
\caption*{\small \textbf{Notes:} This table reports the out-of-sample forecasting performance comparison between the LSTM Ensemble and the ARIMA baseline across four sequential walk-forward folds. Performance is measured by the Root Mean Square Error (RMSE) and Mean Absolute Error (MAE). The Diebold-Mariano (DM) $p$-value ($p$-val) indicates the statistical significance of the predictive accuracy between models. ARIMA $(p, d, q)$ orders are optimized per fold via the Akaike Information Criterion (AIC). Overall Mean values and Standard Deviations (Std Dev) are calculated across the four folds, with the latter provided in parentheses.}
\end{small}
\resizebox{0.9\textwidth}{!}{%
\begin{tabular}{@{}lcccccc@{}}
\toprule
 & \multicolumn{3}{c}{LSTM Ensemble} & \multicolumn{3}{c}{ARIMA (Baseline)} \\
\cmidrule(lr){2-4} \cmidrule(l){5-7}
Fold & RMSE & MAE & DM $p$-val & RMSE & MAE & Order $(p,d,q)$ \\ \midrule
 1 & 0.1007 & 0.0820 & 0.192 & 0.1295 & 0.1067 & (2, 0, 2) \\
 2 & 0.0802 & 0.0607 & 0.813 & 0.0876 & 0.0623 & (2, 0, 2) \\
 3 & 0.0560 & 0.0405 & 0.849 & 0.0671 & 0.0477 & (2, 0, 2) \\
 4 & 0.0669 & 0.0534 & 0.022 & 0.0651 & 0.0497 & (0, 0, 1) \\
\midrule
Overall (Mean) & 0.0759 & 0.0591 & --- & 0.0873 & 0.0666 & --- \\
Overall (Std Dev) & (0.017) & (0.015) & --- & (0.026) & (0.024) & --- \\
\bottomrule
\end{tabular}}
\end{table}

\begin{table}[H]
\centering
\caption{Global Feature Importance for the Retrospectively Selected LSTM (lowest test-set RMSE)}
\label{tab:global_shap}
\begin{small}
\caption*{\small \textbf{Notes:} This table reports the global feature importance for the LSTM forecasting model, quantified using aggregated Mean Absolute SHAP (SHapley Additive exPlanations) values. The variables are ranked based on their average contribution to the model's predictions across four cross-validation folds; values in parentheses represent the standard deviation of these contributions.}
\end{small}
\resizebox{0.99\textwidth}{!}{%
\begin{tabular}{l c c llll}
\toprule
 & Rank & Average & Fold 1 & Fold 2 & Fold 3 & Fold 4  \\
Variable &  &  &  &  &  &   \\
\midrule
\textit{GgleInfl} & 1 & 0.007 (0.007) & 0.014 (0.004) & 0.014 (0.007) & 0.001 (0.001) & 0.001 (0.001) \\
\textit{FFR} & 2 & 0.003 (0.002) & 0.004 (0.003) & 0.003 (0.001) & 0.003 (0.001) & 0.002 (0.001) \\
\textit{5yInflExp} & 3 & 0.003 (0.002) & 0.005 (0.002) & 0.003 (0.001) & 0.001 (0.001) & 0.001 (0.001) \\
\textit{GgleReces} & 4 & 0.003 (0.004) & 0.002 (0.001) & 0.008 (0.003) & 0.000 (0.000) & 0.000 (0.000) \\
\textit{SP500} & 5 & 0.002 (0.002) & 0.003 (0.003) & 0.005 (0.003) & 0.001 (0.001) & 0.001 (0.000) \\
\textit{MPE} & 6 & 0.002 (0.001) & 0.003 (0.001) & 0.003 (0.001) & 0.002 (0.001) & 0.001 (0.001) \\
\textit{Gold} & 7 & 0.002 (0.002) & 0.004 (0.003) & 0.003 (0.001) & 0.001 (0.000) & 0.001 (0.001) \\
\textit{ExchRate} & 8 & 0.002 (0.002) & 0.003 (0.002) & 0.004 (0.002) & 0.001 (0.000) & 0.001 (0.000) \\
\textit{USDollar} & 9 & 0.002 (0.002) & 0.003 (0.002) & 0.003 (0.001) & 0.001 (0.000) & 0.001 (0.000) \\
\textit{Infect} & 10 & 0.002 (0.002) & 0.005 (0.002) & 0.002 (0.001) & 0.000 (0.000) & 0.000 (0.000) \\
\textit{GeopolRisk} & 11 & 0.002 (0.002) & 0.004 (0.002) & 0.001 (0.000) & 0.001 (0.000) & 0.001 (0.000) \\
\textit{PolUncert} & 12 & 0.002 (0.001) & 0.003 (0.002) & 0.002 (0.001) & 0.001 (0.001) & 0.001 (0.000) \\
\textit{GgleClimate} & 13 & 0.002 (0.001) & 0.003 (0.001) & 0.003 (0.001) & 0.000 (0.000) & 0.000 (0.000) \\
\textit{HighYield} & 14 & 0.002 (0.002) & 0.002 (0.001) & 0.004 (0.002) & 0.001 (0.000) & 0.000 (0.000) \\
\textit{Brent} & 15 & 0.002 (0.001) & 0.003 (0.001) & 0.002 (0.001) & 0.001 (0.000) & 0.001 (0.000) \\
\textit{VIX} & 16 & 0.002 (0.001) & 0.003 (0.001) & 0.002 (0.001) & 0.001 (0.001) & 0.001 (0.000) \\
\textit{BTC} & 17 & 0.001 (0.001) & 0.003 (0.001) & 0.002 (0.000) & 0.000 (0.000) & 0.000 (0.000) \\
\textit{NewsSent} & 18 & 0.001 (0.001) & 0.002 (0.001) & 0.002 (0.001) & 0.001 (0.000) & 0.001 (0.000) \\
\textit{JoblessClaim} & 19 & 0.001 (0.001) & 0.002 (0.001) & 0.001 (0.000) & 0.000 (0.000) & 0.000 (0.000) \\
\bottomrule
\end{tabular}}
\end{table}

\begin{table}[ht]
\centering
\caption{LSTM Individual Run Performance}
\label{tab:lstm_metrics}
\caption*{\small \textbf{Notes:} This table reports the out-of-sample forecasting performance of
the individual LSTM runs. Performance is measured using Root Mean Square Error (RMSE) and Mean
Absolute Error (MAE). Within a fold, all runs re-estimate the single configuration selected by
that fold's Optuna search, so the validation loss (Val. Loss) and the hyperparameters --
Optimizer (Opt.) and Lookback (Lbk.) -- are common to the runs and reported once in the fold
header. The number of hidden units is 8 in every fold. Ensemble denotes the average of the ten
runs.}\small
\begin{tabular}{@{}lcc@{}}
\toprule
Run & RMSE & MAE \\ \midrule
\multicolumn{3}{@{}l}{\textit{Fold 1} \quad Loss 0.0049 \quad RMSprop \quad Lbk. 13} \\
\textbf{Ensemble} & 0.1007 & 0.0820 \\
R1  & 0.1555 & 0.1240 \\
R2  & 0.1777 & 0.1462 \\
R3  & 0.3276 & 0.2945 \\
R4  & 0.1204 & 0.1038 \\
R5  & 0.1947 & 0.1703 \\
R6  & 0.4657 & 0.3902 \\
R7  & 0.1670 & 0.1345 \\
R8  & 0.1673 & 0.1360 \\
R9  & 0.1578 & 0.1303 \\
R10 & 0.2227 & 0.1788 \\
\addlinespace[0.5em]
\multicolumn{3}{@{}l}{\textit{Fold 2} \quad Loss 0.0046 \quad RMSprop \quad Lbk. 8} \\
\textbf{Ensemble} & 0.0802 & 0.0607 \\
R1  & 0.0986 & 0.0671 \\
R2  & 0.0984 & 0.0719 \\
R3  & 0.0973 & 0.0773 \\
R4  & 0.0901 & 0.0674 \\
R5  & 0.1374 & 0.0922 \\
R6  & 0.1479 & 0.1136 \\
R7  & 0.1556 & 0.1303 \\
R8  & 0.1395 & 0.1052 \\
R9  & 0.1390 & 0.1137 \\
R10 & 0.2062 & 0.1896 \\
\addlinespace[0.5em]
\multicolumn{3}{@{}l}{\textit{Fold 3} \quad Loss 0.0044 \quad Adam \quad Lbk. 13} \\
\textbf{Ensemble} & 0.0560 & 0.0405 \\
R1  & 0.0644 & 0.0508 \\
R2  & 0.0597 & 0.0446 \\
R3  & 0.0585 & 0.0440 \\
R4  & 0.0624 & 0.0465 \\
R5  & 0.0632 & 0.0498 \\
R6  & 0.0951 & 0.0770 \\
R7  & 0.0569 & 0.0454 \\
R8  & 0.0601 & 0.0422 \\
R9  & 0.0734 & 0.0576 \\
R10 & 0.0753 & 0.0597 \\
\addlinespace[0.5em]
\multicolumn{3}{@{}l}{\textit{Fold 4} \quad  Loss 0.0018 \quad RMSprop \quad Lbk. 13} \\
\textbf{Ensemble} & 0.0669 & 0.0534 \\
R1  & 0.0678 & 0.0525 \\
R2  & 0.0836 & 0.0645 \\
R3  & 0.0640 & 0.0504 \\
R4  & 0.0756 & 0.0616 \\
R5  & 0.0690 & 0.0538 \\
R6  & 0.0664 & 0.0538 \\
R7  & 0.0781 & 0.0596 \\
R8  & 0.0718 & 0.0587 \\
R9  & 0.0857 & 0.0639 \\
R10 & 0.0695 & 0.0548 \\
\bottomrule
\end{tabular}
\end{table}

\subsection{Predictive Validity: Does the MPE Improve the ARIMA?} \label{sec:arimax-mpe}

To provide a complementary and model-independent test of whether the MPE index contains predictive information, we directly examine whether incorporating the MPE into a conventional linear forecasting model improves the ARIMA benchmark.
The design isolates the index as the only difference between the two models. The forecasting pipeline is run twice using the same AIC-selected order, estimation window, rolling one-step-ahead forecasting scheme and test weeks. The first specification is the ARIMA benchmark, while the second is an ARIMAX model that adds the one-week lagged MPE index as an exogenous regressor. The four disjoint test windows are subsequently stacked into a single out-of-sample series of 215 weeks.

Because the ARIMA model is nested within the ARIMAX specification, equal predictive accuracy is assessed using the Clark-West test rather than relying on the Diebold-Mariano test alone. The Clark-West adjustment accounts for the tendency of conventional forecast-comparison tests to favour the more parsimonious model when comparing nested specifications. The test is one-sided, with the null corresponding to equal predictive accuracy and the alternative that the ARIMAX specification provides improved forecasts.

As reported in Table~\ref{tab:arimax_mpe}, adding the MPE index produces an improvement in out-of-sample forecast accuracy. The pooled RMSE decreases from $0.0902$ for the ARIMA benchmark to $0.0895$ for the ARIMAX specification, corresponding to a $0.82\%$ reduction in RMSE and an out-of-sample $R^2$ of $1.62\%$ relative to the ARIMA benchmark. The Clark-West statistic is $2.071$, with a one-sided $p$-value of $0.019$, leading to rejection of equal predictive accuracy at the $5\%$ level. The corresponding Diebold--Mariano statistic is not significant ($p=0.644$), which is consistent with the fact that the models are nested and the Clark-West test is the appropriate primary comparison in this setting.

These results provide direct evidence that the MPE index contains incremental predictive information for Bitcoin returns beyond that captured by the ARIMA benchmark. The magnitude of the improvement is modest, as expected for a weekly sentiment-based measure, but the statistically detectable gain demonstrates that the predictive content of the MPE is not solely a consequence of the flexibility of the LSTM architecture. The ARIMAX results therefore provide an independent linear validation of the information content of the MPE, complementing the nonlinear and interaction-based evidence obtained from the LSTM--SHAP analysis.

\begin{table}[H]
\centering
\caption{Does Adding the MPE Improve the ARIMA?}
\label{tab:arimax_mpe}
\begin{small}
\caption*{\small \textbf{Notes:} Nested one-step-ahead forecast comparison over the
215 pooled out-of-sample weeks. Both models share the same AIC-selected order,
estimation window and rolling forecast scheme, and differ only by the lagged MPE index
$\text{mp\_exp}_{t-1}$ as an exogenous regressor; the one-week lag makes the regressor
observable at the forecast origin. The four disjoint test windows are stacked into a single
series, as individual windows of about 50 weeks are too short to be informative. Gain is the
percentage reduction in RMSE and $R^2_{OS}$ is computed against the ARIMA benchmark.
Clark--West is the appropriate test for nested models, where Diebold--Mariano is biased against
the larger model; it is one-sided against $H_1$: the ARIMAX forecasts better. Test statistics
are reported with $p$-values in parentheses.
$^{*}p<0.10$, $^{**}p<0.05$, $^{***}p<0.01$.}
\end{small}
\resizebox{0.8\textwidth}{!}{%
\begin{tabular}{l cccc}
\toprule
& RMSE ARIMA & RMSE ARIMAX & Gain (\%) & $R^2_{OS}$ (\%) \\
\midrule
\textit{Forecast accuracy} & 0.0902 & 0.0895 & +0.82 & +1.62 \\
\midrule
& \multicolumn{2}{c}{Clark--West} & \multicolumn{2}{c}{Diebold--Mariano} \\
\midrule
\textit{Equal predictive accuracy} & \multicolumn{2}{c}{$2.071^{**}$ (0.0192)} & \multicolumn{2}{c}{$0.462$ (0.6439)} \\
\bottomrule
\end{tabular}}
\end{table}

\end{document}